\documentclass[twocolumn]{aastex631}
\usepackage[utf8]{inputenc}

\usepackage[version=4]{mhchem}
\usepackage{booktabs}
\usepackage{comment}
\usepackage{tabularx}

\usepackage{siunitx}
\usepackage{xcolor}
\usepackage{amsmath}
\usepackage[flushleft]{threeparttable}

\DeclareSIUnit\angstrom{\text {Å}}

\begin{document}

\title{Astrochemical Modeling of Propargyl Radical Chemistry in TMC-1}

\author[0000-0002-4593-518X]{Alex N. Byrne}
\affiliation{Department of Chemistry, Massachusetts Institute of Technology, Cambridge, MA 02139, USA}

\author[0000-0003-2760-2119]{Ci Xue}
\affiliation{Department of Chemistry, Massachusetts Institute of Technology, Cambridge, MA 02139, USA}

\author[0000-0002-0850-7426]{Ilsa R. Cooke}
\affiliation{Department of Chemistry, University of British Columbia, 2036 Main Mall, Vancouver, BC V6T 1Z1, Canada}

\author[0000-0001-9142-0008]{Michael C. McCarthy}
\affiliation{Center for Astrophysics \textbar{} Harvard \& Smithsonian, Cambridge, MA 02138, USA}

\author[0000-0003-1254-4817]{Brett A. McGuire}
\affiliation{Department of Chemistry, Massachusetts Institute of Technology, Cambridge, MA 02139, USA}
\affiliation{National Radio Astronomy Observatory, Charlottesville, VA 22903, USA}

\correspondingauthor{Alex N. Byrne}
\email{lxbyrne@mit.edu}

\begin{abstract}
Recent detections of aromatic species in dark molecular clouds suggest formation pathways may be efficient at very low temperatures and pressures, yet current astrochemical models are unable to account for their derived abundances, which can often deviate from model predictions by several orders of magnitude. The propargyl radical, a highly abundant species in the dark molecular cloud TMC-1, is an important aromatic precursor in combustion flames and possibly interstellar environments. We performed astrochemical modeling of TMC-1 using the three-phase gas-grain code \texttt{NAUTILUS} and an updated chemical network, focused on refining the chemistry of the propargyl radical and related species. The abundance of the propargyl radical has been increased by half an order of magnitude compared to the previous GOTHAM network. This brings it closer in line with observations, but it remains underestimated by two orders of magnitude compared to its observed value. Predicted abundances for the chemically related \ce{C4H3N} isomers within an order of magnitude of observed values corroborate the high efficiency of CN addition to closed-shell hydrocarbons under dark molecular cloud conditions. The results of our modeling provide insight into the chemical processes of the propargyl radical in dark molecular clouds and highlight the importance of resonance-stabilized radicals in PAH formation.
\end{abstract}

\section{Introduction}

Polycyclic aromatic hydrocarbons (PAHs) are the probable carriers of the unidentified infrared emission bands pervasive in our and other galaxies \citep{tielens_interstellar_2008}, as well as potentially some of the diffuse interstellar bands that can be observed in the infrared, visible, and ultraviolet (UV) spectra of the interstellar medium (ISM) \citep{duley_polycyclic_2006}. Despite the ubiquity of these molecules in the ISM, their formation pathways remain ambiguous. One suggested formation route is through high-temperature growth processes in the circumstellar envelopes of evolved stars followed by injection into the ISM \citep{kaiser_aromatic_2021, tielens_interstellar_2008}. However, in the diffuse ISM and photodissociation regions, carbonaceous material on dust grains and in the gas phase is exposed to UV photon irradiation and shocks \citep{berne_top-down_2015}. It has been assumed that small PAHs of less than $20-30$ atoms often cannot internally redistribute energy from the absorption of UV photons in such regions or re-radiate it back out before they are destroyed, resulting in short lifetimes \citep{chabot_coulomb_2019}. Recent observations of the Taurus Molecular Cloud (TMC-1) reported the detection of benzonitrile \citep{mcguire_detection_2018}, the first six-membered aromatic ring detected via radio astronomy, followed by detection of the small PAHs 1- and 2-cyanonaphthalene, indene, and 2-cyanoindene \citep{mcguire_detection_2021,burkhardt_discovery_2021,sita_discovery_2022}. While the discovery of aromatic molecules in dark molecular clouds appears to imply the existence of efficient low-temperature formation routes, current astrochemical models dramatically fail to reproduce their observed abundances. This gap in our understanding of aromatic chemistry highlights the need for further exploration of the formation and chemical evolution of these molecules. 

\begin{figure*}
    \centering
    \includegraphics[scale=0.75]{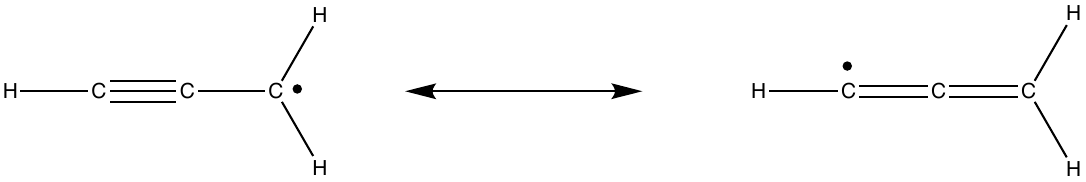}
    \caption{Resonance structures of \ce{CH2CCH}.}
    \label{fig:propresonance}
\end{figure*}

Several theoretical and experimental studies have proposed gas-phase formation pathways to benzene (\ce{C6H6}) from small hydrocarbons \citep{jones_formation_2011, caster_kinetic_2019, hebrard_photochemical_2006}.
In combustion flames, benzene, as well as the phenyl radical, are suggested to be efficiently formed from the recombination of two propargyl radicals (\ce{CH2CCH}) \citep{miller_kinetic_1992} 
\begin{equation}
    \label{eqa:ch2cch-ch2cch}
        \ce{CH2CCH + CH2CCH -> C6H6}/\ce{C6H5 + H}.
\end{equation}
\ce{CH2CCH} is a small, resonance-stabilized hydrocarbon radical (as shown in Figure~\ref{fig:propresonance}) that can form in flames from the insertion of \ce{\textsuperscript{1}CH2} into the C-H bond of \ce{C2H2}, where \ce{\textsuperscript{1}CH2} denotes an electronically excited methylene radical. \citep{miller_kinetic_1992}. Computational studies of the \ce{C6H6} potential energy surface have shown that two propargyl radicals can barrierlessly combine and undergo decomposition to form o-benzyne (\ce{c-C6H4}) + \ce{H2} or phenyl radical (\ce{C6H5}) + \ce{H}, with the latter pathway expected to be the dominant bimolecular pathway \citep{miller_recombination_2003}. Experimental studies at high temperatures have revealed the formation of o-benzyne as well as the \ce{C6H6} isomers benzene, fulvene, 1,5-hexadiyne, and 2-ethynyl-1,3-butadiene \citep{zhao_gas-phase_2021}. The product branching ratios were found to heavily depend on temperature and pressure, providing critical information about formation of aromatic species in combustion flames and in dense environments such as Titan’s atmosphere \citep{zhao_gas-phase_2021}. In contrast, the \ce{C6H5 + H} channel is expected to prevail in cold and less dense environments such as TMC-1,  whereas the \ce{c-C6H4} channel is expected to be negligible. While previous astrochemical studies have placed an emphasis on benzene as the first aromatic ring formed en route to PAH formation {\citep{jones_formation_2011, caster_kinetic_2019}}, these studies of \ce{CH2CCH} recombination indicate that the phenyl radical may be {equally} important. In order to better model and understand aromatic chemistry of interstellar environments, a greater emphasis must be placed on alternative formation pathways involving the \ce{C6H5}, beginning with \ce{CH2CCH} recombination.

The detection of \ce{CH2CCH} in TMC-1 at millimeter wavelengths has further spurred astrochemical interest in the molecule \citep{agundez_discovery_2021,agundez_detection_2022}. An observed abundance of $1.0\times10^{-8}$ with respect to molecular hydrogen makes this one of the most abundant radicals in TMC-1, with a \ce{CH3CCH}/\ce{CH2CCH} ratio about equal to one \citep{agundez_detection_2022}. In addition to the detection of \ce{CH2CCH}, the authors performed an astrochemical modelling study of the radical and its closed-shell counterparts. They concluded that the major contributions to the production of \ce{CH2CCH} include the neutral-neutral reaction between ethylene (\ce{C2H4}) and atomic carbon in the gas phase, 
\begin{equation}
    \label{eqa:c2h4-c}
        \ce{C2H4  +   C  -> CH2CCH + H},
\end{equation}
and the dissociative recombination (DR) reactions of 3-carbon cations, 
\begin{equation}
    \label{eqa:c3hn}
        \ce{C3H$_n$+ + \textit{e}- -> CH2CCH + $(n-3)$H},
\end{equation}
where $n=$ 4, 5, and 6. The major destruction pathways are through reactions with neutral atoms such as \ce{C}, \ce{O}, and \ce{N}. While these models reproduce the observed abundance of the closed-shell species \ce{CH3CCH} well, the modeled propargyl abundance is underestimated by over two orders of magnitude. The presence of low-temperature kinetic measurements for a number of relevant neutral-neutral reactions is a significant benefit toward understanding the chemistry of this radical under molecular cloud conditions \citep{slagle_kinetics_1991, canosa_reactions_1997,  chastaing_neutralneutral_1999, chastaing_rate_2000, canosa_experimental_2007}, but the sizeable difference between observed and modeled abundances indicates that existing gas-phase reactions may need to be re-evaluated. In particular, experimental studies of the reactions between \ce{CH2CCH} and neutral atoms at low temperatures are rare. Moreover, a main limitation of the astrochemical model presented in \citet{agundez_discovery_2021} is the lack of consideration for grain chemistry. \citet{2016MolAs...3....1H} revealed that surface hydrogenation of 3-carbon hydrocarbons on interstellar grains is critical for the formation of \ce{CH2CCH} and its closed shell counterparts. 

The high abundance of \ce{CH2CCH} despite its open-shell nature and its close connection to aromatic species highlights it as an important species in the chemical evolution of molecular clouds, yet previous astrochemical knowledge on this species was incomplete. Detailed kinetic studies at low temperature provide the most relevant information for modeling the chemistry of dark molecular clouds and other interstellar regions \citep{hebrard_how_2009}, but the sheer number of possible reactions and difficulty in replicating interstellar conditions makes this a formidable problem. In order to investigate the chemical evolution of \ce{CH2CCH} in the ISM, we performed astrochemical modeling under dark cloud conditions with an updated and expanded chemical network. In Section \ref{sec:Model}, we describe the modeling code and physical conditions used as well as the modifications made to the chemical network. The results of the modeling efforts toward \ce{CH2CCH} and related species are presented in Section \ref{sec:Results}. Finally, we discuss these results in Section \ref{sec:Discussion} along with comparisons to previous works and future directions.

\section{Methods}
\label{sec:Model}

The astrochemical modeling code utilized is \texttt{NAUTILUS}, a three-phase gas-grain rate equation model \citep{ruaud_gas_2016}. The gas phase, grain surface, and grain mantle are treated individually with their own parameters and chemical reactions. Additional physical processes of diffusion between grain surface and mantle, diffusion within grain surface and mantle, adsorption of gas-phase species onto the grain, and desorption of grain surface species into the gas phase {are} considered. {Additionally, the \texttt{NAUTILUS} model considers photodissociation and photodesorption by external UV photons and cosmic-ray induced UV photons, thermal evaporation, evaporation via cosmic-ray stochastic heating, and chemical desorption. We used the default 1\% efficiency for the chemical desorption mechanism and peak grain temperature of 70 K for $1\times10^{-5}$ years for the cosmic ray heating mechanism. More information on these mechanisms and their implementations can be found in the \texttt{NAUTILUS} documentation (https://forge.oasu.u-bordeaux.fr/LAB/astrochem-tools/pnautilus/-/blob/master/pnautilus\_documentation.pdf) as well as \citet{ruaud_gas_2016} and the references therein.} 

The physical conditions chosen were those of a standard cold molecular cloud, namely a gas density of \SI{2e4}{\per\cubic\centi\metre} \citep{snell_determination_1982}, a kinetic temperature of 10 K for both gas and dust \citep{pratap_study_1997}, a cosmic ray ionization rate of \SI{1.3e-17}{\per\second} \citep{spitzer_heating_1968}, and a visual extinction of 10 mag \citep{rodriguez-baras_gas_2021}. The initial elemental abundances used, which are listed in Table \ref{tab:init_abundances} along with their references, were the low metal abundances of \citet{graedel_kinetic_1982} with a few changes. In particular, the ratio of initial carbon abundance to initial oxygen abundance (C/O ratio) has been found to have a noticeable effect on the modeled chemistry, with long carbon chains being the most affected \citep{hincelin_oxygen_2011}. It has been found that a C/O ratio of 1.1 due to a lowered initial abundance of atomic oxygen significantly improves the general agreement between modeled and observed abundances of carbon-chain molecules in TMC-1 \citep{xue_detection_2020, loomis_investigation_2021}. {Additionally, a depletion factor of 20 with respect to the solar abundance for sulfur has been used according to a recent study based on the GEMS survey \citep{fuente_gas_2023}.}

\begin{table*}[htb]
    \centering
    \caption{\textbf{Initial atomic/molecular abundances with reference to hydrogen}}
    \label{tab:init_abundances}
    \centering
    \begin{tabular}{@{} l @{} c c c c c}
    \toprule
    Species & $n/n_H$ & Reference & Species & $n/n_H$ & Reference \\
    \midrule
    \ce{H2} & 0.499 & \citet{ruaud_gas_2016} & \ce{F} & \num{6.68e-9} & \citet{neufeld_chemistry_2005} \\
    \ce{He} & 0.09 & \citet{wakelam_polycyclic_2008} & \ce{Cl} & \num{1.00e-9} & \citet{graedel_kinetic_1982} \\
    \ce{C} & \num{1.70e-4} & \citet{jenkins_unified_2009} & \ce{Si} & \num{8.00e-9} & \citet{graedel_kinetic_1982} \\
    \ce{N} & \num{6.20e-5} & \citet{jenkins_unified_2009} & \ce{Na} & \num{2.00e-9} & \citet{graedel_kinetic_1982}\\
    \ce{O} & \num{1.55e-4} & \citet{xue_detection_2020} & \ce{Mg} & \num{7.00e-9} & \citet{graedel_kinetic_1982} \\
    \ce{S} & {\num{7.50e-7}} & {\citet{fuente_gas_2023}} & \ce{Fe} & \num{3.00e-9} & \citet{graedel_kinetic_1982} \\
    \ce{P} & \num{2.00e-10} & \citet{graedel_kinetic_1982} & & & \\
    \bottomrule
    \end{tabular}
\end{table*}

Based on kida.uva.2014 \citep{wakelam_2014_2015}, the chemical network was extended to include aromatic and carbon-chain networks in the GOTHAM (Green Bank Telescope Observations of TMC-1: Hunting for Aromatic Molecules) project \citep{mcguire_detection_2018, xue_detection_2020,mcguire_early_2020, mccarthy_interstellar_2021, mcguire_detection_2021, loomis_investigation_2021}. To compare and contrast, we present the model results of \ce{CH2CCH} and its related species with both the default network used in previous GOTHAM analyses (without any of the changes new to this work, hereafter referred to as GOTHAM DR1) and a modified network, as detailed below \footnote{A Zenodo repository containing the input files for our updated network can be found here: https://zenodo.org/record/8320140.}

\subsection{Formation Mechanisms of \ce{CH2CCH}}

To better represent the chemical evolution of \ce{CH2CCH} and its closed shell counterparts in cold molecular clouds, an extensive literature search and analysis was performed to identify new reactions and to examine existing reactions. In general, low temperature and density laboratory kinetic studies provide the most accurate rate information for astrochemical models \citep{hebrard_how_2009}. In the absence of these, theoretical data and/or high-temperature experimental data are taken into account with caution when extrapolating high-temperature kinetic data to low temperatures such as 10 K. For example, rate constants estimated from high-temperature data could be inaccurate by multiple orders of magnitude if there is a U-shaped temperature dependence \citep{jimenez_first_2016}. Since previous astrochemical models underestimated the peak abundance of \ce{CH2CCH}, a special focus was placed on the production mechanisms of this species.

\subsubsection{Dissociative Recombination}
\label{subsubsec:DR}

DR reactions of molecular cations with electrons are essential to astrochemistry due to their high efficiency under low-temperature and low-density conditions \citep{geppert_dissociative_2004,florescu-mitchell_dissociative_2006}. In our current chemical model, we considered four different species that produce \ce{CH2CCH} via dissociative recombination (DR), namely the DR reactions of \ce{C3H4+}, \ce{C3H5+}, \ce{C5H5+}, and \ce{C4H7+}. The \ce{C3H4+} DR reaction was investigated experimentally by \citet{geppert_dissociative_2004}, finding a large preference for the formation of \ce{CH2CCH}. The branching ratios determined from this experiment were adopted here, along with an overall rate constant from \citet{florescu-mitchell_dissociative_2006}. \citet{angelova_branching_2004} studied the DR reaction of \ce{C3H5+} and found that products containing \ce{C3} species account for 86.7\% of products, with \ce{C2}-containing products accounting for the remaining 13.3\%. The individual branching ratios and overall rate constant were determined by \citet{loison_interstellar_2017} using information from photo-dissociation studies of the neutral species. The last two species, \ce{C5H5+} and \ce{C4H7+}, have not received experimental study and have been assumed to produce \ce{CH2CCH} with branching ratios of 50\% and 65\% respectively {based on enthalpies of reaction}. A full list of changes to dissociative recombination reactions implemented in our updated model is shown in {Table~\ref{tab:DR} in the Appendix}.

\subsubsection{Ion-Neutral Reactions}
\label{subsec:ion-mol}
In addition to DR reactions, {reactions} between ions and neutral molecules account for a significant portion of interstellar chemistry due to relatively large reaction rates, the absence of activation energies, and a variety of abundant interstellar ions \citep{friedman_ion-molecule_1968}. Rate constants for ion-molecule reactions can be estimated according to capture theories such as {Equations~\ref{eqa:x} - \ref{eqa:ionpol2}, the Su-Chesnavich formulae \citep{su_parametrization_1982, woon_quantum_2009}. 

\begin{equation}
    x = \dfrac{\mu_D}{\sqrt{2 \alpha k_B T}}
    \label{eqa:x}
\end{equation}

\begin{equation}
    k_D = 2\pi{}e\sqrt{\dfrac{\alpha}{\mu}}\left(0.4767x + 0.6200\right)
    \label{eqa:ionpol1}
\end{equation}

\begin{equation}
    k_D = 2\pi{}e\sqrt{\dfrac{\alpha}{\mu}}\left(\dfrac{\left(x + 0.5090\right)^{2}}{10.526} + 0.9754\right)
    \label{eqa:ionpol2}
\end{equation}

These expressions are an empirical fit to classical trajectory calculations where {$k_D$ is the ion-neutral rate constant,} $\mu_D$ is the dipole moment, $\alpha$ is the polarizability, $k_B$ is the Boltzmann constant, $e$ is the charge of an electron, and $\mu$ is the reduced mass. The parameter $x$ is a unitless value that determines which expression is used for calculation of a rate coefficient. When $x\geq 2$, $k_D$ increases linearly with $x$ and Equation~\ref{eqa:ionpol1} is used. When $x<2$, the relationship between $k_D$ and $x$ is quadratic and Equation~\ref{eqa:ionpol2} is used. If $x=0$, the expression is reduced to the Langevin rate,
\begin{equation}
\label{eqa:langevin}
    k_L = 2\pi{}e\sqrt{\dfrac{\alpha}{\mu}},
\end{equation}
the classical rate constant for the reaction between an ion and a non-polar neutral molecule. For many polar molecules, this value is greater than 2 over the temperature range of interest ($10-300~\rm{K}$). For \ce{CH2CCH}, however, $x<2$ over this temperature range due to its small dipole moment of 0.150 D \citep{kupper_free_2002}.

In our reaction network, three neutral species, namely \ce{CNCH2CCH}, \ce{CNCHCHCCH}, and \ce{CNCHCCH}, can generate \ce{CH2CCH} via reaction with abundant cations. For \ce{CNCH2CCH} and \ce{CNCHCHCCH}, it was assumed that reactions with \ce{H+}, \ce{H3+}, \ce{He+}, \ce{C+}, \ce{HCO+}, and \ce{H3O+} all form \ce{CH2CCH}. For \ce{CNCHCCH}, only reactions with \ce{H+} and \ce{H3+} were considered to contribute to the formation of \ce{CH2CCH}, as \ce{He+} and \ce{C+} do not contain hydrogen and reactions with \ce{HCO+} and \ce{H3O+} were assumed to produce \ce{l-C3H2} instead. Ion-neutral rate coefficients were re-evaluated for each of these species using Equation~\ref{eqa:ionpol1} and the $\mu_D$ and $\alpha$ values in Table~\ref{tab:dipoles}. For species without experimental or calculated polarizabilities, the rate constant was estimated by adding a negative temperature dependence to the polarizability-independent term. The results of these calculations are listed in {Table~\ref{tab:Ion-neutral2} in the Appendix}. 

In our network, \ce{CNCH2CCH} and \ce{CNCHCHCCH} can be formed directly or indirectly from reactions involving \ce{CH2CCH}. In order to ensure that these ion-neutral reactions were not generating a loop of net-zero loss and arbitrarily inflating the abundance of \ce{CH2CCH}, we tested the effect of these reactions on modeled abundances by varying only the products. These alternate products were taken from \citet{quan_possible_2007}.

\begin{table*}[htb]
    \centering
    \caption{\textbf{Polarizabilities and Dipole Moments for Updated Ion-Neutral Reactions}}
    \label{tab:dipoles}
    \begin{threeparttable}
    \centering
    \begin{tabular}{@{} l @{} c c c}
    \toprule
    Molecule & Polarizability (\unit{\cubic\angstrom}) & Dipole Moment (D) & Reference\tnote{1} \\
    \midrule
    \ce{CH2CCH} & 5.556 & 0.150 & \citet{woon_quantum_2009}, \citet{kupper_free_2002} \\
    \ce{CNCH2CCH} & 7.175 & 3.61 & \citet{woon_quantum_2009}, \citet{mcnaughton_microwave_1988} \\
    \ce{CH2CCHCN} & 8.136 & 4.28 & \citet{woon_quantum_2009}, \citet{bouchy_microwave_1973} \\
    \ce{CNCHCHCCH} &  & 5.00 & \citet{lee_discovery_2021} \\
    \ce{CNCHCCH} &  & 5.47 & \citet{cabezas_laboratory_2022} \\
    \bottomrule
    \end{tabular}

    \begin{tablenotes}
    \item[1] References for polarizability are listed first, followed by references for dipole moment, unless there is no polarizability value.
    \end{tablenotes}

    \end{threeparttable}
\end{table*}

\subsubsection{Neutral-neutral reactions}

While collisions between two neutral molecules may not be as efficient at low temperatures due to a lack of strong attractive forces and the possibility of energy barriers, it has been shown that some of these reactions play a crucial role in the chemistry of molecular clouds \citep{chastaing_neutralneutral_1999, carty_low_2001}.
The reaction
\begin{equation}
    \label{eqa:ch-c2h4}
        \ce{CH + C2H4 -> CH3CCH + H}
\end{equation}
has been studied experimentally by \citet{goulay_cyclic_2009} using time-resolved vacuum ultraviolet ionization mass spectrometry at room temperature, finding that 30\% of the detected products were \ce{CH3CCH}. Kinetic studies of this reaction were performed by \citet{canosa_reactions_1997} between $23$ and $298~\rm{K}$ using the CRESU experimental setup. This technique, developed by \citet{rowe_study_1998}, uses Laval nozzles to create a uniform flow of gas at temperatures as low as $10~\rm{K}$ without precooling, making it ideal for studying the kinetics of astrochemical reactions \citep{cooke_experimental_2019, hays_collisional_2022}. We therefore updated this reaction to reflect the experimental branching ratio and $23~\rm{K}$ rate. Additionally, the radical-radical reaction 
\begin{equation}
    \label{eqa:cch-ch3}
        \ce{CCH + CH3 -> CH2CCH + H}
\end{equation}
was added to the network with a rate estimated by \citet{loison_interstellar_2017}. Finally, we have incorporated the rate for the reaction
\begin{equation}
    \label{eqa:c2-ch4}
        \ce{C2 + CH4 -> CH2CCH + H}
\end{equation}
as measured experimentally at $24\rm~{K}$ by \citet{canosa_experimental_2007}, again using the CRESU technique. The rate coefficients are listed in {Tables~\ref{tab:Neutral-neutral1} and \ref{tab:Neutral-neutral2} in the Appendix}.

\subsection{Destruction Mechanisms of \ce{CH2CCH}}

For a full chemical description of the \ce{CH2CCH} and its counterparts, it is also necessary to investigate the relevant destruction mechanisms. As with the production mechanisms, an extensive literature search was performed with a priority toward low-temperature kinetic experiments.

\subsubsection{Ion-neutral reactions}

Analogous to many other polar neutral species discussed in Section~\ref{subsec:ion-mol}, \ce{CH2CCH} is assumed to be destroyed mainly by reacting upon collision with cations in dense molecular clouds. The reaction rates are well-described by the Su-Chesnavich formalism. In addition to abundant cations such as \ce{H3+} and \ce{HCO+}, a range of complex hydrocarbon cations such as \ce{C2H2+} and \ce{C3H4+} were also considered. These rates have been recalculated as required using Equation \ref{eqa:ionpol2} with the dipole and polarizability values from Table \ref{tab:dipoles} and are summarized in {Table~\ref{tab:Ion-neutral} in the Appendix}.

\subsubsection{Neutral-neutral reactions}
\label{sec:neutral-neutral-rxn}
In addition to cations, \ce{CH3CCH} and \ce{CH2CCH} can be destroyed by reaction with atoms and small molecules at low temperatures. The reaction between \ce{CH3CCH} and \ce{C} has been studied at room temperature by \citet{loison_reaction_2004} with a fast-flow reactor and resonance fluorescence and down to $15~\rm{K}$ by \citet{chastaing_neutralneutral_1999} with the CRESU apparatus. \citet{loison_reaction_2004} observed a hydrogen production ratio of 85\% corresponding to the \ce{H + C4H3} product channel, with the remaining 15\% most likely attributed to the \ce{H2 + C4H2} channel. These branching ratios, along with the $15~\rm{K}$ reaction rate, were adopted into our chemical network.

For \ce{CH2CCH}, reactions with \ce{H}, \ce{O}, \ce{N}, and \ce{OH} are of interest. Theoretical studies of the \ce{C3H4} potential energy surface have shown the presence of activation barriers for the reactions
\begin{equation}
    \label{eqa:c-C3H2}
        \ce{CH2CCH + H -> c-C3H2}\ce{ + H2}
\end{equation} and 
\begin{equation}
    \label{eqa:l-C3H2}
        \ce{CH2CCH + H -> l-C3H2}\ce{ + H2}.
\end{equation}
As such, they are unlikely to be viable mechanisms under cold dense cloud conditions \citep{miller_multiple-well_2003}. The radiative-association reaction 
\begin{equation}
    \label{eqa:h-ch2cch}
        \ce{CH2CCH + H -> CH3CCH + h$\nu$},
\end{equation}
however, is barrierless and feasible under interstellar conditions \citep{hebrard_photochemistry_2013}. As such, the rate and branching ratio for this reaction have been updated according to \citet{loison_interstellar_2017}. The \ce{CH2CCH + O} reaction has been studied from $295-750~\rm{K}$ using a heatable flow reactor and photoionization mass spectrometry, but did not show any temperature dependence in this range \citep{slagle_kinetics_1991}. The branching ratios for the products have been assigned based on theoretical studies \citep{loison_interstellar_2017}. Due to a lack of experimental or theoretical data, the rate and branching ratios for the reaction of \ce{CH2CCH} with \ce{N} has been estimated based on the \ce{N + C2H3} reaction. \citep{loison_interstellar_2017}. For the \ce{CH2CCH + OH} reaction, we use a temperature-independent estimate from {\citet{hansen_isomer-specific_2009}} for the overall rate, with assumed 50/50 branching ratios between the two product channels. These reactions have all been implemented in our updated chemical network and {are} listed with the rate coefficients in {Tables~\ref{tab:Neutral-neutral1} and \ref{tab:Neutral-neutral2} in the Appendix}.

The self-recombination between propargyl radicals, Reaction~\ref{eqa:ch2cch-ch2cch}, has been the subject of multiple kinetic studies at room temperature and above. \citet{atkinson_rate_1999}, \citet{fahr_kinetics_2000}, and \citet{desain_infrared_2003} have all measured the rate constant of this reaction at room temperature and for pressures of $2.25-100$ Torr, 50 Torr, and 16 Torr respectively. The results of these three experiments agree well upon a room-temperature rate constant of $\sim$\SI{4.0e-11}{\cubic\centi\metre\per\second} at the high pressure limit. There has not yet been any experimental low-temperature study of this propargyl recombination, however a theoretical study by \citet{georgievskii_association_2007} used variable reaction coordinate transition state theory (VRC-TST) to determine the temperature dependence of this reaction in the high-pressure limit. These calculations agree well with the aforementioned experiments and suggest a slight negative temperature dependence, leading to a rate constant that is only slightly larger ($\sim$\SI{6.0e-11}{\cubic\centi\metre\per\second}) at 10 K. We adopt the experimental value of $4\times10^{-11}$ into our network as an estimate. The branching ratios under conditions relevant to TMC-1 are not known either, however, \citet{miller_recombination_2003} estimated that \ce{C6H5 + H} is the dominant bimolecular exit channel due to a barrierless reverse association and low energy transition state, contrary to the \ce{C6H4 + H2} exit channel. In our network, we assumed that \ce{C6H5 + H} are the only products of this reaction, and thus implement the reaction

\begin{equation}
    \label{eqa:ch2cch-c6h5}
        \ce{CH2CCH + CH2CCH -> C6H5 + H}.
\end{equation}

\subsection{Incorporation of \ce{CH2CCH2}}

Of the two linear closed-shell isomers of \ce{C3H4}, \ce{CH3CCH} and \ce{CH2CCH2}, only the former species was previously considered in the chemical network. As with \ce{CH3CCH}, \ce{CH2CCH2} is a closed-shell counterpart of \ce{CH2CCH}, and thus these species are likely linked chemically. Our model has been updated to produce \ce{CH2CCH2} through three reaction paths, including the DR reaction
\begin{align}
    \label{eqa:C3H5+_CH2CCH2}
        \ce{C3H5+ + e- & -> CH2CCH2 + H},
\end{align}
and the neutral-neutral reactions 
\begin{equation}
    \label{eqa:CH2CCH_CH2CCH2}
        \ce{H + CH2CCH -> CH2CCH2 + Photon}
\end{equation}
and 
\begin{equation}
    \label{eqa:C2H4_CH2CCH2}
        \ce{C2H4 + CH -> CH2CCH2}\ce{ + H}
\end{equation}
\citep{goulay_cyclic_2009,loison_interstellar_2017, canosa_reactions_1997}. This latter reaction also has a \ce{CH3CCH + H} product channel that was already included in the network. For the radiative-association of \ce{CH2CCH} with H, we assume 50/50 branching ratios for the \ce{C3H4} due to the lack of low-temperature information, in line with \citep{loison_interstellar_2017}.

Destruction pathways of \ce{CH2CCH2} include reaction with atomic carbon, as measured down to $15~\rm{K}$ by \citet{chastaing_neutralneutral_1999}, reaction with \ce{CCH} as measured down to $63~\rm{K}$ by \citet{carty_low_2001}, and reaction with \ce{CH} as measured down to $77~\rm{K}$ by \citet{daugey_kinetic_2005}. Additional ion-neutral destruction pathways have also been included \citep{loison_interstellar_2017}. Some interstellar species are capable of isomerization with assistance from collisions with \ce{H} and \ce{H2}. However, the inter-conversion of the two \ce{C3H4} species has barriers in both directions of a few kcal/mol, enough to make them negligible under TMC-1 conditions \cite{narendrapurapu_combustion_2011}.

\subsection{Reactions with \ce{CN} radicals}

\begin{figure*}
    \centering
    \includegraphics[width=.3\textwidth]{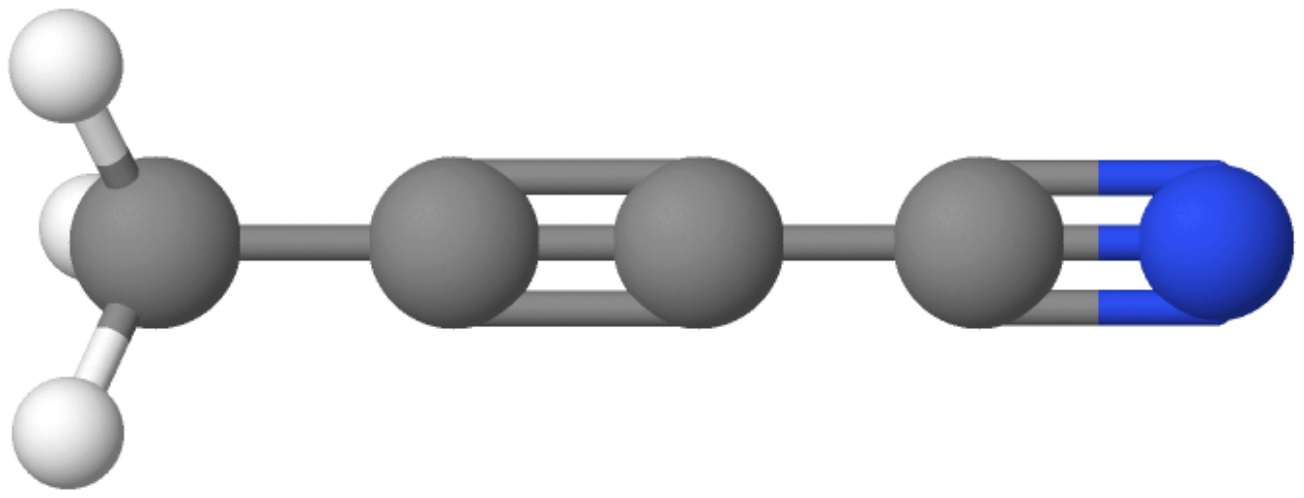}\hfill
    \includegraphics[width=.3\textwidth]{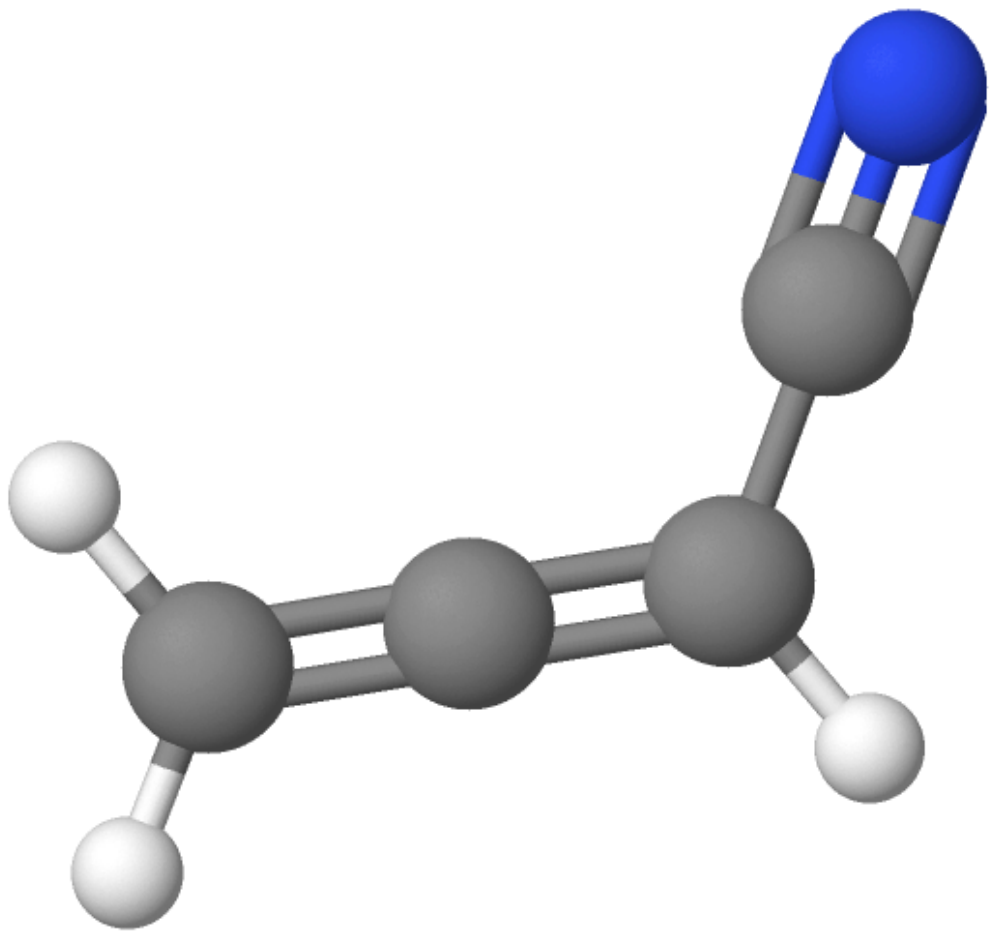}\hfill
    \includegraphics[width=.3\textwidth]{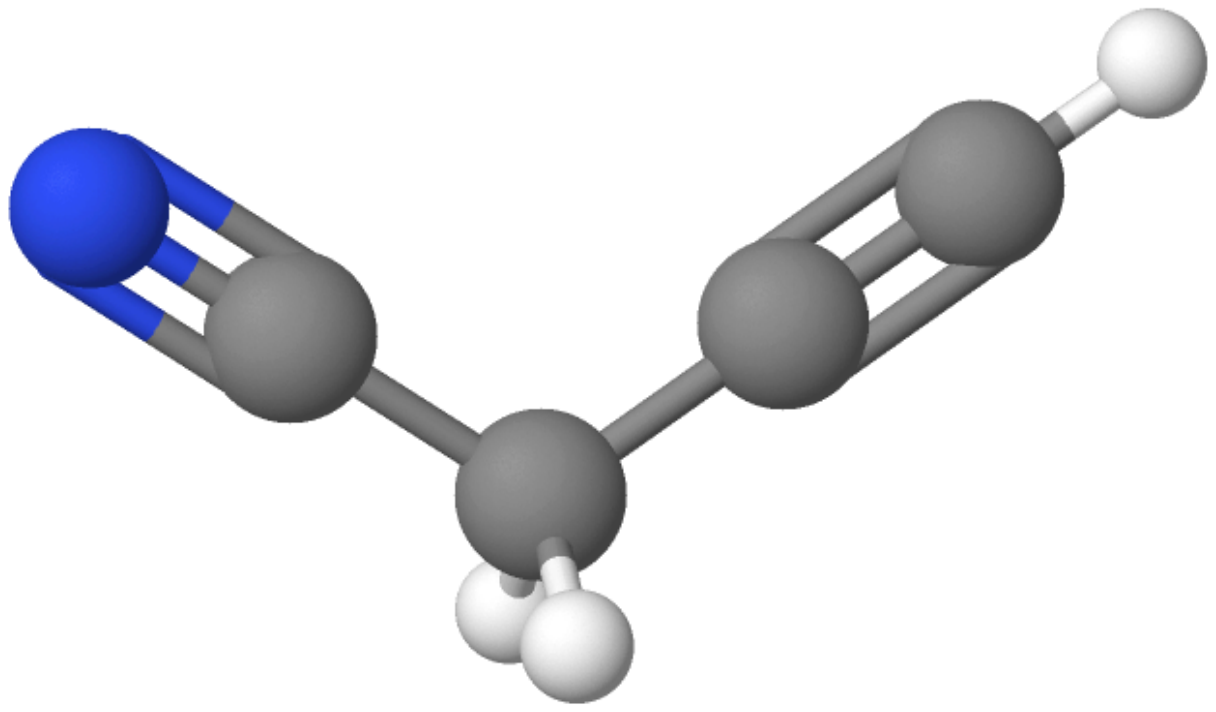}\hfill
    \caption{Chemical structures of the three isomers of \ce{C4H3N} that have been detected in TMC-1: methylcyanoacetylene (left), cyanoallene (middle), and propargyl cyanide (right).}
    \label{fig:C4H3N_isomers}
\end{figure*}

Laboratory studies at low temperature and density suggest that the formation of CN-substituted hydrocarbons via reaction with CN radicals is efficient under dark molecular cloud conditions \citep{carty_low_2001,cooke_benzonitrile_2020}. There are three isomers of \ce{C4H3N}, propargyl cyanide (\ce{CNCH2CCH}), cyanoallene (\ce{CH2CCHCN}), and methylcyanoacetylene (\ce{CH3C3N}) as shown in Figure~\ref{fig:C4H3N_isomers}, all of which have been detected in TMC-1 \citep{broten_detection_1984,lovas_hyperfine_2006,mcguire_early_2020}. In particular, \citet{mcguire_early_2020} assumed \ce{CNCH2CCH} to be formed as the major product from the reaction between \ce{CN} and \ce{CH3CCH}. However, \citet{balucani_formation_2002} estimated a 50/50 branching ratio of \ce{CH2CCHCN}/\ce{CH3C3N} based on crossed molecular beam experiments and electronic structure calculations. Additionally, \citet{abeysekera_product_2015} investigated this reaction via chirped-pulse/uniform flow microwave spectroscopy and further confirmed that \ce{CH3C3N} is the only \ce{C4H3N} isomer formed, along with \ce{HC3N} and \ce{CH2CCH},
\begin{align}
    \label{eqa:CN+CH3CCH}
        \ce{CN + CH3CCH & -> HC3N + CH3} & 66\% \nonumber \\
                        &\ce{ -> CH3C3N + H} & 22\% \nonumber \\
                        & \ce{-> HCN + CH2CCH.} & 12\%
\end{align} 
We, therefore, modified this reaction based on the branching ratios from \citet{abeysekera_product_2015} and the rate from \citet{carty_low_2001}.

The reaction between \ce{CN} and \ce{CH2CCH2} can also form \ce{C4H3N} isomers. Using the same procedure as the \ce{CN + CH3CCH} reaction, \citet{balucani_formation_2002} suggested that \ce{CH2CCHCN} is the major product of this reaction, with \ce{CNCH2CCH} being the minor product,
\begin{align}
    \label{eqa:CN+CH2CCH2}
        \ce{CN + CH2CCH2 & -> CH2CCHCN + H} & 90\% \nonumber \\
                         &\ce{ -> CNCH2CCH + H.} & 10\% 
\end{align}
This reaction was introduced to our network with the above branching ratios from \citet{balucani_formation_2002} and the rate from \citet{carty_low_2001}. As \ce{CH2CCHCN} had previously not been included in our network, destruction pathways via reaction with abundant interstellar ions were included analogous to \ce{CNCH2CCH}. The rates for these reactions were calculated using Equation~\ref{eqa:ionpol1} and the dipole and polarizability in Table \ref{tab:dipoles}. As with \ce{CNCH2CCH}, we tested alternate ion-neutral products from \citet{quan_possible_2007} to avoid arbitrarily inflating the abundance of \ce{CH2CCH} and related species.

There is limited information on the reaction between \ce{CN} and \ce{CH2CCH}. Although \citet{cabezas_discovery_2021} proposed 3-cyano propargyl radical (\ce{CH2C3N}) as the main product, radical-radical reactions involving \ce{CN} may not follow the pattern of CN addition, H elimination being the main product channel \citep{decker_2003}. In contrast, the recombination channel seems to be more likely. A recent theoretical study found that radiative association reactions between radical species can proceed rapidly at low temperatures, such as that of TMC-1 \citep{tennis_radiative_2021}. As such, the reaction
\begin{equation}
    \label{eqa:CN+CH2CCH}
        \ce{CN + CH2CCH -> CNCH2CCH + h$\nu$}
\end{equation}
has been added, with \ce{CNCH2CCH} assumed as the sole product and an estimated rate constant of \SI{1.0e-10}{\cubic\centi\metre\per\second}, in line with the calculated rate constants for the \ce{CH3 + CH3O} reaction \citep{tennis_radiative_2021}. It is likely that the CN-addition, H-elimination channel to form cyano propargyl radicals is in competition, however the cyano propargyl radicals are not of direct concern to this study. Instead we are more interested in determining whether the inclusion of this reaction can reproduce the observed abundance of \ce{CNCH2CCH}. More experimental and theoretical work is necessary to confirm whether the closed-shell \ce{CNCH2CCH} can form in this reaction.

\subsection{Grain Chemistry}

In addition to the various gas-phase chemistry, grain-surface processes, in particular, hydrogenation reactions, are also considered in cold interstellar regions. Even at 10 K, light \ce{H} atoms can move efficiently on grain surfaces and reactions with activation barriers can occur through quantum mechanical tunneling. \citet{hickson_methylacetylene_2016} and \citet{loison_interstellar_2017} investigated the chemistry of \ce{C3H_n} species in cold dense clouds, including hydrogenation of these species on grain surfaces. These authors suggested that such hydrogenation processes can begin with \ce{C3} and proceed all the way to \ce{C3H8}, with the majority of these reactions having no activation barrier. Specifically, the hydrogenation reactions of \ce{C3H2} isomers to form \ce{CH2CCH},
\begin{equation}
    \label{eqa:H-l-C3H2}
        \ce{s-H + s-l-C3H2 -> s-CH2CCH}
\end{equation}
and
\begin{equation}
    \label{eqa:H-c-C3H2}
        \ce{s-H + s-c-C3H2 -> s-CH2CCH},
\end{equation}
were found to have no barrier (here `s-' denotes a grain-surface species). Likewise, using quantum chemical calculations \citet{miller_multiple-well_2003} found the hydrogenation of \ce{CH2CCH} to \ce{CH3CCH} and \ce{CH2CCH2},
{\begin{align}
    \label{eqa:sCH2CCH}
        \ce{s-H + s-CH2CCH & -> s-CH3CCH} & 50\% \nonumber \\
                           &\ce{ -> s-CH2CCH2} & 50\%
\end{align}}
to be barrierless for the two \ce{C3H4} isomers, suggesting that these species can be efficiently formed on grain surfaces. These grain processes of the \ce{C3H_n} species were thus incorporated into our network according to \citet{hickson_methylacetylene_2016} and \citet{loison_interstellar_2017}, and a full list {can be found in Table~\ref{tab:Grain} in the Appendix}. 

As well as grain surface reactions, desorption energies of relevant species were updated according to recent experimental and theoretical studies. Accurate desorption energies for species with efficient grain-surface formation pathways are vital for describing the release of these species into the gas phase.  Recent temperature-programmed desorption (TPD) experiments on amorphous water ices reported a value of 4400 K for the desorption energy of \ce{CH3CCH} on compact water ice \citep{behmard_desorption_2019}. In this study, we used this value for the desorption energy of \ce{CH3CCH} in our model and assumed \ce{CH2CCH2} has the same desorption energy as that of \ce{CH3CCH}. \citet{wakelam_binding_2017} estimated desorption energies for a variety of interstellar species on amorphous solid water using electronic structure calculations, including the \ce{CH2CCH} radical (3300 K). Additionally, \citet{villadsen_predicting_2022} predicted binding energies of astrochemically-relevant molecules by applying Gaussian process regression to a training set of experimental TPD energies, finding generally good agreement between experiment and predictions. We incorporated their predicted value of 9520 K for the desorption energy of all three \ce{C4H3N} isomers on water surfaces into our model. 

\subsection{Sensitivity Analysis}
In order to gauge the sensitivity of our modeling results to the rate constants of certain reactions, we have performed a sensitivity analysis on a select number of reactions involving \ce{CH2CCH}, specifically \ce{CH2CCH} reacting with H, C, O, N, CN, OH, and itself. {The only formation pathway tested in this manner was Reaction~\ref{eqa:cch-ch3}, as the rate constants for other formation pathways of \ce{CH2CCH} are well-constrained via low temperature kinetic measurements of capture theory.} For each reaction, the rate constant was multiplied by values ranging from 0.01 to 100, with separate models performed for each modification. Additionally, the removal of each reaction from the network was tested. This technique allows us to determine the effects that different rate constants have on the modeled abundances of interstellar species, which is particularly useful for reactions that lack dedicated low-temperature studies. However, it is important to note that only one reaction is modified at a time, and thus we do not obtain information about how two reactions may be correlated.

{Similarly, we have performed a sensitivity analysis on the effect of the C/O ratio on \ce{CH2CCH}, \ce{CH3CCH}, \ce{CH2CCH2}. As stated previously, we assume a C/O ratio of 1.1 in our model in line with previous GOTHAM studies \citep{xue_detection_2020, burkhardt_ubiquitous_2021, loomis_investigation_2021, mccarthy_interstellar_2021, mcguire_detection_2021}, as this ratio improves modeling results for a variety of carbon-bearing species, especially the cyanopolyynes. Additionally, \citet{hincelin_oxygen_2011} and \citet{wakelam_effect_2006} both investigated the effect of C/O ratio on astrochemical models of TMC-1 and found better overall agreement to observations with a larger C/O ratio. However, some models of TMC-1 use a significantly lower C/O ratio of 0.7 \citep{ruaud_gas_2016, loison_interstellar_2017}, while other astronomical sources may have different chemical conditions and thus different C/O ratios. We have tested values of the C/O ratio ranging from 0.7 to 1.1 in increments of 1.1 by varying the initial abundance of elemental oxygen.}

\section{Results}
\label{sec:Results}
In order to consistently analyze the results of our models, we calculated best fit times by minimizing the {mean absolute difference} between the log observed abundances and the log modeled abundances for species of interest (\ce{CH2CCH}, \ce{CH3CCH}, \ce{CNCH2CCH}, \ce{CH3C3N}, and \ce{CH2CCHCN}). {This gives a best-fit time of {$4.739\times10^5$ years}. The cyanonaphthalene isomers were excluded from this calculation as the modeled abundances of these species are still $\sim6$ orders of magnitude below observations.} This is in agreement with a best fit time of $\sim5\times10^5$ years from previous GOTHAM modeling studies \citep{xue_detection_2020, loomis_investigation_2021}.

{The modeled column densities of select species at this best-fit time can be seen in Table~\ref{tab:abundances}, along with observed values and confidence levels. These latter values were calculated using the method introduced by \citet{garrod_non-thermal_2007} and are a metric of confidence that the best-fit abundance is in agreement with the observed abundance, assuming a log-normal distribution centered on the observed value with a standard deviation of one. We also calculated the mean confidence level excluding the cyanonaphthalene isomers, as the agreement for these species is very poor and would heavily skew the mean.}

In astrochemical models of dark molecular clouds, it can also be assumed that some elements, typically carbon, sulfur, silicon, phosphorus, chlorine, and the metals, begin as cations in their initial state \citep{ruaud_gas_2016}. If the initial elemental abundances as described in Table~\ref{tab:init_abundances} are kept but the previously mentioned species are assumed to begin as cations, the resulting best-fit abundances {increase slightly by factors of 1.21 or less}. The relative contributions of production and destruction pathways are affected, such as Reaction~\ref{eqa:c2h4-c} becoming less prevalent to the production of \ce{CH2CCH} at early times. The similarity in abundances at later times may be due to the efficient conversion of \ce{C+} to \ce{C} via electron capture and charge transfer.

\begin{table*}[htb]
    \caption{\textbf{Modeled and observed column densities for select species}}
    \label{tab:abundances}
    \begin{threeparttable}
    \centering
    \begin{tabular}{l c c c c}
    \toprule
    Molecule & Best-Fit\tnote{1} (\unit{\per\square\centi\metre}) & Observed\tnote{2} (\unit{\per\square\centi\metre}) & {Confidence Level\tnote{3}} & Reference\\
    \midrule
    \ce{CH2CCH} & {\num{1.67e12}} & \num{1.0e14} & {\num{0.0757}} & \citet{agundez_detection_2022} \\
    \ce{CH3CCH} & {\num{4.01e12}} & \num{1.15e14} & {\num{0.145}} & \citet{gratier_new_2016} \\
    \ce{CH2CCH2} & {\num{5.20e12}} &  &  & \\
    \ce{CH3C3N} & {\num{3.36e12}} & \num{8.66e11} & {\num{0.556}} & \citet{siebert_ch3-terminated_2022} \\
    \ce{CH2CCHCN} & {\num{8.48e11}} & \num{2.7e12} & {\num{0.615}} & \citet{marcelino_study_2021} \\
    \ce{CNCH2CCH} & {\num{2.00e11}} & \num{9.2e11} & {\num{0.507}} & \citet{mcguire_early_2020} \\
    \ce{C10H7CN} & {\num{4.44e6}} & \num{7.35e11} & {\num{1.80e-7}} & \citet{mcguire_detection_2021} \\
    \ce{C10CNH7} & {\num{4.93e6}} & \num{7.05e11} & {\num{2.53e-7}} & \citet{mcguire_detection_2021} \\
    \ce{C6H5} & {\num{2.54e11}} & & & \\
    \midrule
    {Mean confidence level\tnote{4}} & & & {\num{0.380}} & \\
    \bottomrule
    \end{tabular}

    \begin{tablenotes}
    \item[1] Modeled column density at the best-fit time of {$4.739\times10^5$} years.
    \item[2] Species with {no value in this column have not been detected in TMC-1}.
    \item[3] {Calculated according to \citet{garrod_non-thermal_2007}. Species with no value in this column have not been detected in TMC-1.}
    \item[4] The cyanonaphthalene isomers (\ce{C10H7CN} and \ce{C10CNH7}) were not included in this calculation.
    \end{tablenotes}

    \end{threeparttable}
\end{table*}

\subsection{\ce{C3H_n} Hydrocarbons}

In Figure~\ref{fig:hydrocarbon_abundances}, the modeled abundances of \ce{CH2CCH}, \ce{CH3CCH}, and \ce{CH2CCH2} are plotted as a function of time, as well as their observed column densities in TMC-1. \citet{gratier_new_2016} constrained the column density of \ce{CH3CCH} as \SI{1.15e14}{\per\square\centi\metre} based on observations of the TMC-1 cyanopolyyne peak using the Nobeyama \SI{45}{\metre} telescope. \citet{agundez_discovery_2021} first detected \ce{CH2CCH} at a wavelength of \SI{8}{\milli\metre} with a column density of \SI{8.7e13}{\per\square\centi\metre}, and later used observations at \SI{3}{\milli\metre} to revise the column density to \SI{1.0e14}{\per\square\centi\metre} \citep{agundez_detection_2022}. \ce{CH2CCH2} has not been detected in TMC-1, and thus there is no observed value for this species. {The observed column densities, as well as best-fit modeled column densities can be seen in Table~\ref{tab:abundances}. Our updated model gives best-fit abundances of \ce{CH2CCH} and \ce{CH2CCH2} that are {$\sim60$} and $\sim29$ times lower than observations, respectively.} Compared to the model results with the GOTHAM DR1 network, the modeled abundance of \ce{CH2CCH} at {$4.739\times10^5$} years has been increased by {a factor of 2.65}. The removal of Reactions \ref{eqa:c-C3H2} and \ref{eqa:l-C3H2} resulted in the largest increase to the modeled \ce{CH2CCH} abundance, followed by the addition of the neutral-neutral Reactions \ref{eqa:cch-ch3} and \ref{eqa:c2-ch4}. The modeled abundance of \ce{CH3CCH} at this time has only been increased by a factor of 1.41 compared to the previous model results. The column densities of both \ce{CH2CCH} and \ce{CH3CCH} are still under-predicted by between 1 and 2 orders of magnitude, although the \ce{CH3CCH} abundance is almost within the 1$\sigma$ confidence interval.

\begin{figure}[hbt!]
    \centering
    \includegraphics[width=\columnwidth]{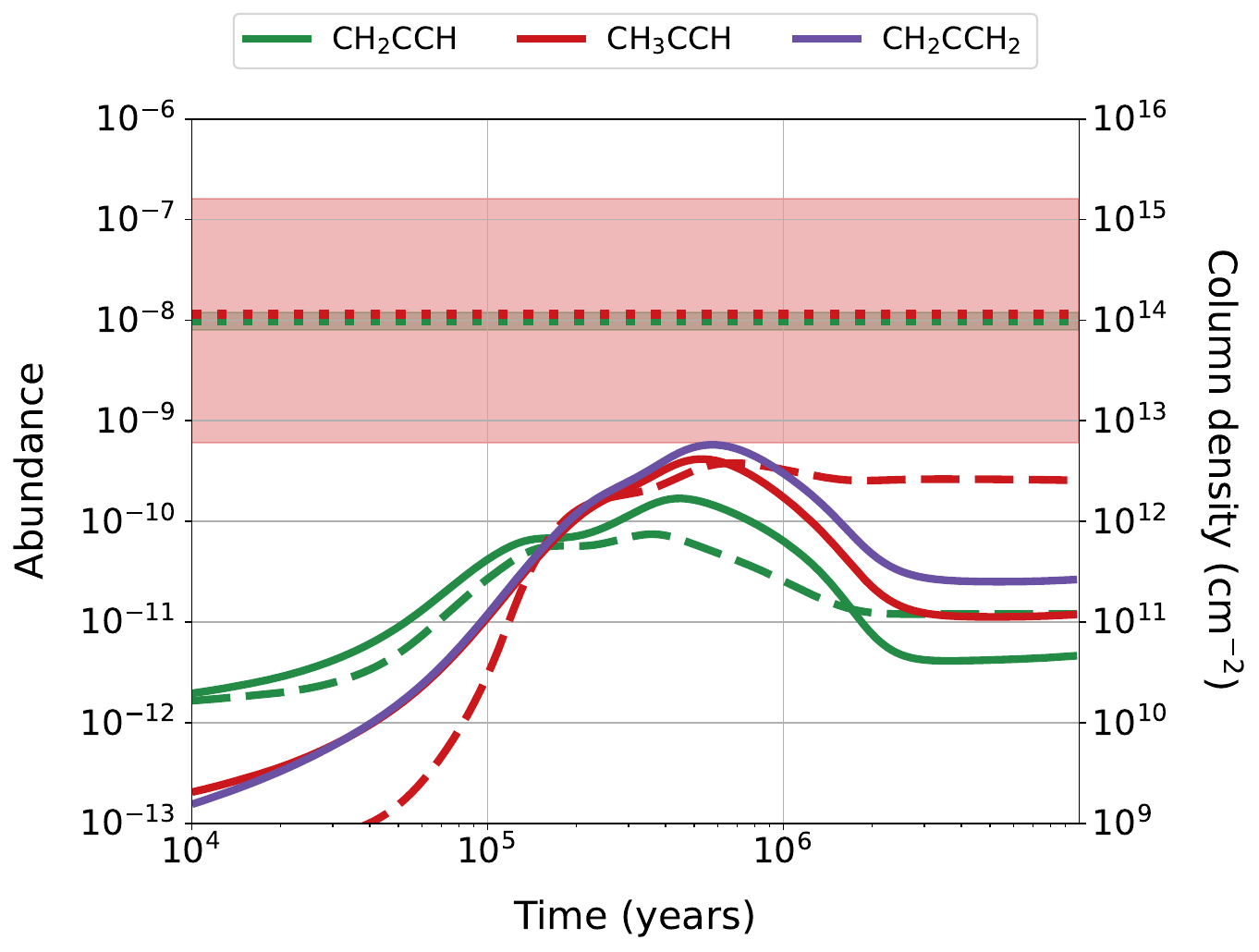}
    \caption{Modeled abundances and column densities of propargyl radical ({green}), methylacetylene ({red}), and allene ({purple}) as a function of time. The dotted lines represent the observed column densities in TMC-1, with the shaded areas signifying an error of 1 $\sigma$. The dashed lines represent modeled abundances obtained from GOTHAM DR1{, and the solid lines are the modeled abundances with the updated model presented in this work}.}
    \label{fig:hydrocarbon_abundances}
\end{figure}

In our updated model, the early-time production of \ce{CH2CCH} is dominated by the reaction between atomic carbon and \ce{C2H4} (Reaction~\ref{eqa:c2h4-c}), as shown in Figure~\ref{fig:ch2cch_prod_rates}. Grain-surface hydrogenation of both isomers of \ce{C3H2} (Reactions \ref{eqa:H-l-C3H2} and \ref{eqa:H-c-C3H2}) followed by chemical desorption also contributes significantly to \ce{CH2CCH} formation due to the large abundance of atomic hydrogen and its high mobility on grains. After $\sim10^4$ years, the rates of other \ce{CH2CCH} formation pathways become sizeable due to the build up of more complex hydrocarbons. In addition to the aforementioned reactions, the reaction between diatomic carbon and methane (Reaction~\ref{eqa:c2-ch4}) and the dissociative recombinations of \ce{C5H5+} and \ce{C3H5+} become significant sources of \ce{CH2CCH} around $10^5$ years. Despite a similar rate constant to the dissociative recombination of \ce{C5H5+}, we find that the dissociative recombination of \ce{C3H5+} is unable to efficiently form \ce{CH2CCH} at all times. Likewise, the dissociative recombination of \ce{C3H4+} has the greatest rate constant of these three reactions but still exhibits a lower rate than the dissociative recombination of \ce{C5H5+}. This is due to an approximately 10-fold difference in abundance between \ce{C5H5+} and the \ce{C3H$_n$+} cations. Similarly, the neutral-neutral production pathways of \ce{CH2CCH} generally outpace the dissociative recombination pathways, despite greater rate constants in the latter set of reactions, due to much larger modeled abundances of small neutral hydrocarbons. \ce{CH2CCH} is primarily destroyed via reaction with O, C, and N at early times (before $10^5$ years) and via reaction with H at later times (after {$3\times10^5$ years}).

\begin{figure}[hbt!]
    \centering
    \includegraphics[width=\columnwidth]{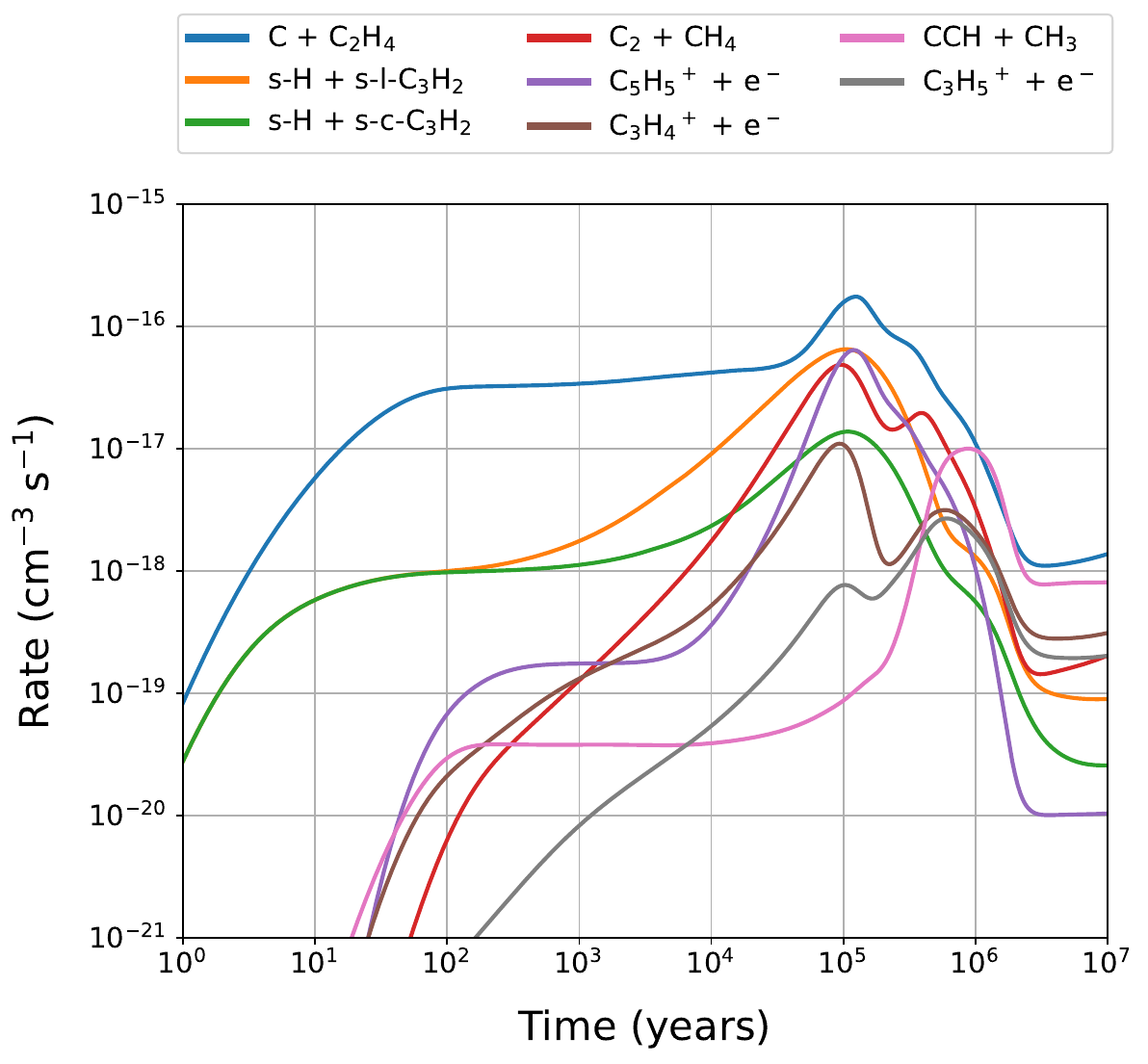}
    \caption{Reaction rates for key \ce{CH2CCH} formation pathways as a function of time. Each color represents a different reaction, with the reactants given in the legend (`s-' denotes a grain species). For more information on these reactions, including product channels, refer to Tables \ref{tab:DR}-\ref{tab:Grain} in the Appendix.}
    \label{fig:ch2cch_prod_rates}
\end{figure}

For the first $\sim10^3$ years of the model, \ce{CH3CCH} is primarily formed by the grain-surface reaction
\begin{equation}
    \label{CH_C2H3}
        \ce{s-CH + s-C2H3 -> CH3CCH},
\end{equation}
where energy from reaction exothermicity ejects a portion of the products into the gas phase. At $10^4$ years and later, \ce{CH3CCH} is predominantly formed from Reactions \ref{eqa:h-ch2cch} and \ref{eqa:sCH2CCH}. Similarly, Reactions \ref{eqa:CH2CCH_CH2CCH2} and \ref{eqa:sCH2CCH} constitute the main formation pathways of \ce{CH2CCH2} at all times. Despite the low rate constant, the gas-phase radiative association between \ce{H} and \ce{CH2CCH} is able to efficiently form both \ce{C3H4} isomers due to the large abundance of \ce{H} and increased abundance of \ce{CH2CCH}. Likewise, the analogous grain-surface hydrogenations are able to form large amounts of \ce{C3H4} in both the gas phase and on grain surfaces, resulting in high grain-surface abundances for \ce{CH3CCH} and \ce{CH2CCH2}. {The rates of these reactions as a function of time can be seen in Figures \ref{fig:ch3cch_prod_rates} and \ref{fig:ch2cch2_prod_rates}.} Destruction of both \ce{C3H4} isomers occurs mainly through a combination of reaction with \ce{C}, which dominates before $10^5$ years, and reaction with \ce{H3+}, which dominates shortly after $10^5$ years. Between $10^5$ and $10^6$ years, reaction with CN radical significantly contributes to the destruction of \ce{CH3CCH} and \ce{CH2CCH2}. The rates of key destruction mechanisms for \ce{CH2CCH}, \ce{CH3CCH}, and \ce{CH2CCH2} as a function of time are displayed in Figure~\ref{fig:dest_rates} in the Appendix.

\begin{figure}[hbt!]
    \centering
    \includegraphics[width=\columnwidth]{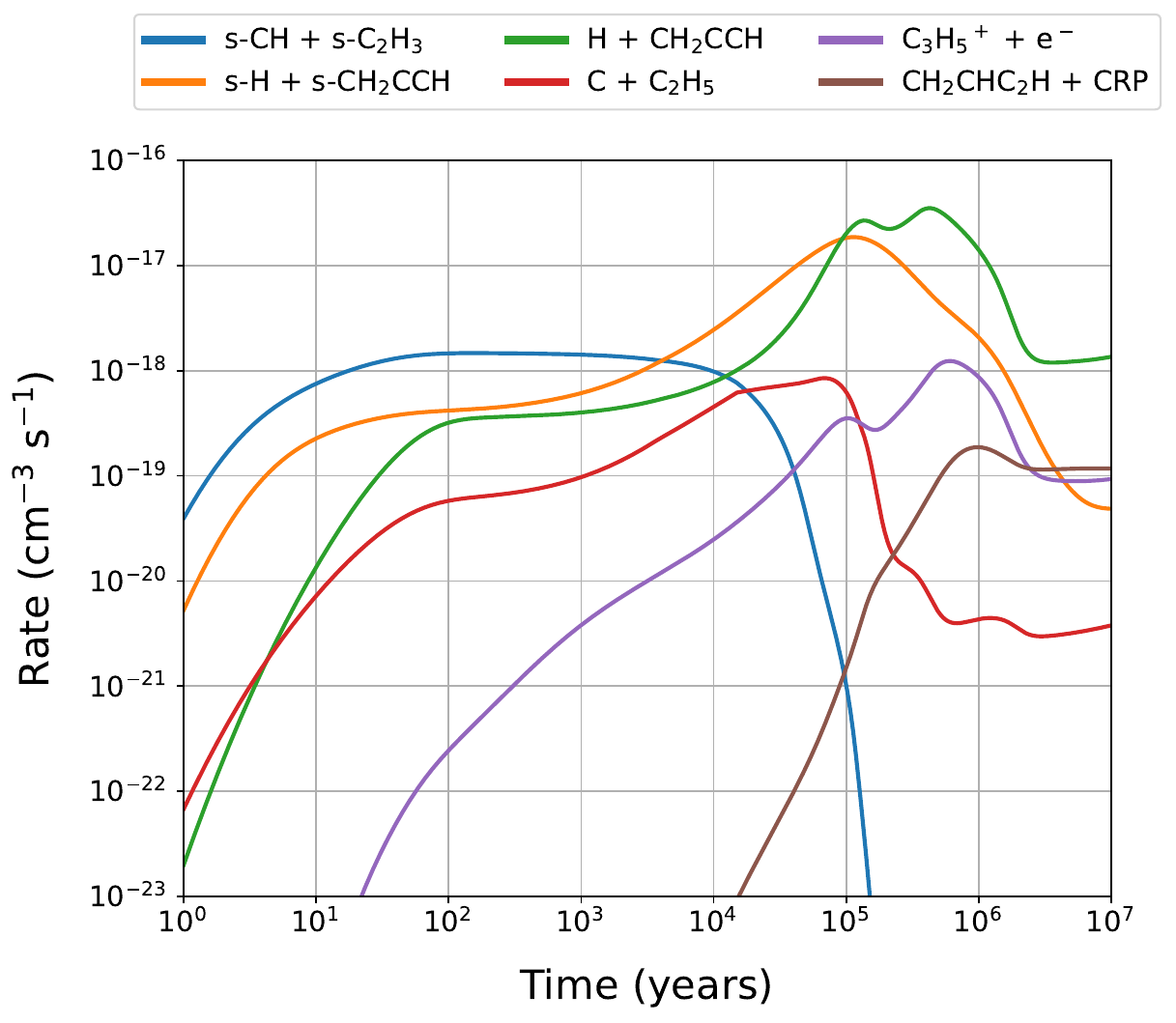}
    \caption{Reaction rates for key \ce{CH3CCH} formation pathways as a function of time. Each color represents a different reaction, with the reactants given in the legend (`s-' denotes a grain species, and CRP denotes a cosmic-ray induced photon). For more information on these reactions, including product channels, refer to Tables \ref{tab:DR}-\ref{tab:Grain} in the Appendix.}
    \label{fig:ch3cch_prod_rates}
\end{figure}

\begin{figure}[hbt!]
    \centering
    \includegraphics[width=\columnwidth]{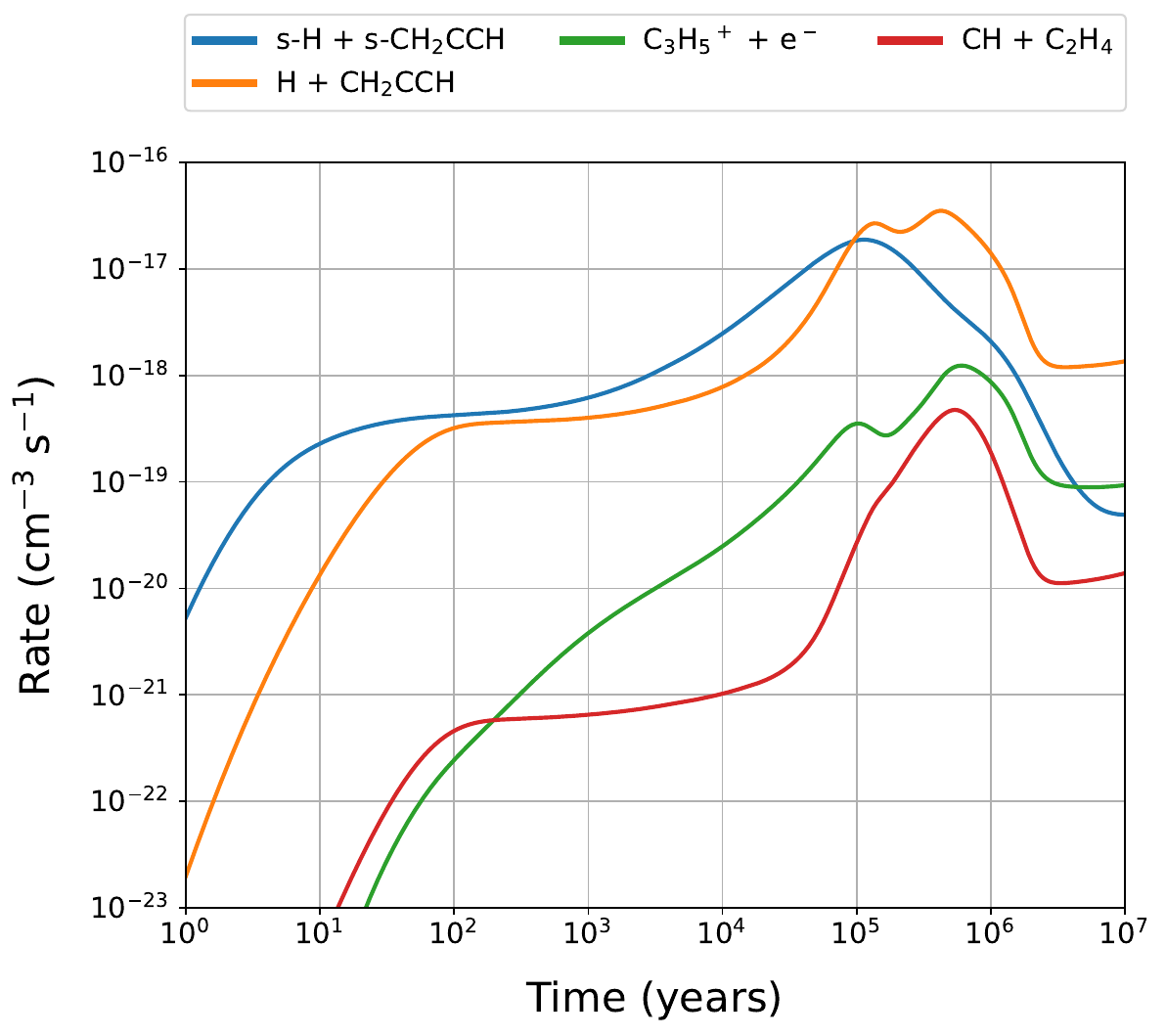}
    \caption{Reaction rates for key \ce{CH2CCH2} formation pathways as a function of time. Each color represents a different reaction, with the reactants given in the legend (`s-' denotes a grain species). For more information on these reactions, including product channels, refer to Tables \ref{tab:DR}-\ref{tab:Grain} in the Appendix.}
    \label{fig:ch2cch2_prod_rates}
\end{figure}

\subsection{CN-Substituted Species}
\begin{figure}[hbt!]
    \centering
    \includegraphics[width=\columnwidth]{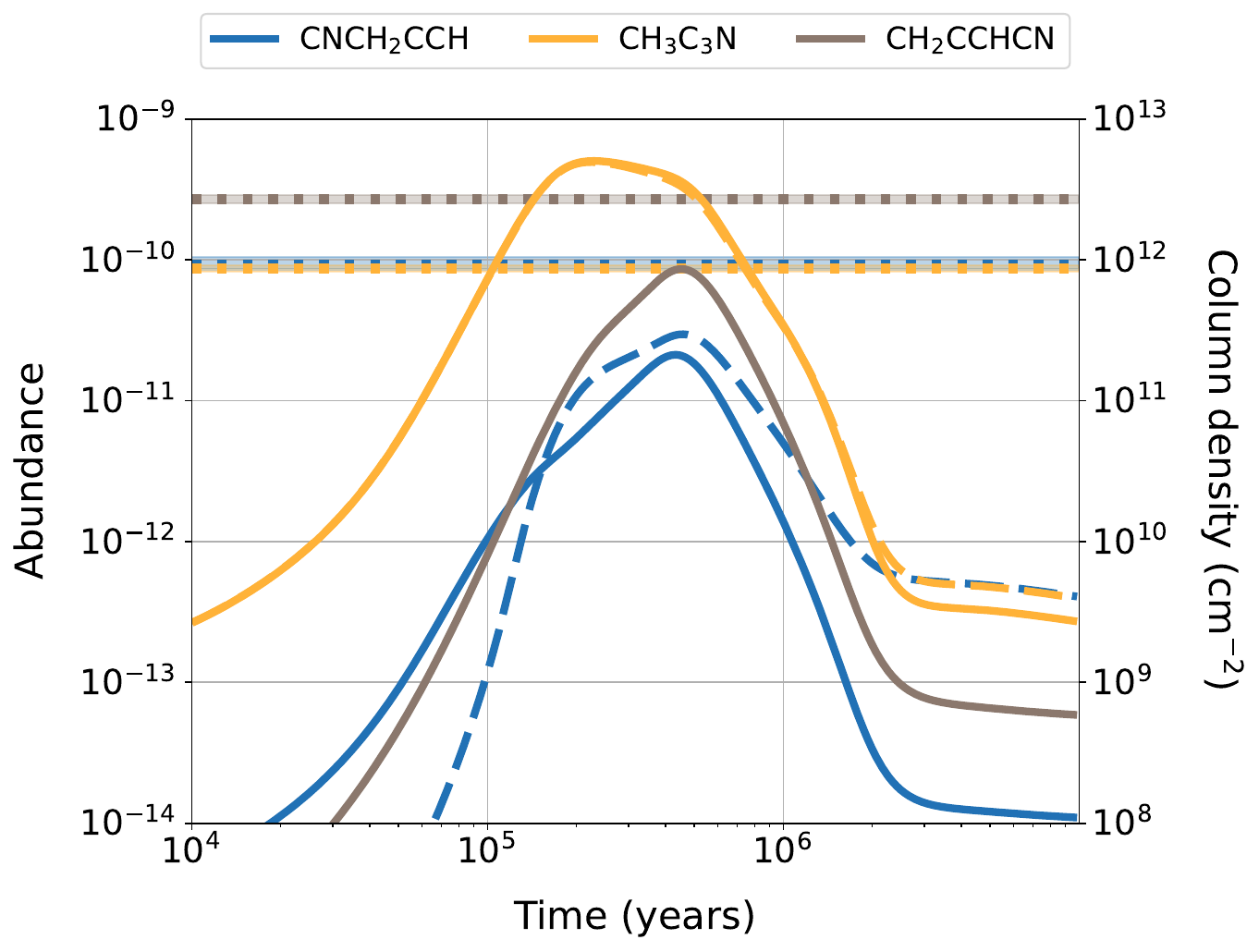}
    \caption{Modeled abundances and column densities of propargyl cyanide ({blue}), methylcyanoacetylene ({yellow}), and cyanoallene ({brown}) as a function of time. The dotted lines represent the observed column densities in TMC-1, with the shaded areas signifying an error of 1 $\sigma$. The dashed lines represent modeled abundances from GOTHAM DR1{, and the solid lines are the modeled abundances with the updated model presented in this work}.}
    \label{fig:nitrile_abundances}
\end{figure}

Considering the strong chemical link between pure hydrocarbons and their CN-substituted derivatives in TMC-1, we are interested in how the updated chemical network affects the formation of \ce{C4H3N} isomers. The modeled abundances of these species as a function of time are shown in Figure~\ref{fig:nitrile_abundances}. The best-fit column densities of these species as calculated by the model are listed in Table \ref{tab:abundances}, along with their observed column densities. The best-fit column density of \ce{CH2CCHCN} is predicted particularly well by our model to within a factor of $\sim3.2$, whereas the best-fit column density for \ce{CNCH2CCH} is only a factor of {$\sim4.4$} lower than the observed value. Finally, our model over-predicted the column density of \ce{CH3C3N} by a factor of {$\sim3.8$}. Compared to the GOTHAM DR-1 model, the \ce{CH3C3N} abundance does not change significantly despite the incorporation of a new formation route in Reaction \ref{eqa:CN+CH3CCH}, suggesting that the contribution from this reaction is only minor. 

The decrease in the modeled abundance of \ce{CNCH2CCH} stems from the change in its formation pathways, particularly the removal of Reaction \ref{eqa:CN+CH3CCH} as a formation route to this species. The addition of new production routes in Reaction \ref{eqa:CN+CH2CCH} and Reaction~\ref{eqa:CN+CH2CCH2} are able to nearly reproduce the GOTHAM DR-1 abundance, despite \ce{CNCH2CCH} being only a minor product of the latter reaction. {Before $3\times10^5$ years, the formation of \ce{CNCH2CCH} is dominated by Reaction~\ref{eqa:CN+CH2CCH}, and following this time Reaction~\ref{eqa:CN+CH2CCH2} becomes the dominant formation pathway}. As mentioned in Section~\ref{sec:Model}, we tested models with alternate products for \ce{CNCH2CCH} and \ce{CH2CCHCN} ion-neutral pathways to determine their effects on the abundances of species considered here. When alternate product channels were used, the peak abundance of the directly affected species, \ce{CH2CCH}, decreased by a factor less than 1.15. Likewise, the peak abundances of the \ce{C3H4} and \ce{C4H3N} isomers changed by factors of less than 1.2. Considering the almost negligible differences in results, the products of these ion-neutral reactions do not seem to significantly affect the species of interest.

\subsection{Aromatic Species}
In addition to the previously mentioned species, we also examined {\ce{C6H5} and the two cyanonaphthalene isomers to determine the effect of the propargyl recombination reaction on the formation of these species.} These abundances were plotted as a function of time in Figure~\ref{fig:aromatic_abundances}, and the best-fit modeled abundances can be seen in Table \ref{tab:abundances}. The astrochemical model still severely underpredicts the column densities of {1- and 2-cyanonaphthalene (\ce{C10H7CN}, \ce{C10CNH7}), although the modeled abundancse of the cyanonaphthalenes have been increased by about a factor of 1.7 at {the best-fit time} for both isomers}. The two \ce{C10H7CN} isomers are primarily produced from reaction between naphthalene (\ce{C10H8}) and \ce{CN}, while \ce{C10H8}’s main production pathway is the reaction between \ce{C6H5} and \ce{CH2CHC2H}. The reaction of \ce{CN} with benzene to form \ce{C6H5CN} is barrierless and efficient at low temperatures, and the analogous reaction with \ce{C10H8} to form \ce{C10H7CN} is expected to be as well \citep{cooke_benzonitrile_2020, mcguire_detection_2021}. Thus it is the formation of \ce{C10H8} that limits production of \ce{C10H7CN}. Our addition of \ce{CH2CCH2} as a species and its reaction with \ce{CH} to form \ce{CH2CHC2H} leads to an increase in \ce{C10H8}, and an increase in \ce{C10H7CN} as a result. 

\begin{figure}[hbt!]
    \centering
    \includegraphics[width=\columnwidth]{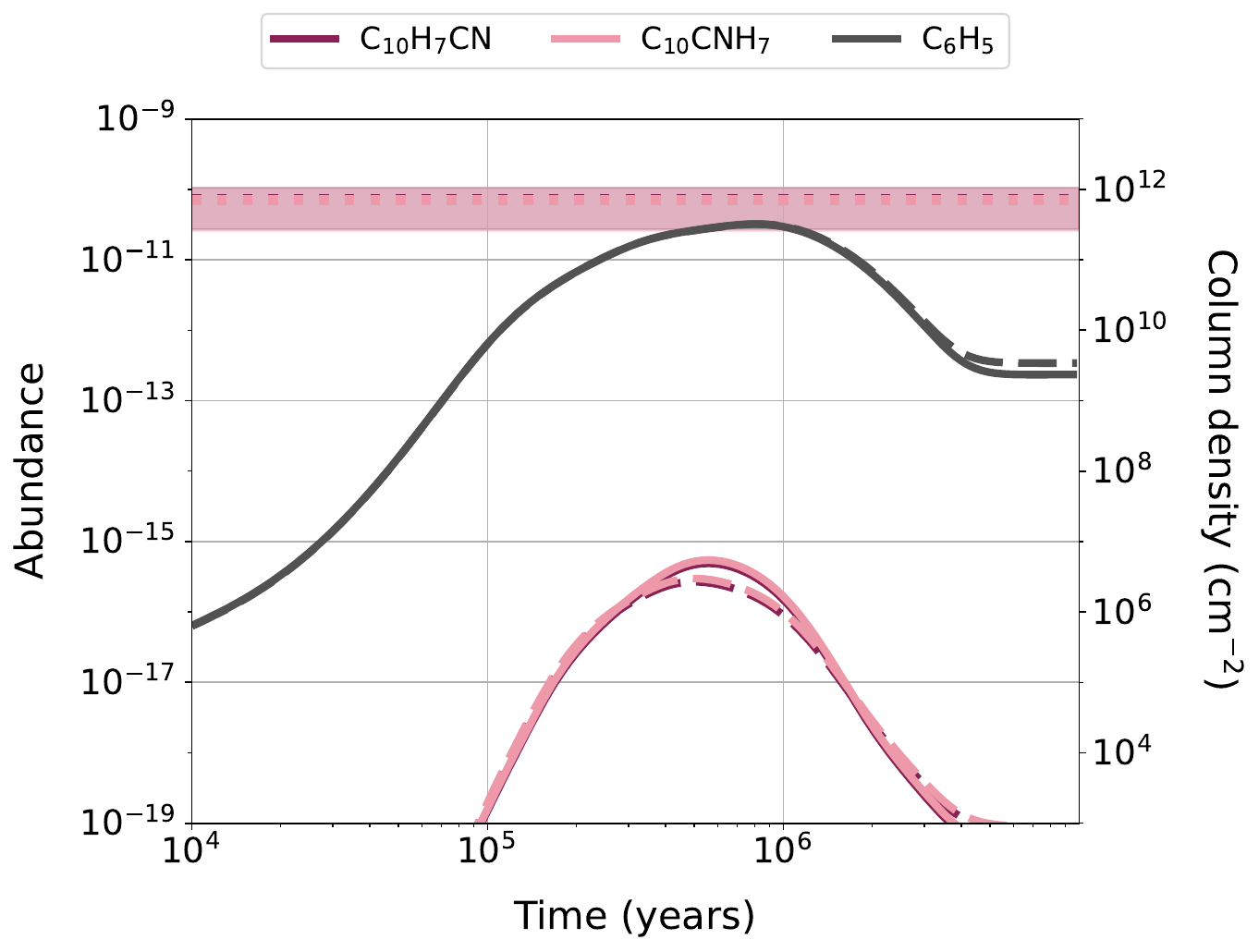}
    \caption{Modeled abundances and column densities of 1-cyanonaphthalene (\ce{C10H7CN}, maroon), 2-cyanonaphthalene (\ce{C10CNH7}, pink) and \ce{C6H5} (gray) as a function of time. The dotted lines represent the observed column densities in TMC-1, with the shaded areas signifying an error of 1 $\sigma$. The dashed lines represent modeled abundances from GOTHAM DR1, and the solid lines are the modeled abundances with the updated model presented in this work. As the formation pathways and observed TMC-1 abundances of the two cyanonaphthalene isomers are very similar, the plotted curves for one isomer are overlapping with those of the other.}
    \label{fig:aromatic_abundances}
\end{figure}

\begin{figure}[hbt!]
    \centering
    \includegraphics[width=\columnwidth]{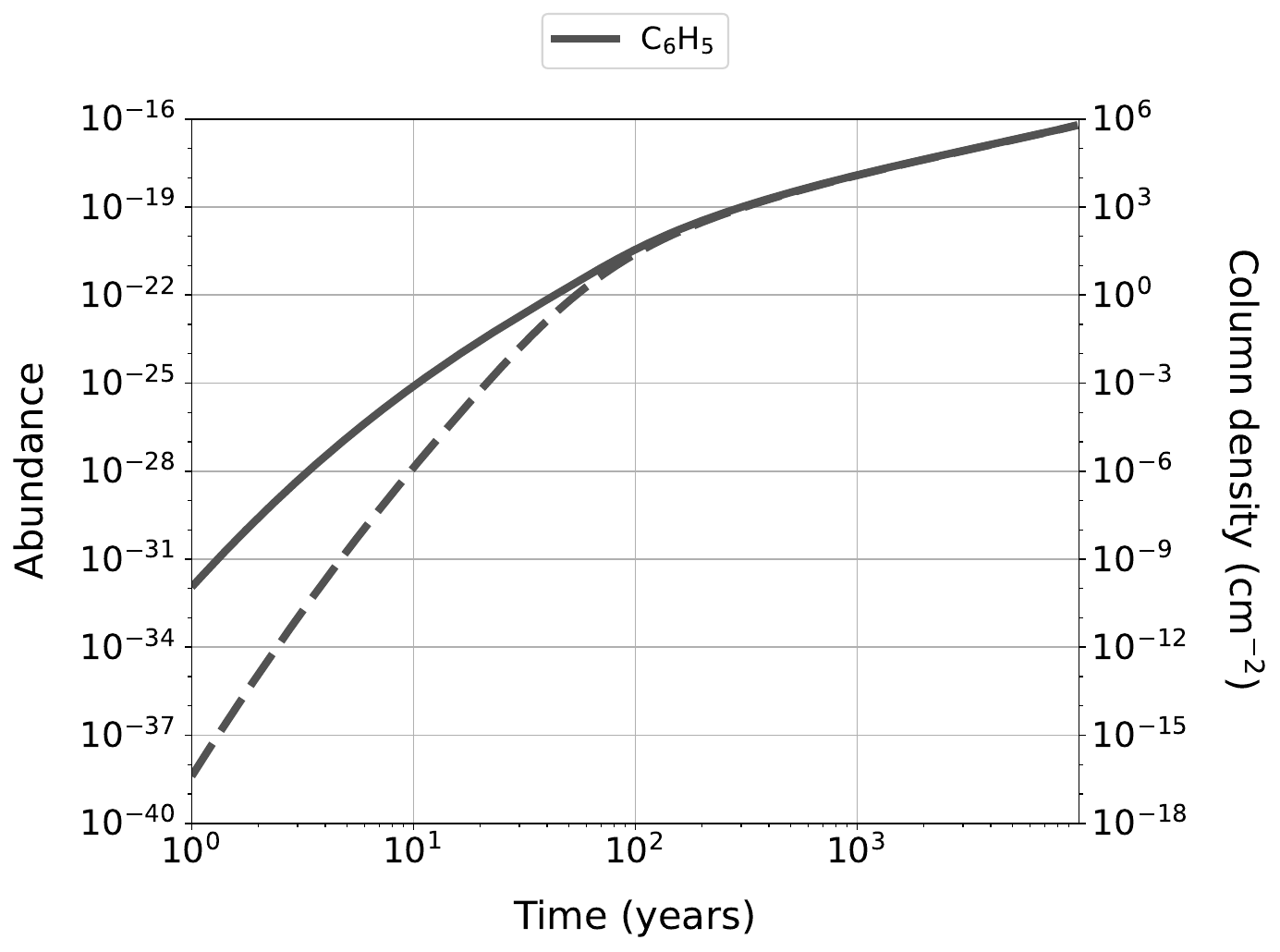}
    \caption{Modeled abundances and column densities of the phenyl radical from 1 to $10^4$ years. The dashed line represents modeled abundances from GOTHAM DR1 and the solid line is the modeled abundances of the updated model presented in this work.}
    \label{fig:phenyl_early}
\end{figure}

The modeled abundance of \ce{C6H5} has not been changed by the addition of \ce{CH2CCH} recombination (Reaction \ref{eqa:ch2cch-c6h5}), except for at early times. At these times a significant increase in \ce{C6H5} abundance is observed, but this difference is made up for by a time of 100 years. To understand why, we must look at the other production mechanisms for \ce{C6H5} included in our network. These are dissociation of benzene via cosmic ray induced photons or external photons and hydrogen abstraction from benzene via \ce{OH} and \ce{H}. Activation barriers to hydrogen abstraction and a large visual extinction make dissociation by cosmic ray induced photons the major mechanism. At early times, the modeled benzene abundance is over 10 orders of magnitude lower than the modeled propargyl abundance. However, the benzene abundance quickly increases, and this difference becomes only six orders of magnitude at 50 years. At this point, the unimolecular decomposition becomes fast enough to outpace the bimolecular recombination of \ce{CH2CCH}. Although \ce{CH2CCH} abundance has been increased at most times, the resulting rate is not large enough to exceed the rate for \ce{C6H6} dissociation. {This can be seen in Figure~\ref{fig:phenyl_early}, where the modeled abundance of the phenyl radical is significantly increased in our new model compared to GOTHAM DR1 at early times but the GOTHAM DR1 abundance makes up the difference by approximately 100 years.} 

\subsection{Sensitivity Studies of Key Reactions and C/O Ratio}
\begin{figure}[hbt!]
    \centering
    \includegraphics[width=\columnwidth]{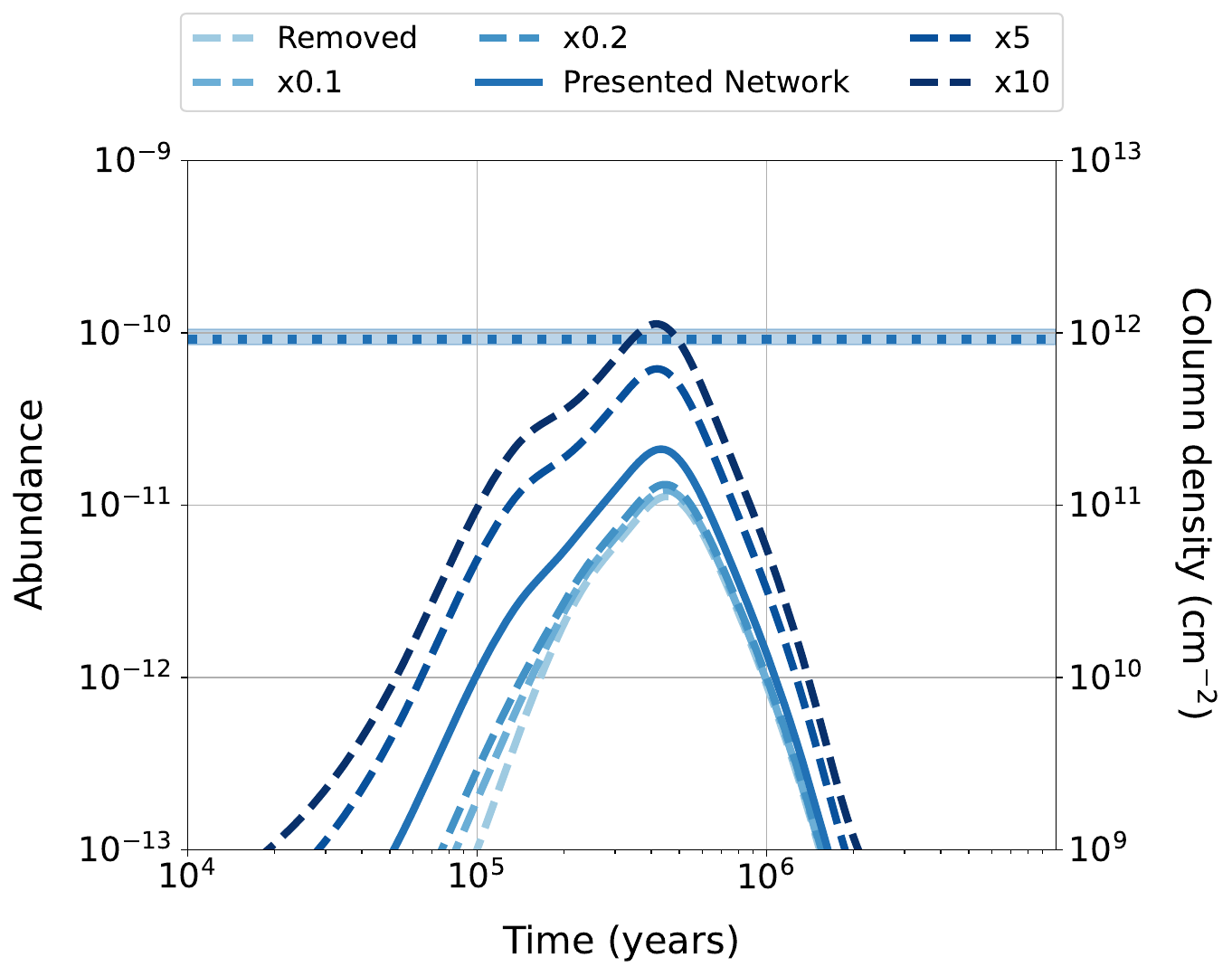}
    \caption{Modeled abundances and column densities of \ce{CNCH2CCH} for different values of the \ce{CN + CH2CCH} rate constant. The dashed lines are networks where this rate constant has been modified by the factors in the legend. The base network (presented in this paper) is shown for comparison as a solid line, as well as a network where this reaction is removed entirely. The color gradient corresponds with the magnitude of the factor of change. As with the other previous figures, a dotted horizontal line is plotted representing the observed abundance of \ce{CNCH2CCH} in TMC-1, with the shaded region signifying an error of 1 $\sigma$.}
    \label{fig:cnch2cch_cn_sens}
\end{figure}

In addition to the astrochemical modeling performed using the updated chemical network, we have also performed sensitivity studies on this chemical network to investigate the importance of several reactions to the chemistry of \ce{CH2CCH}. The results of these studies on Reaction~\ref{eqa:CN+CH2CCH} can be seen in Figure~\ref{fig:cnch2cch_cn_sens}. When the rate constant for this reaction is increased by one order of magnitude to a value of \SI{1.0e-9}{\cubic\centi\metre\per\second}, the best-fit abundance of \ce{CNCH2CCH} increases by a factor of almost 5, essentially reproducing the observed value. The best-fit abundance continues to increase beyond the observed value as the rate constant is further raised. If this reaction is removed from the network entirely, the best-fit abundance of \ce{CNCH2CCH} only decreases by a factor of 1.7 compared to the presented network. In contrast, there is no noticeable change in the modeled abundance of \ce{CH2CCH}, even when this rate constant is increased by two orders of magnitude or removed from the network entirely.

This procedure has been repeated for the reactions of \ce{CH2CCH} with C, H, N, O, OH, and \ce{CH2CCH}{, as well as for the reaction \ce{CCH + CH3 -> CH2CCH + H}}. 
The modeled abundances of \ce{CH2CCH}, \ce{CH3CCH}, and \ce{CH2CCH2} for different rate constants of the \ce{CH2CCH + H} reaction are shown in Figures \ref{fig:ch2cch_h_sens}, \ref{fig:ch3cch_h_sens}, and \ref{fig:ch2cch2_h_sens}. Both increases and decreases to this rate constant are shown to have large effects on modeled abundances of all three species, with the most noticeable variation in \ce{CH2CCH} abundance. Figures~\ref{fig:SI_sens} and \ref{fig:prop_recomb_sens} in the Appendix display the modeled abundances of \ce{CH2CCH} in response to variations in the remaining rate constants. For the reactions of \ce{CH2CCH} with itself and OH, none of the tested changes significantly alter the modeled abundance of \ce{CH2CCH}. Furthermore, the best-fit abundance of \ce{C6H5} only changes slightly when the rate constant of the propargyl recombination is modified. Lowering the rate constant or removing the reactions of \ce{CH2CCH} with C, N, and O also do not change the modeled abundance of \ce{CH2CCH}, however increasing the rate constants of these reactions does significantly lower the best-fit abundance. Likewise, removing  Reaction~\ref{eqa:cch-ch3} (\ce{CCH + CH3}) does not have a significant effect on \ce{CH2CCH} abundance, while increasing this rate leads to a slight increase in \ce{CH2CCH} abundance at a later time.

\begin{figure}[hbt!]
    \centering
    \includegraphics[width=\columnwidth]{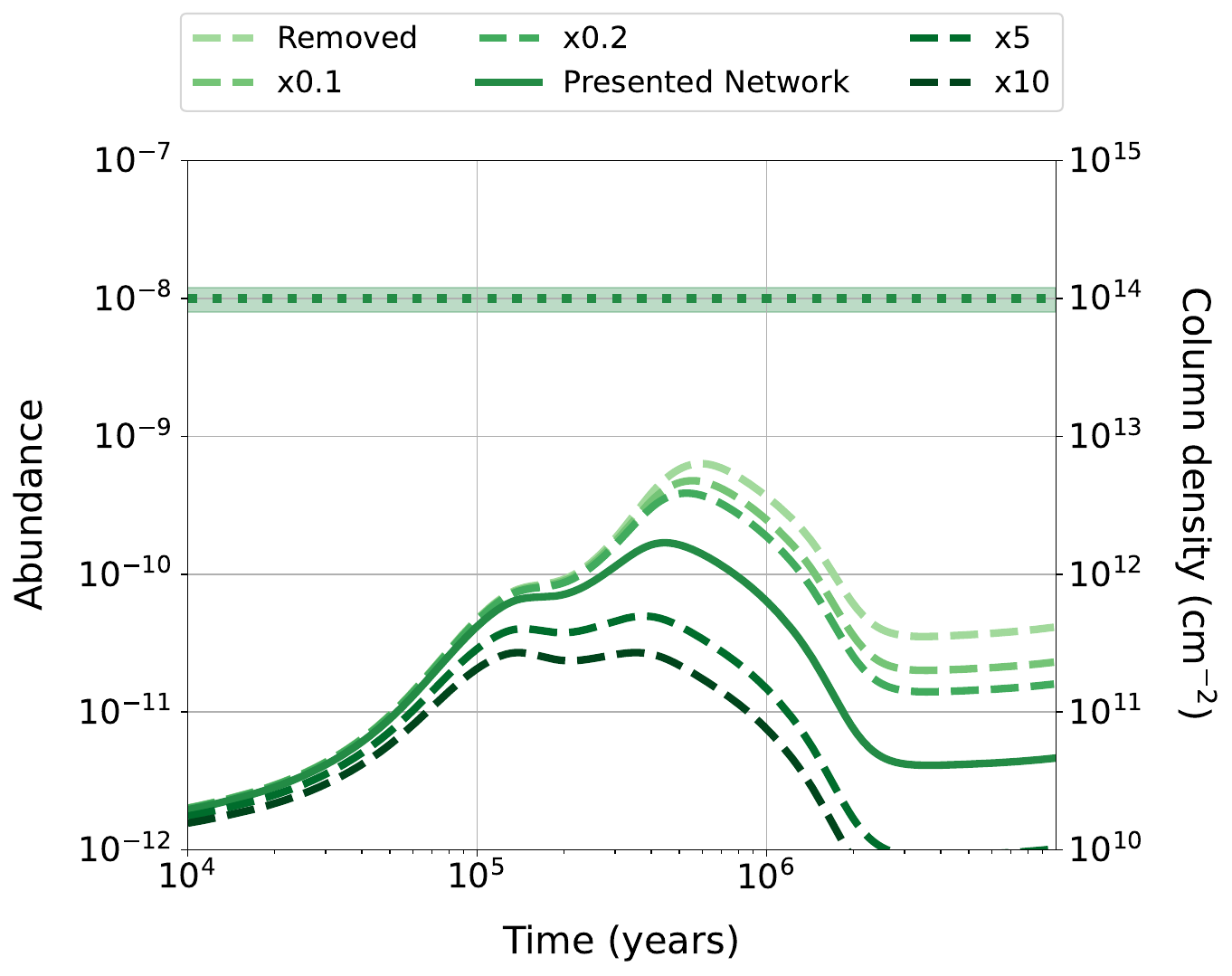}
    \caption{Modeled abundances and column densities of \ce{CH2CCH} for different values of the \ce{CH2CCH + H} rate constant. {The dashed lines are networks where this rate constant has been modified by the factors in the legend. The base network (presented in this paper) is shown for comparison as a solid line, as well as a network where this reaction is removed entirely. The color gradient corresponds with the magnitude of the factor of change. The dotted line represents the observed column density in TMC-1, with the shaded area signifying an error of 1 $\sigma$.}}
    \label{fig:ch2cch_h_sens}
\end{figure}

\begin{figure}[hbt!]
    \centering
    \includegraphics[width=\columnwidth]{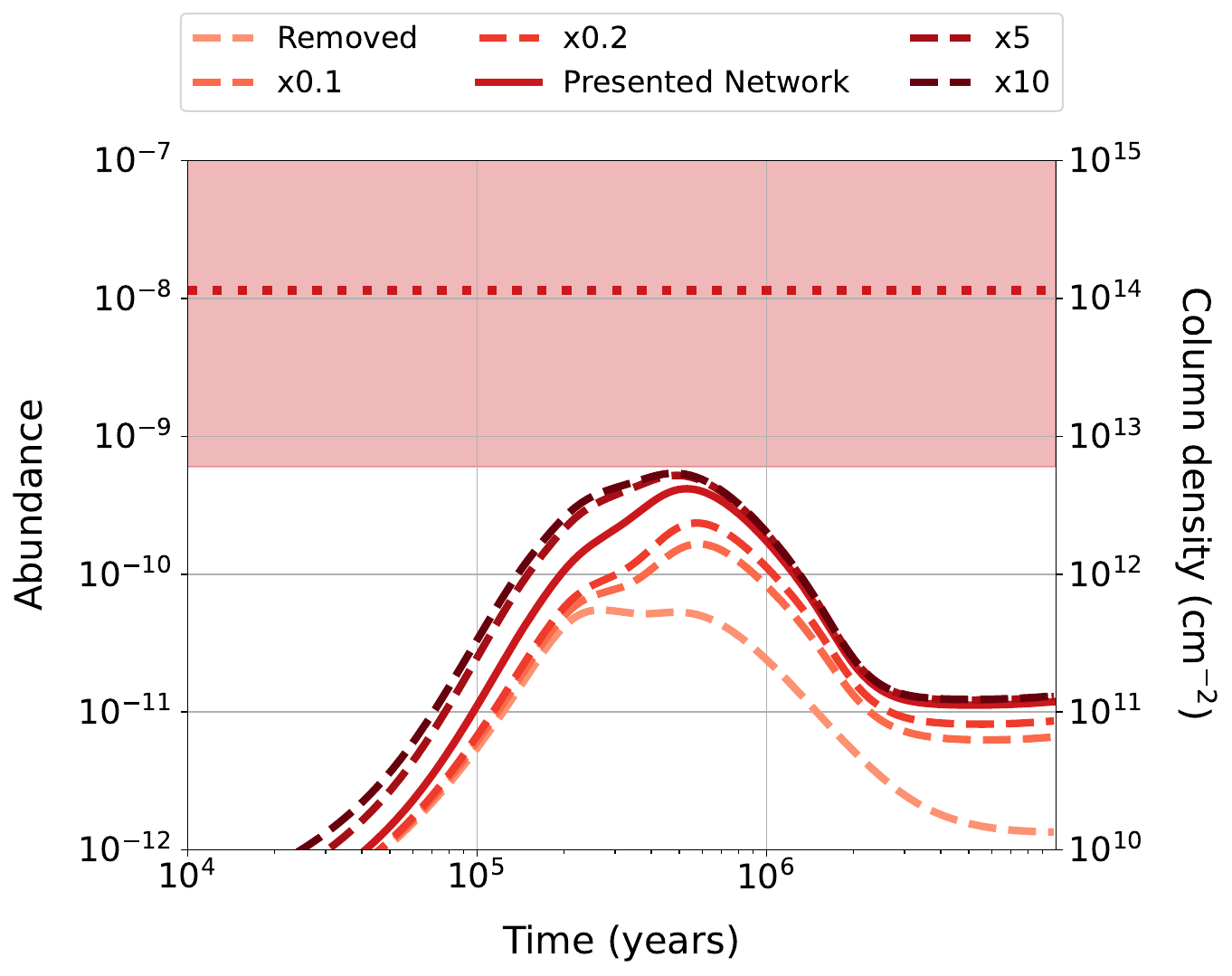}
    \caption{Modeled abundances and column densities of \ce{CH3CCH} for different values of the \ce{CH2CCH + H} rate constant. {The dashed lines are networks where this rate constant has been modified by the factors in the legend. The base network (presented in this paper) is shown for comparison as a solid line, as well as a network where this reaction is removed entirely. The color gradient corresponds with the magnitude of the factor of change. The dotted line represents the observed column density in TMC-1, with the shaded area signifying an error of 1 $\sigma$.}}
    \label{fig:ch3cch_h_sens}
\end{figure}

\begin{figure}[hbt!]
    \centering
    \includegraphics[width=\columnwidth]{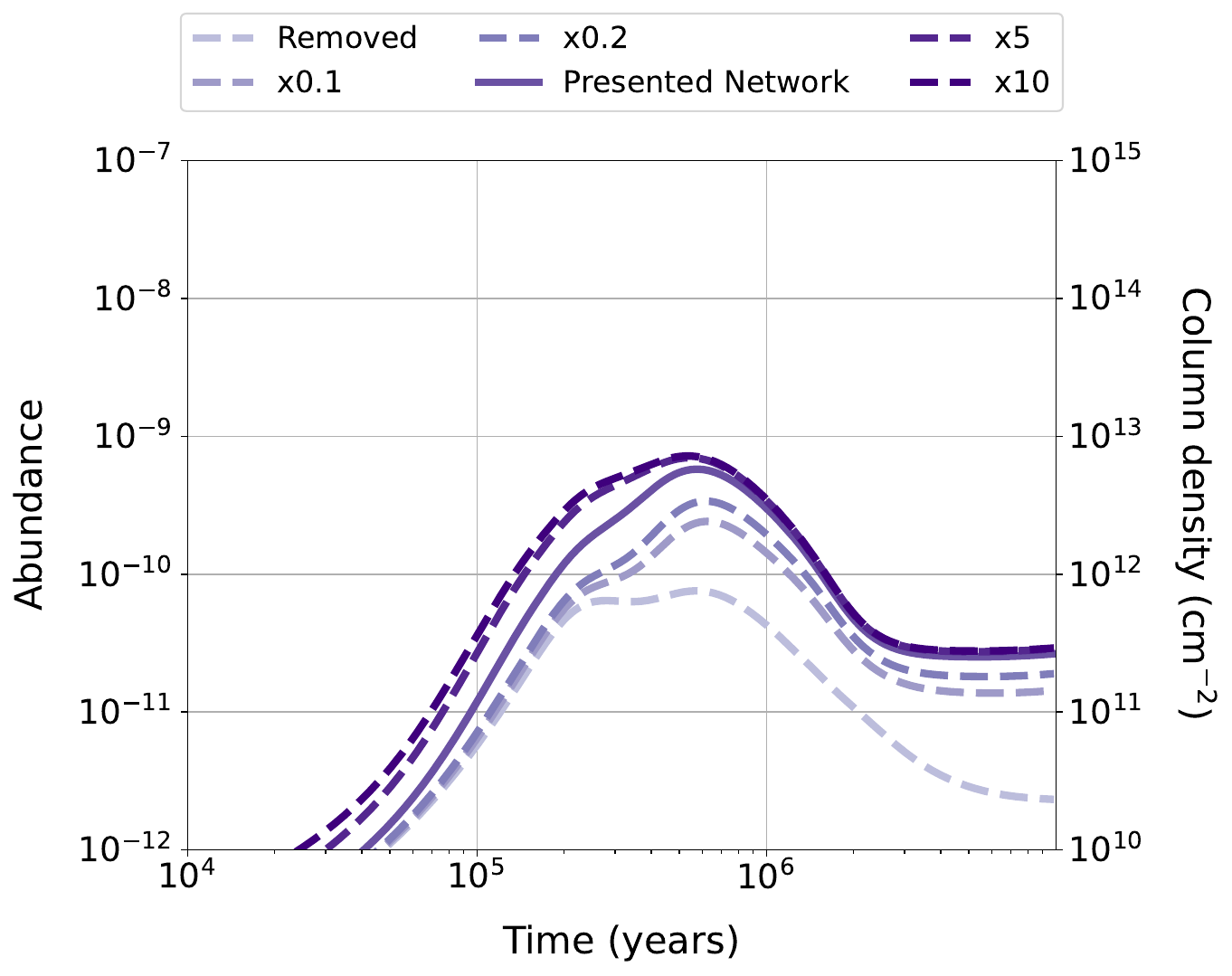}
    \caption{Modeled abundances and column densities of \ce{CH2CCH2} for different values of the \ce{CH2CCH + H} rate constant. {The dashed lines are networks where this rate constant has been modified by the factors in the legend. The base network (presented in this paper) is shown for comparison as a solid line, as well as a network where this reaction is removed entirely. The color gradient corresponds with the magnitude of the factor of change.}}
    \label{fig:ch2cch2_h_sens}
\end{figure}

{The results of our C/O sensitivity tests can be seen in Figure~\ref{fig:COtest} in the Appendix. As the C/O ratio is lowered from 1.1 to 0.7, the late-time abundances ($\sim5\times10^5$ years) of \ce{CH2CCH}, \ce{CH3CCH}, and \ce{CH2CCH2} significantly decrease, with smaller decreases in early-time ($\sim1-2\times10^5$ years) abundances. In particular, the abundance of \ce{CH2CCH} at our best fit time decreases by a factor of 22 when the C/O ratio is lowered from 1.1 to 0.7. The maximum modeled column density of \ce{CH2CCH} using a C/O ratio of 0.7 is almost a factor of 7 lower than the best-fit abundance using a C/O ratio of 1.1. Likewise, the mean confidence level of \ce{CH2CCH}, \ce{CH3CCH}, \ce{CH2CCH2}, and the \ce{C4H3N} isomers decreases from 0.385 to 0.174 when the C/O ratio is changed from 1.1 to 0.7. These results indicate that \ce{CH2CCH} and its closed-shell relatives are somewhat sensitive to the C/O ratio, with a larger value favoring the formation of these species.}

\section{Discussion}
\label{sec:Discussion}

\subsection{Dark Molecular Cloud Chemistry of \ce{C3H3} and \ce{C3H4} Species}

As noted by \citet{agundez_discovery_2021} in their initial detection, \ce{CH2CCH} is one of the most abundant radicals detected in TMC-1, with a derived column density only slightly below that of the closed-shell \ce{CH3CCH}. As the authors remark, this radical is an example of a resonance-stabilized radical, and as such the delocalization of the unpaired electron is expected to lower its reactivity compared to radicals without resonance. In addition to the detection, the authors use astrochemical modeling to investigate the major formation and destruction mechanisms of \ce{CH2CCH} and its closed-shell relatives. They achieve a peak modeled abundance of $\sim1\times10^{-10}$with respect to \ce{H2} similar to our model, but at an earlier time of $2\times10^5$ years. They also obtain much larger modeled abundances of \ce{CH3CCH} and \ce{CH2CCH2} on the order of $10^{-8}$ at a time of $\sim5\times10^5$ years, close to the TMC-1 observed value for \ce{CH3CCH}. 

In terms of production pathways, our models agree that Reaction~\ref{eqa:c2h4-c} is the dominant pathway toward forming \ce{CH2CCH}. The kinetics of this reaction were measured twice from 295 K to 15 K by \citet{chastaing_neutralneutral_1999} and \citet{chastaing_rate_2001} using the CRESU technique and two different product detection techniques. In both cases, the data was fit via a non-linear least-squares fit to a modified Arrhenius form, yielding a temperature-dependent rate constant of $k(T) = 3.1\times10^{-10}(T/298~\rm{K})^{-0.07}$ cm$^3$ molecule$^{-1}$ s$^{-1}$ from the first set of measurements and $k(T) = 3.0\pm0.4\times10^{-10}(T/298~\rm{K})^{-0.11\pm0.07}$ cm$^3$ molecule$^{-1}$ s$^{-1}$ from a fit combining both sets of measurements. The strong agreement between the two sets of measurements and relatively low errors in fit parameters and individual rate constant measurements suggest this reaction is fast at 10 K with a rate constant near $4\times10^{-10}$ cm$^3$ molecule$^{-1}$ s$^{-1}$. The KIDA database and our model presently use the temperature dependent fit from \citet{chastaing_neutralneutral_1999}, while \citet{agundez_discovery_2021} use a slightly smaller, temperature-independent value of $3.1\times10^{-10}$ cm$^3$ molecule$^{-1}$ s$^{-1}$. A number of experimental and theoretical studies \citep{le_ab_2001, bergeat_reaction_2001, geppert_combined_2003, chin_exploring_2012, mandal_dynamics_2018} indicate that \ce{CH2CCH + H} is the only major product channel, resulting from more than 90\% of reactions.

However, \citet{agundez_discovery_2021} also observe that \ce{CH3CCH} and \ce{CH2CCH2} are primarily formed from the dissociative recombinations of \ce{C3H_n+} cations. In our model, the dissociative recombination of \ce{C3H5+} is inefficient compared to gas-phase and grain-surface hydrogenations of \ce{CH2CCH}, and \ce{C3H_n+} cations larger than \ce{C3H5+} are not included in the network. This suggests significant differences in chemical networks, particularly regarding the chemistry of \ce{C3H_n} species. Inclusion of larger neutral and cationic \ce{C3H_n} species and a thorough investigation of their formation paths in cold molecular clouds would benefit modeling of \ce{CH2CCH} and the \ce{C3H4} isomers.

The model of \citet{agundez_discovery_2021} agrees with our model that the major destruction mechanisms of \ce{CH2CCH} are with atoms such as C, N, and O, although they do not consider the radiative association of \ce{CH2CCH} with H. The authors propose that \ce{CH2CCH} could be reproduced by the model if reaction with O and N atoms are removed, or alternatively if the initial C/O ratio is above one. Despite an initial C/O ratio of 1.1, we obtain a similar maximum abundance of \ce{CH2CCH} to their initial model. Additionally, our sensitivity analysis of \ce{CH2CCH} reactions shows that the maximum abundance of \ce{CH2CCH} is not significantly affected by the removal of any individual reaction except for \ce{CH2CCH + H}. This barrierless reaction is estimated to be fairly efficient at low temperature based on a semi-empirical approach involving reaction exothermicity and molecule size, however there is significant uncertainty in the rate constant \citep{hebrard_photochemistry_2013}. We find that \ce{CH2CCH}, \ce{CH3CCH}, and \ce{CH2CCH2} abundances are all sensitive to this rate constant, as this reaction is a major destruction mechanism for the first species and production mechanism for the latter two. A decrease in the rate constant would lead to a significantly larger modeled abundance for \ce{CH2CCH}, which could then be compounded by removal of the \ce{CH2CCH + O} and \ce{CH2CCH + N} reactions. However, such a scenario could result in significant decreases to the modeled abundances of \ce{CH3CCH} and \ce{CH2CCH2}. Potential energy surface calculations on the \ce{CH2CCH + O} and \ce{CH2CCH + N} reactions, as well as an experimental measurement of the \ce{CH2CCH + H} radiative association rate constant and branching ratios, would significantly improve our understanding of propargyl radical chemistry in cold molecular clouds.

Neither our model nor the model of \citet{agundez_discovery_2021} considers the reaction between \ce{CCH} and \ce{CH3} to be a major formation reaction for \ce{CH2CCH}. In our model, the rate for this reaction is slow compared to many other \ce{CH2CCH} production mechanisms at most times. It isn't until $\sim10^6$ years, after the expected age of TMC-1, that this rate becomes large enough for the reaction to significant to \ce{CH2CCH} formation. Both reactants experience significant increases in modeled abundance, over an order of magnitude for \ce{CH3} and slightly less for \ce{CCH}, from $10^5$ years to $10^6$ years. As the major destruction pathways for these species are with C, N, and O atoms, these increases in abundance may be due to the drop in neutral atom abundances during this time period. The rate constant for this reaction was estimated to be $1\times10^{-10}$ molecule$^{-1}$ s$^{-1}$ as an approximate value for a radical-radical association followed by H elimination, but without any experimental or theoretical studies there is significant uncertainty in this value. Our sensitivity analysis of the rate constant for this reaction reveals that an increase in this rate constant could result in a moderate increase to modeled \ce{CH2CCH} abundance, but not until after $6\times10^5$ years. Decreasing this rate constant or removing this reaction only leads to a slight decrease in \ce{CH2CCH} abundance after this time, and a negligible decrease before then.

\subsection{\ce{C4H3N} Isomers and \ce{CH2CCH2} Abundance Constraint}

{In cold molecular clouds, CN radical is thought to react efficiently with many unsaturated hydrocarbons via a CN-addition H-elimination pathway \citep{balucani_formation_2000, carty_low_2001, cooke_benzonitrile_2020}. The ability of our model to reproduce the observed column densities of \ce{CH2CCHCN} and \ce{CNCH2CCH} within a factor of 5 further demonstrates the importance of this mechanism in forming interstellar CN-derivatives. The column density of \ce{CH3C3N} is also reproduced by our model within a factor of 5, however the major mechanism for this species is the dissociative recombination of \ce{H3C4NH+}. In addition to a 10\% branching ratio from the Reaction~\ref{eqa:CN+CH2CCH2}, we assumed a radiative association reaction between \ce{CH2CCH} and \ce{CN} to produce \ce{CNCH2CCH} with a rate constant of \SI{1.0e-10}{\per\square\centi\metre}. Our sensitivity analysis of this reaction reveals that an order of magnitude increase in this rate constant could reproduce the observed \ce{CNCH2CCH}, however such a large rate constant is unreasonable for an association reaction between two neutral species. Conversely, lowering this rate constant or removing the reaction entirely only slightly lowers the best-fit abundance, resulting in a value still within an order of magnitude compared to observations. As the major formation route to \ce{CNCH2CCH} and \ce{CH2CCHCN} is through \ce{CH2CCH2}, it is possible that a deficiency in modeled \ce{CH2CCH2} is responsible for the deficiencies in these two \ce{C4H3N} isomers.}

{\citet{quan_possible_2007} have performed astrochemical modeling on \ce{CH2CCHCN} and \ce{CH3C3N}, testing the effects of different \ce{CH3CCH}/\ce{C3H4} ratios on abundances at $10^5$ years. Using branching ratios of 50/50 \ce{CH2CCHCN}/\ce{CH3C3N} for \ce{CN + CH3CCH} and 90/10 \ce{CH2CCHCN}/\ce{CNCH2CCH} for \ce{CN + CH2CCH2}, the authors find the best agreement with observations when the \ce{CH3CCH} abundance is 35\% of the total \ce{C3H4} abundance. More recent branching ratio measurements for \ce{CN + CH3CCH} suggest that \ce{CH2CCHCN} is not formed significantly at low temperature \citep{abeysekera_product_2015}, and the detection of \ce{CNCH2CCH} allows for the inclusion of the third \ce{C3H4N} isomer in the analysis. More recently, \citet{marcelino_study_2021} examined all three \ce{C4H3N} isomers using astrochemical modeling, testing both sets of experimental branching ratios for \ce{CN + CH3CCH}. The authors obtain very similar results to our own model, finding good agreement with observations for all three species. As in our study, they find that the dissociative recombination of \ce{H3C4NH+} is mainly responsible for the formation of \ce{CH3C3N}. They also find that the choice in branching ratios for \ce{CN + CH3CCH} does not notably affect the resulting modeled abundances.}

Since \ce{CH2CCH2} is a symmetric hydrocarbon and does not possess a permanent dipole moment, it is not possible to detect this species via radio astronomy. However, a rough estimate to its TMC-1 abundance can be made using our modeled CH/CN ratio and the observed abundance of \ce{CH2CCHCN}. Taking the ratio of \ce{CH2CCH2} to \ce{CH2CCHCN} at our best-fit time, 6.1, and multiplying it by the observed column density of \ce{CH2CCHCN}, we obtain an extrapolated \ce{CH2CCH2} column density of \SI{1.64e13}{\per\square\centi\metre}. This value is almost an order of magnitude lower than the observed column density of \ce{CH3CCH} and would suggest a difference in interstellar chemistry between these species that is not present in our model, although it is within the one $\sigma$ uncertainty of the observed \ce{CH3CCH} column density. It is important to note that modeled CH/CN ratios are sensitive to the model parameters used, such as the rate constant for the CN-addition H-elimination pathway \cite{sita_discovery_2022}, as well as the time of choice. Although the \ce{CN + CH2CCH2} rate constant has been measured at low temperature \citep{carty_low_2001}, the branching ratios have not. Additionally, CH/CN ratios could be significantly different in other interstellar regions with different physical and chemical conditions. Thus this method of column density approximation is only a rough estimate based on certain modeling parameters and our time of best fit. A detailed analysis of CH/CN ratios in TMC-1, as well as a more constrained column density for \ce{CH3CCH}, would be beneficial toward understanding the relationship between these two isomers and their CN-substituted derivatives.

\subsection{Implications for PAH formation}

\ce{C6H5} is an aromatic resonance-stabilized radical and a key precursor to formation of PAHs in cold molecular clouds. The Hydrogen Abstraction-Vinylacetylene Addition (HAVA) mechanism, which begins with the addition of \ce{C6H5} to vinylacetylene followed by rearrangement and hydrogen loss, is one viable low-temperature formation pathway to \ce{C10H8} and larger PAHs \citep{parker_low_2012, kaiser_aromatic_2021}. This mechanism is the major generator of \ce{C10H8} in our chemical network. Likewise, indene can be formed via reactions of \ce{C6H5} with small hydrocarbons, although these reactions are so far limited to high-temperature \citep{doddipatla_low-temperature_2021, kaiser_aromatic_2021}. An understanding of the production of the \ce{C6H5} in TMC-1 is critical for describing the formation of \ce{C10H8}, indene, and other PAHs in cold molecular clouds. {A major motivation of modeling \ce{CH2CCH} was to assess its role as an interstellar aromatic precursor, however addition of the \ce{CH2CCH} recombination reaction resulted in no significant change to \ce{C6H5} abundance after 100 years.} Based on the sensitivity analysis of this reaction, increasing this rate constant by an order of magnitude does not appreciably change the best-fit abundance of \ce{C6H5}, nor would this increase agree with the temperature-dependence calculated by \citet{georgievskii_association_2007}. Thus the abundance of \ce{CH2CCH} appears to be the limiting factor. As the \ce{CH2CCH} recombination reaction is second order in \ce{CH2CCH} abundance, an underproduction of \ce{CH2CCH} by almost two orders of magnitude would lead to a recombination rate that is lowered by almost four orders of magnitude. Continuing to improve our understanding of propargyl radical chemistry is necessary to assess the role of this species in PAH formation. Additionally, a detection of \ce{C6H5} in TMC-1 and a derived column density would allow us to better gauge our knowledge of interstellar aromatic chemistry.

Additional reactions of resonance-stabilized radicals, including \ce{CH2CCH}, may be important to the formation of aromatic species in dark molecular clouds. In combustion flames, \ce{CH2CCH} is thought to react with allyl radicals (\ce{C3H5}) to form fulvene (\ce{C5H4CH2}),
\begin{equation}
    \label{eqa:C3H5}
        \ce{CH2CCH + C3H5 -> C4H4CH2 + H + H},
\end{equation}
followed by isomerization to benzene \citep{miller_reactions_2010}. The authors calculated the potential energy surface for this system and found it to be a barrierless addition with two exothermic, bimolecular exit channels corresponding to two isomers of the hydrofulvenyl radical (\ce{C5H5CH2}). They also predicted the rate constant for this reaction to be on the order of $10^{-10}$ \unit{\cubic\centi\metre\per\second} at low temperature, independent of the pressure. Moreover, calculations on the \ce{C6H5 + CH2CCH} potential energy surface have been performed by \citet{morozov_theoretical_2020}, resulting in two barrierless sites for addition of \ce{CH2CCH} to \ce{C6H5}. The addition of \ce{CH2CCH} via its \ce{CH2} terminal yields 3-phenyl-1-propyne (\ce{C6H5CH2CCH}), which can decompose via H-loss directly to form \ce{C6H5CHCCH} or after isomerization to form the indenyl radical (\ce{C9H7})
\begin{align}
\label{phenylpropyne}
    \ce{CH2CCH + C6H5} & \ce{-> C6H5CH2CCH} \nonumber \\
    \ce{C6H5CH2CCH}    & \ce{-> C6H5CHCCH + H} \nonumber \\
                       & \ce{-> C9H7 + H.}
\end{align}
Addition of \ce{CH2CCH} via its \ce{CH} terminal results in phenylallene (\ce{C6H5CHCCH2}), which can also isomerize to \ce{C9H7} or immediately decompose to form two monocyclic H-loss products
\begin{align}
\label{phenylallene}
    \ce{CH2CCH + C6H5} & \ce{-> C6H5CHCCH2} \nonumber \\ 
    \ce{C6H5CHCCH2}    & \ce{-> C6H5CCCH2 + H} \nonumber \\
                       & \ce{ -> C6H5CCCH2 + H} \nonumber \\
                       & \ce{ -> C6H5CHCCH + H.} 
\end{align}
The mixture of these products is not known at temperatures relevant to dark molecular clouds, nor is the rate constant, but this reaction presents a promising formation route for monocyclic and bicyclic aromatic species. The radical-radical reaction between \ce{CH2CCH} and the benzyl radical (\ce{C7H7}) may be yet another key process to PAH formation in dark molecular clouds. A recent theoretical study \citep{krasnoukhov_formation_2022} on the \ce{C7H7 + CH2CCH} potential energy surface has shown that \ce{CH2CCH} can add to the \ce{-CH2} group of \ce{C7H7} without barrier, leading to a variety of exothermic bimolecular exit channels. Possible products consist of bicyclic aromatic species such as methyleneindanyl radicals (\ce{C9H7CH2}), methyleneindenes (\ce{C9H6CH2}), H-naphthalenyl (\ce{C10H9}), and naphthalene (\ce{C10H8}). In the high temperature environments of asymptotic branch stars, \ce{C7H7} is formed from H-abstraction of toluene (\ce{C7H8}), which itself is formed from benzene or \ce{C6H5} \citep{krasnoukhov_formation_2022}. It is currently not known whether \ce{C7H8} or \ce{C7H7} can be formed efficiently under dark molecular cloud conditions. Likewise, computational studies of fulvenallene ({\ce{C7H6}}) and the corresponding fulvenallenyl radical (\ce{C7H5}) in combustion environments suggest that the \ce{C7H5} radical can be formed from H-abstraction of \ce{C7H6} and further react to form PAHs, such as naphthalene via reaction with \ce{CH2CCH} \citep{da_silva_c7h5_2009}. Fulvenallene has been detected recently in TMC-1, with the reaction between cyclopentadiene and the ethynyl radical proposed as a possible formation pathway \citep{cernicharo_discovery_2022}. Further investigation of these larger resonance-stabilized radicals under dark molecular cloud conditions will greatly benefit the understanding of PAH formation in these regions.

\section{Conclusions}
We performed astrochemical modeling on the propargyl radical, \ce{CH2CCH}, and related species using the \texttt{NAUTILUS} code and an updated chemical network. We also incorporate two new species, \ce{CH2CCH2} and \ce{CH2CCHCN}, into the network. We find that the predicted abundance of \ce{CH2CCH} is improved by about an order of magnitude, but is still almost two orders of magnitude below the observed value. The \ce{C4H3N} isomers are found to be formed primarily from \ce{CH3CCH}, \ce{CH2CCH2}, and \ce{CH2CCH}. We obtain predicted abundances for these \ce{C4H3N} isomers within one order of magnitude of the observed values, suggesting that our chemical model is able to account for the major formation mechanisms of these species. We do not observe any significant improvements for aromatic species, but \ce{CH2CCH} remains a potential precursor to aromatic species in dark molecular clouds. Further studies of interstellar \ce{C3H_n} and resonance-stabilized radical chemistry under conditions relevant to dark molecular clouds would provide key information for improved astrochemical modeling of the species presented here. 

\section{Acknowledgements}
We thank V. Wakelam for use of the \texttt{NAUTILUS} v1.1 code. B. A. M. and C. X. gratefully acknowledge support of National Science Foundation grant AST-2205126. A. N. B thanks Dr. Troy Van Voorhis for his support and advice.

\clearpage

\bibliographystyle{aasjournal}
\bibliography{propargyl}

\appendix
\section{Dissociative Recombination Reactions}
\label{apx:DR}
In Table \ref{tab:DR}, new and updated rate coefficients {($k$)} are listed for the dissociative recombination of \ce{C3H4+} and \ce{C3H5+}, two reactions that produce \ce{CH2CCH}. As stated in Section 2.1.1 of the main text, the branching ratios for \ce{C3H4+} were experimentally determined by \citet{geppert_dissociative_2004}, while the branching ratios for \ce{C3H5+} were estimated by \citet{loison_interstellar_2017} based on experimental results by \citet{angelova_branching_2004} and photodissociation studies of the \ce{C3H4} isomers. As the exact branching ratios for this latter species are estimates, there is still significant uncertainty regarding these values. The reactions with a listed rate constant of 0 are ones that have been removed from the chemical network. 

\begin{table}[htbp]
    \scriptsize
    \caption{\textbf{Related DR Reactions}}
    \label{tab:DR}
    \begin{threeparttable}
    \centering
    \begin{tabularx}{\textwidth}{@{}
    l
    S[table-format=1.2e-2, table-alignment-mode=none]
    S[table-format=-1.2, table-alignment-mode=none]
    S[table-format=4, table-alignment-mode=none]
    S[table-format=1.2e-2]
    l @{}}
    \toprule
    {Reactions} & {$\alpha$ {(\unit{\cubic\centi\metre\per\second})}} & {$\beta$} & {$\gamma$ (\unit{\kelvin})} & {k (10 \unit{\kelvin}) (\unit{\cubic\centi\metre\per\second})\tnote{1}} & References \\
    \midrule
    {\ce{C3H4+ + e- -> H + CH2CCH}} & 6.00e-07 & -0.70 & 0 & 6.49e-06 & {Branching ratios and rate constant from \citet{loison_interstellar_2017}},\\
    {\ce{C3H4+ + e- -> CCH + CH3}} & 1.00e-08 & -0.70 & 0 & 1.08e-07 & {based on branching ratios from \citet{geppert_dissociative_2004} and a} \\
    {\ce{C3H4+ + e- -> CH2 + C2H2}} & 4.00e-08 & -0.70 & 0 & 4.33e-07 & {lower rate constant in line with other CRYRING} \\
    {\ce{C3H4+ + e- -> H + H + c-C3H2}} & 1.00e-08 & -0.70 & 0 & 1.08e-07 & {measurements \citep{larsson_rate_2005, fournier_novel_2013}.} \\
    {\ce{C3H4+ + e- -> H + H + l-C3H2}} & 1.00e-08 & -0.70 & 0 & 1.08e-07 & \\
    {\ce{C3H4+ + e- -> H2 + l-C3H2}} & 0 & 0 & 0 & 0 & \\
    {\ce{C3H4+ + e- -> H2 + c-C3H2}} & 0 & 0 & 0 & 0 & \\
    {\ce{C3H4+ + e- -> CH + C2H3}} & 0 & 0 & 0 & 0 &  \\    
    &&&&&\\
    {\ce{C3H5+ + e- -> H2 + CH2CCH}} & 4.20e-08 & -0.70 & 0 & 4.54e-07 & {Branching ratios and rate constant from \citet{loison_interstellar_2017},} \\
    {\ce{C3H5+ + e- -> H + CH3CCH}} & 7.00e-08 & -0.70 & 0 & 7.57e-07 & {assuming same overall rate as \ce{C3H4+} DR. 87\% \ce{C3} channel} \\
    {\ce{C3H5+ + e- -> H + H + CH2CCH}} & 1.10e-07 & -0.70 & 0 & 1.19e-06 & {and 13\% \ce{C2 + C} channel from \citet{angelova_branching_2004},} \\
    {\ce{C3H5+ + e- -> H + H2 + c-C3H2}} & 1.00e-07 & -0.70 & 0 & 1.08e-06 & {with exact branching ratios estimated from \ce{C3H4}} \\
    {\ce{C3H5+ + e- -> H + H2 + l-C3H2}} & 5.00e-08 & -0.70 & 0 & 5.41e-07 & {photodissociation studies.} \\
    {\ce{C3H5+ + e- -> C2H2 + CH3}} & 4.70e-08 & -0.70 & 0 & 5.08e-07 & \\
    {\ce{C3H5+ + e- ->  H + CH2 + C2H2}} & 4.70e-08 & -0.70 & 0 & 5.08e-07 &  \\
    {\ce{C3H5+ + e- -> H + CH2CCH2}} & 1.10e-07 & -0.70 & 0 & 1.19e-06 &  \\
    \bottomrule
    \end{tabularx}
    
    \begin{tablenotes}
        \item[1] The rate constants of dissociative recombination reactions are computed using the modified Arrhenius equation: $k = \alpha\left(T/300\right)^{\beta}e^{-\gamma/T}${, where $\alpha$ is a temperature-independent prefactor, $\beta$ is a unitless temperature-dependence parameter, and $\gamma$ is the activation energy in K.}
    \end{tablenotes}
    
    \end{threeparttable}
\end{table}

\clearpage

\section{Ion-Molecule Reactions}
\label{apx:Ion-neutral}
Table \ref{tab:Ion-neutral} lists the updated rate coefficients {($k$)} for the reactions of \ce{CH2CCH} with interstellar cations while table \ref{tab:Ion-neutral2} lists those of four species chemically related to \ce{CH2CCH} with abundant interstellar cations. The rate coefficients were calculated using the Su-Chesnavich equations and the dipole and polarizability from Table~\ref{tab:dipoles}. In the case of multiple product channels for one reaction, the branching ratios are assumed to be an even split among all product channels. In actuality this is likely not the case, however these branching ratios do not have a notable effect on our model as these reactions primarily act as destruction mechanisms.

\begin{table}[htbp]
    \footnotesize
    \caption{\textbf{Ion-Molecule Reactions of Propargyl Radical}}
    \label{tab:Ion-neutral}
    \begin{threeparttable}
    \centering
    \begin{tabularx}{\textwidth}{@{}
    l
    S[table-format=1.3e-2, table-alignment-mode=none]
    S[table-format=-1.3e-2, table-alignment-mode=none]
    S[table-format=1.4, table-alignment-mode=none]
    S[table-format=1.2e-2]
    c @{}}
    \toprule
    {Reactions} & {$\alpha$} & {$\beta$} & {$\gamma$ (\unit{\kelvin})} & {k (10 \unit{\kelvin}) (\unit{\cubic\centi\metre\per\second})} & Formula\tnote{1}\\
    \midrule
    \ce{CH2CCH + H+ -> H2 + c-C3H2+} & 0.25 & 5.570e-9 & 0.2211 & 1.66e-9 & 5 \\
    \ce{CH2CCH + H+ -> H2 + l-C3H2+} & 0.25 & 5.570e-9 & 0.2211 & 1.66e-9 & 5 \\
    \ce{CH2CCH + H+ -> H + c-C3H3+} & 0.25 & 5.570e-9 & 0.2211 & 1.66e-9 & 5 \\
    \ce{CH2CCH + H+ -> H + l-C3H3+} & 0.25 & 5.570e-9 & 0.2211 & 1.66e-9 & 5 \\
    \ce{CH2CCH + He+ -> H + He + H2 + C3+} & 0.25 & 2.888e-9 & 0.2211 & 8.64e-10 & 5 \\
    \ce{CH2CCH + He+ -> He + H2 + C3H+} & 0.25 & 2.888e-9 & 0.2211 & 8.64e-10 & 5 \\
    \ce{CH2CCH + He+ -> H + He + c-C3H2+} & 0.25 & 2.888e-9 & 0.2211 & 8.64e-10 & 5 \\
    \ce{CH2CCH + He+ -> H + He + l-C3H2+} & 0.25 & 2.888e-9 & 0.2211 & 8.64e-10 & 5 \\
    \ce{CH2CCH + C+ -> H2 + C4H+} & 0.25 & 1.816e-9 & 0.2211 & 5.43e-10 & 5 \\ 
    \ce{CH2CCH + C+ -> C + l-C3H3+} & 0.25 & 1.816e-9 & 0.2211 & 5.43e-10 & 5 \\
    \ce{CH2CCH + C+ -> H + C4H2+} & 0.25 & 1.816e-9 & 0.2211 & 5.43e-10 & 5 \\
    \ce{CH2CCH + C+ -> C + c-C3H3+} & 0.25 & 1.816e-9 & 0.2211 & 5.43e-10 & 5 \\
    \ce{CH2CCH + Si+ -> H + SiC3H2+} & 1 & 1.362e-9 & 0.2211 & 1.63e-9 & 5 \\
    \ce{CH2CCH + H3+ -> H2 + C3H4+} & 1 & 3.296e-9 & 0.2211 & 3.95e-9 & 5 \\
    \ce{CH2CCH + HCO+ -> CO + C3H4+} & 1 & 1.349e-9 & 0.2211 & 1.61e-9 & 5 \\
    \ce{CH2CCH + H3O+ -> H2O + C3H4+} & 1 & 1.539e-9 & 0.2211 & 1.84e-9 & 5 \\
    \ce{CH2CCH + CH3+ -> H + C4H5+} & 1 & 1.671e-9 & 0.2211 & 2.00e-9 & 5 \\
    \ce{CH2CCH + C2H2+ -> H2 + C5H3+} & 0.25 & 1.393e-9 & 0.2211 & 4.17e-10 & 5 \\
    \ce{CH2CCH + C2H2+ -> H + C5H4+} & 0.25 & 1.393e-9 & 0.2211 & 4.17e-10 & 5 \\
    \ce{CH2CCH + C2H2+ -> C2H2 + c-C3H3+} & 0.25 & 1.393e-9 & 0.2211 & 4.17e-10 & 5 \\
    \ce{CH2CCH + C2H2+ -> C2H2 + l-C3H3+} & 0.25 & 1.393e-9 & 0.2211 & 4.17e-10 & 5 \\
    \ce{CH2CCH + C2H3+ -> C2H2 + C3H4+} & 0.33 & 1.377e-9 & 0.2211 & 5.50e-10 & 5 \\
    \ce{CH2CCH + C2H3+ -> H2 + C5H4+} & 0.33 & 1.377e-9 & 0.2211 & 5.50e-10 & 5 \\
    \ce{CH2CCH + C2H3+ -> H + C5H5+} & 0.33 & 1.377e-9 & 0.2211 & 5.50e-10 & 5 \\
    \ce{CH2CCH + c-C3H2+ -> H + C6H4+} & 1 & 1.254e-9 & 0.2211 & 1.50e-9 & 5 \\
    \ce{CH2CCH + l-C3H2+ -> H + C6H4+} & 1 & 1.254e-9 & 0.2211 & 1.50e-9 & 5 \\
    \ce{CH2CCH + C4H+ -> H + C7H3+} & 1 & 1.180e-9 & 0.2211 & 1.41e-9 & 5 \\
    \ce{CH2CCH + C2H4+ -> C2H4 + c-C3H3+} & 0.25 & 1.362e-9 & 0.2211 & 4.08e-10 & 5 \\
    \ce{CH2CCH + C2H4+ -> C2H4 + l-C3H3+} & 0.25 & 1.362e-9 & 0.2211 & 4.08e-10 & 5 \\
    \ce{CH2CCH + C2H4+ -> CH4 + C4H3+} & 0.25 & 1.362e-9 & 0.2211 & 4.08e-10 & 5 \\
    \ce{CH2CCH + C2H4+ -> H2 + C5H5+} & 0.25 & 1.362e-9 & 0.2211 & 4.08e-10 & 5 \\
    \ce{CH2CCH + c-C3H3+ -> H2 + C6H4+} & 0.50 & 1.246e-9 & 0.2211 & 7.46e-10 & 5 \\
    \ce{CH2CCH + c-C3H3+ -> H + C6H5+} & 0.50 & 1.246e-9 & 0.2211 & 7.46e-10 & 5 \\
    \ce{CH2CCH + l-C3H3+ -> H2 + C6H4+} & 0.50 & 1.246e-9 & 0.2211 & 7.46e-10 & 5 \\
    \ce{CH2CCH + l-C3H3+ -> H + C6H5+} & 0.50 & 1.246e-9 & 0.2211 & 7.46e-10 & 5 \\
    \ce{CH2CCH + C4H2+ -> H2 + C7H3+} & 0.50 & 1.175e-9 & 0.2211 & 7.03e-10 & 5 \\
    \ce{CH2CCH + C4H2+ -> H + C7H4+} & 0.50 & 1.175e-9 & 0.2211 & 7.03e-10 & 5\\
    \ce{CH2CCH + C4H3+ -> H2 + C7H4+} & 0.50 & 1.170e-9 & 0.2211 & 7.00e-10 & 5\\
    \ce{CH2CCH + C4H3+ -> H + C7H5+} & 0.50 & 1.170e-9 & 0.2211 & 7.00e-10 & 5\\
    \ce{CH2CCH + C5H2+ -> H2 + C8H3+} & 0.50 & 1.124e-9 & 0.2211 & 6.73e-10 & 5 \\
    \ce{CH2CCH + C5H2+ -> H + C8H4+} & 0.50 & 1.124e-9 & 0.2211 & 6.73e-10 & 5 \\
    \ce{CH2CCH + C6H2+ -> H2 + C9H3+} & 0.50 & 1.088e-9 & 0.2211 & 6.52e-10 & 5 \\
    \ce{CH2CCH + C6H2+ -> H2 + C9H3+} & 0.50 & 1.088e-9 & 0.2211 & 6.52e-10 & 5 \\
    \bottomrule
    \end{tabularx}
    
    \begin{tablenotes}
        \item[1] The rate constants for reactions with formula 5 are calculated using the ion-polar formula 2, 
        $k = \alpha \beta \left(1+0.0967\gamma \left(\dfrac{300}{T}\right)^{1/2}+\dfrac{\gamma^2}{10.526}\dfrac{300}{T}\right)$. $\alpha$ is the branching ratio, $\beta$ is the Langevin rate in units of \unit{\cubic\centi\metre\per\second} given by the expression $\beta=2\pi e \sqrt{\dfrac{\alpha}{\mu}}$, and $\gamma=x=\dfrac{\mu_D}{\sqrt{2 \alpha k_B T}}$ is a unitless correction term based on the dipole moment and polarizability of the neutral species. See Table 2 of the main text for the dipole moment and polarizability values used in these calculations.
    \end{tablenotes}
    
    \end{threeparttable}
\end{table}

\begin{table}[htbp]
    \small
    \caption{\textbf{Ion-Molecule Reactions of CN-substituted species}}
    \label{tab:Ion-neutral2}
    \begin{threeparttable}
    \centering
    \begin{tabularx}{1.05\textwidth}{@{}
    l
    S[table-format=1.3e-2, table-alignment-mode=none]
    S[table-format=-1.3e-2, table-alignment-mode=none]
    S[table-format=1.4, table-alignment-mode=none]
    S[table-format=1.2e-2]
    c @{}}
    \toprule
    {Reactions} & {$\alpha$} & {$\beta$} & {$\gamma$ (\unit{\kelvin})} & {k (10 \unit{\kelvin}) (\unit{\cubic\centi\metre\per\second})} & Formula\tnote{1}\textsuperscript{,}\tnote{2}\\
    \midrule
\sidehead{1-Cyano Propargyl (\ce{CNCHCCH})}
    \ce{CNCHCCH + H+ -> CN+ + CH2CCH} & 1.597e-8 & -0.5 & 0 & 8.75e-8 & 3 \\
    \ce{CNCHCCH + C+ -> C2N+ + l-C3H2} & 4.985e-9 & -0.5 & 0 & 2.73e-8 & 3 \\
    \ce{CNCHCCH + He+ -> He + CN+ + l-C3H2} & 8.167e-9 & -0.5 & 0 & 4.47e-8 & 3 \\
    \ce{CNCHCCH + H3+ -> H2 + CN+ + CH2CCH} & 9.361e-9 & -0.5 & 0 & 5.13e-8 & 3 \\
    \ce{CNCHCCH + HCO+ -> HCO + CN+ + l-C3H2} & 3.547e-9 & -0.5 & 0 & 1.94e-8 & 3 \\
    \ce{CNCHCCH + H3O+ -> H2O + HCN+ + l-C3H2} & 4.140e-9 & -0.5 & 0 & 2.27e-8 & 3 \\
\sidehead{Propargyl Cyanide (\ce{CNCH2CCH})}
    \ce{CNCH2CCH + H+ -> HCN+ + CH2CCH} & 1.00 & 6.299e-9 & 4.683 & 8.09e-8 & 4 \\
    \ce{CNCH2CCH + He+ -> He + CN+ + CH2CCH} & 1.00 & 3.220e-9 & 4.683 & 4.14e-8 & 4 \\
    \ce{CNCH2CCH + C+ -> C2N+ + CH2CCH} & 1.00 & 1.964e-9 & 4.683 & 2.52e-8 & 4 \\
    \ce{CNCH2CCH + H3+ -> H2 + HCN+ + CH2CCH} & 1.00 & 3.691e-9 & 4.683 & 4.74e-8 & 4 \\
    \ce{CNCH2CCH + HCO+ -> HCO + CN+ + CH2CCH} & 1.00 & 1.396e-9 & 4.683 & 1.79e-8 & 4 \\
    \ce{CNCH2CCH + H3O+ -> H2O + HCN+ + CH2CCH} & 1.00 & 1.630e-9 & 4.683 & 2.09e-8 & 4 \\
 \sidehead{Cyanoallene (\ce{CH2CCHCN})}
    \ce{CH2CCHCN + H+ -> HCN+ + CH2CCH} & 1.00 & 6.707e-09 & 5.213 & 9.55e-08 & 4 \\
    \ce{CH2CCHCN + He+ -> He + CN+ + CH2CCH} & 1.00 & 3.429e-09 & 5.213 & 4.88e-08 & 4 \\
    \ce{CH2CCHCN + C+ -> C2N+ + CH2CCH} & 1.00 & 2.091e-09 & 5.213 & 2.98e-08 & 4 \\
    \ce{CH2CCHCN + H3+ -> H2 + HCN+ + CH2CCH} & 1.00 & 3.931e-09 & 5.213 & 5.59e-08 & 4 \\
    \ce{CH2CCHCN + HCO+ -> HCO + CN+ + CH2CCH} & 1.00 & 1.486e-09 & 5.213 & 2.12e-08 & 4 \\
    \ce{CH2CCHCN + H3O+ -> H2O + HCN+ + CH2CCH} & 1.00 & 1.736e-09 & 5.213 & 2.47e-08 & 4 \\
 \sidehead{Cyanovinylacetylene (\ce{CNCHCHCCH})}
    \ce{C4H3CN + H+ -> HCN+ + CH2CCH + C} & 1.649e-8 & -0.5 & 0 & 9.03e-8 & 3 \\
    \ce{C4H3CN + He+ -> He + CN+ + CH2CCH + C} & 8.404e-9 & -0.5 & 0 & 4.60e-8 & 3 \\
    \ce{C4H3CN + C+ -> C2N+ + CH2CCH + C} & 5.086e-9 & -0.5 & 0 & 2.79e-8 & 3 \\
    \ce{C4H3CN + H3+ -> CH2 + HCN+ + CH2CCH} & 9.644e-9 & -0.5 & 0 & 5.28e-8 & 3 \\
    \ce{C4H3CN + HCO+ -> HCO + CN+ + CH2CCH + C} & 3.570e-9 & -0.5 & 0 & 1.96e-8 & 3 \\
    \ce{C4H3CN + H3O+ -> H2O + HCN+ + CH2CCH + C} & 4.198e-9 & -0.5 & 0 & 2.30e-8 & 3 \\
    \bottomrule
    \end{tabularx}
    
    \begin{tablenotes}
        \item[1] The rate constants for reactions with formula 4 are calculated using the ion-polar formula 1, $k = \alpha \beta \left(0.62+0.4767\gamma \left(\dfrac{300}{T}\right)^{1/2}\right)$, where $\alpha$ is the branching ratio, $\beta=2\pi e \sqrt{\dfrac{\alpha}{\mu}}$ is the Langevin rate {in units of \unit{\cubic\centi\metre\per\second}}, and $\gamma=x=\dfrac{\mu_D}{\sqrt{2 \alpha k_B T}}$ is a {unitless} correction term based on the dipole moment and polarizability of the neutral species.

        \item[2] Reactions with formula 3 are calculated using the modified Arrhenius formula, $k = \alpha\left(T/300\right)^{\beta}e^{-\gamma/T}$. Here, $\alpha$ {is in units of \unit{\cubic\centi\metre\per\second} and} is calculated as a portion of the Langevin rate multiplied by the dipole-correction term $x$, $\alpha=0.47672\times 2 \pi e \sqrt{\dfrac{\alpha}{\mu}}\hspace{1 pt}x$. {$\beta$ is unitless temperature-dependence term and $\gamma$ is the activation energy in K.}
    \end{tablenotes}
    
    \end{threeparttable}
\end{table}

\clearpage

\section{Neutral-Neutral Reactions}
\label{sec:neutral-neutral}
Tables~\ref{tab:Neutral-neutral1} and \ref{tab:Neutral-neutral2} contain new and updated rate coefficients for neutral-neutral gas-phase reactions involving \ce{CH2CCH2}, \ce{CH3CCH}, and \ce{CH2CCH}. Table~\ref{tab:Grain} contains new and updated rate information for grain reactions pertaining to \ce{C3H_n} species.

\begin{table}[htbp]
    \scriptsize
    \caption{\textbf{Related Neutral-Neutral Reactions}}
    \label{tab:Neutral-neutral1}
    \begin{threeparttable}
    \centering
    \begin{tabularx}{\textwidth}{@{}
    l
    S[table-format=-1.2e-2, table-alignment-mode=none]
    S[table-format=-1.2, table-alignment-mode=none]
    S[table-format=4, table-alignment-mode=none]
    S[table-format=1.2e-2, table-alignment-mode=none]
    l@{}}
    \toprule
    {Reactions\tnote{1}} & {$\alpha$ (\unit{\cubic\centi\metre\per\second})} & {$\beta$} & {$\gamma$ (\unit{\kelvin})} & {k (10 \unit{\kelvin}) (\unit{\cubic\centi\metre\per\second})\tnote{2}} & Reference \\
    \midrule
    \sidehead{\ce{CH2CCH2} Reactions}
    {*\ce{C2H4 + CH -> H + CH2CCH2}} & 2.74e-10 & 0 & 0 & 2.74e-10 & 23 K rate constant from \citet{canosa_reactions_1997} with \\
    &&&&& 70\% branching ratio from \citet{goulay_cyclic_2009}. \\
    &&&&& \\
    {*\ce{H + CH2CCH -> CH2CCH2 + Photon}} & 1.00e-13 & -1.50 & 0 & 1.64e-11 & Rate constant and branching ratios from \\
    &&&&& \citet{loison_interstellar_2017}, based on \citet{hebrard_photochemistry_2013} \\
    &&&&& with a slightly larger rate and 50\% branching ratio. \\
    &&&&& \\
    {*\ce{C + CH2CCH2 -> H + C4H3}} & 3.50e-10 & -0.01 & 0 & 3.62e-10 & Rate constant from \citet{chastaing_rate_2000} and room\\
    &&&&& temperature products from \citet{loison_reaction_2004}. \\
    &&&&& \\
    *\ce{CCH + CH2CCH2 -> CH3C4H + H} & 3.00e-10 & 0 & 0 & 3.00e-10 & Rate constant from \citet{loison_interstellar_2017} based \\
    &&&&& {on measurements by \citet{carty_low_2001}.} \\
    &&&&& \\
    *\ce{CH + CH2CCH2 -> CH2CHC2H + H}\tnote{3} & 4.20e-10 & 0 & 0 & 4.20e-10 & Rate constant from \citet{daugey_kinetic_2005}. \\
    &&&&& \\
    *\ce{CN + CH2CCH2 -> CH2CCHCN + H} & 3.690e-10 & 0 & 0 & 3.690e-10 & Temperature-independent rate constant recommended \\
    *\ce{CN + CH2CCH2 -> CNCH2CCH + H} & 4.10e-11 & 0 & 0 & 3.00e-10 & by \citet{carty_low_2001} with branching ratios \\
    &&&&& {from \citet{balucani_formation_2002}.} \\
    \sidehead{\ce{CH3CCH} Reactions}
    {\ce{CH + C2H4 -> CH3CCH + H}} & 1.12e-10 & 0 & 0 & 1.12e-10 & 23 K rate constant from \citet{canosa_reactions_1997} with \\
    &&&&& 30\% branching ratio from \citet{goulay_cyclic_2009}. \\
    &&&&& \\
    *\ce{H + CH2CCH -> CH3CCH + Photon} & 1.00e-13 & -1.50 & 0 & 1.64e-11 & Rate constant and branching ratios from \\
    &&&&& \citet{loison_interstellar_2017}, based on \citet{hebrard_photochemistry_2013} \\ 
    &&&&& with a slightly larger rate and 50\% branching ratio. \\
    &&&&& \\
    {\ce{C + CH3CCH -> H + C4H3}} & 2.30e-10 & -0.11 & 0 & 3.34e-10 & Rate constant from \citet{chastaing_rate_2000} and  \\
    {\ce{C + CH3CCH -> H2 + C4H2}} & 4.05e-11 & -0.11 & 0 & 5.88e-11 & branching ratios from \citet{loison_reaction_2004}, \\
    &&&&& assuming 15\% \ce{H2} loss channel. \\ 
    &&&&& \\
    \ce{CH + CH3CCH -> CH2CHC2H + H}\tnote{3} & 4.20e-10 & 0 & 0 & 4.20e-10 & Rate constant from \citet{daugey_kinetic_2005}. \\
    &&&&& \\
    *\ce{CCH + CH3CCH -> CH3C4H + H} & 3.00e-10 & 0 & 0 & 3.00e-10 & Rate constant from \citet{loison_interstellar_2017} based \\
    &&&&& on measurements by \citet{carty_low_2001}. \\
    &&&&& \\
    *\ce{CN + CH3CCH -> HC3N + CH3} & 2.71e-10 & 0 & 0 & 2.706e-10 & {Rate constant from \citet{carty_low_2001} and } \\
    *\ce{CN + CH3CCH -> CH3C3N + H} & 9.020e-11 & 0 & 0 & 9.020e-11 & {branching ratios from \citet{abeysekera_product_2015}.} \\
    *\ce{CN + CH3CCH -> HCN + CH2CCH} & 4.920e-11 & 0 & 0 & 4.920e-11 & \\
    \bottomrule
    \end{tabularx}

\begin{tablenotes}

    \item[1] An asterisk preceding a reaction denotes that the reaction has been added to the network rather than modified.
    
    \item[2] The rates of the listed neutral-neutral reactions are computed using the modified Arrhenius equation: $k = \alpha\left(T/300\right)^{\beta}e^{-\gamma/T}$ {, where $\alpha$ is a temperature-independent prefactor, $\beta$ is a unitless temperature-dependence parameter, and $\gamma$ is the activation energy in K.}
    
    \item[3] The \ce{CH + CH3CCH} and \ce{CH + CH2CCH2} reactions have been shown by \citet{goulay_cyclic_2009} to yield \ce{CH2CCCH2}, \ce{CH2CHC2H}, and other isomers of \ce{C4H4}. Since \ce{CH2CHC2H} is the only \ce{C4H4} isomer  currently included in the network, all of the pathways for these reactions have been combined into the formation of this isomer.

\end{tablenotes}

\end{threeparttable}
\end{table}

\begin{table}[htbp]
    \scriptsize
    \caption{\textbf{Related Neutral-Neutral Reactions Continued}}
    \label{tab:Neutral-neutral2}
    \begin{threeparttable}
    \centering
    \begin{tabularx}{\textwidth}{@{}
    l
    S[table-format=-1.2e-2, table-alignment-mode=none]
    S[table-format=-1.2, table-alignment-mode=none]
    S[table-format=4, table-alignment-mode=none]
    S[table-format=1.2e-2, table-alignment-mode=none]
    l@{}}
    \toprule
    {Reactions\tnote{1}} & {$\alpha$ {(\unit{\cubic\centi\metre\per\second})}} & {$\beta$} & {$\gamma$ (\unit{\kelvin})} & {k (10 \unit{\kelvin}) (\unit{\cubic\centi\metre\per\second})\tnote{2}} & Reference\\
    \midrule
    \sidehead{\ce{CH2CCH} Reactions}
    {*\ce{CCH + CH3 -> H + CH2CCH}} & 1.00e-10 & 0 & 0 & 1.00e-10 & {Rate estimated by \citet{loison_interstellar_2017} with} \\
    &&&&& {H-elimination assumed as major product channel.} \\
    &&&&& \\
    {*\ce{C2 + CH4 -> CH2CCH + H}} & 1.30e-11 & 0 & 0 & 1.30e-11 & {23 K rate constant from \citet{canosa_experimental_2007},} \\
    &&&&& {assuming H-elimination product channel.} \\
    &&&&& \\
    {\ce{C + CH2CCH -> H + C4H2}} & 2.00e-10 & 0 & 0 & 2.00e-10 & {Estimated in \citet{loison_interstellar_2017} based on} \\
    &&&&& {capture theory approach described in \citet{loison_gas-phase_2014}.} \\
    &&&&& \\
    {\ce{N + CH2CCH -> H2 + HC3N}} & 5.00e-11 & 0 & 0 & 5.00e-11 & {Rate and products estimated by \citet{loison_interstellar_2017}} \\
    {*\ce{N + CH2CCH -> C2H2 + HCN}} & 5.00e-11 & 0 & 0 & 5.00e-10 & {based on \ce{N + C2H3} reaction.} \\
    &&&&& \\
    {\ce{O + CH2CCH -> CO + C2H3}} & 5.00e-11 & 0 & 0 & 5.00e-11 & {Temperature-independent Rate constant from} \\
    {\ce{O + CH2CCH -> HCO + C2H2}} & 2.00e-11 & 0 & 0 & 2.00e-11 & {\citet{slagle_kinetics_1991}, with branching ratios estimated} \\
    {*\ce{O + CH2CCH -> OH + c-C3H2}} & 1.00e-11 & 0 & 0 & 1.00e-11 & {by \citet{loison_interstellar_2017} based on calculations} \\
    {*\ce{O + CH2CCH -> HCCCHO + H}}\tnote{3} & 1.60e-10 & 0 & 0 & 1.60e-10 & {by \citet{lee_ab_2006}.} \\
    &&&&& \\
    {*\ce{OH + CH2CCH -> C2H3 + HCO}} & 6.00e-11 & 0 & 0 & 6.00e-11 & {Rate constant from \citet{hansen_isomer-specific_2009} and 50/50} \\
    {*\ce{OH + CH2CCH -> C2H4 + CO}} & 6.00e-11 & 0 & 0 & 6.00e-11 & {branching ratios assumed from \citet{loison_interstellar_2017}.} \\
    &&&&& \\
    *\ce{CN + CH2CCH -> CNCH2CCH + Photon} & 1.00e-10 & 0 & 0 & 1.00e-10 & {Estimated from \citet{tennis_radiative_2021}} \\
    *\ce{CH2CCH + CH2CCH -> C6H5 + H} & 4.00e-11 & 0 & 0 & 4.00e-11 & {Room temperature rate constant from}\\
    &&&&& {\citet{atkinson_rate_1999}, \citet{fahr_kinetics_2000}, and} \\
    &&&&& {\citet{desain_infrared_2003}.} \\
    \sidehead{Removed Reactions}
    \ce{CH2CCH + H -> H2 + c-C3H2} & 0 & 0 & 0 & 0 & {Removed due to sizeable energy barrier as calculated} \\
    \ce{CH2CCH + H -> H2 + l-C3H2} & 0 & 0 & 0 & 0 & {by \citet{miller_multiple-well_2003}} \\
    \bottomrule
\end{tabularx}

\begin{tablenotes}

    \item[1] An asterisk preceding a reaction denotes that the reaction has been added to the network rather than modified.
    
    \item[2] The rates of the listed neutral-neutral reactions are computed using the modified Arrhenius equation: $k = \alpha\left(T/300\right)^{\beta}e^{-\gamma/T}$ {, where $\alpha$ is a temperature-independent prefactor, $\beta$ is a unitless temperature-dependence parameter, and $\gamma$ is the activation energy in K.}
    
    \item[3] Since the molecule HCCCHO is currently not included in our network, we substitute \ce{C3O + H + H2} as the products for this reaction. These do not lead directly back to \ce{CH2CCH}, avoiding a net zero change in abundance.

\end{tablenotes}

\end{threeparttable}
\end{table}

\begin{table}[htbp]
    \small
    \caption{\textbf{Related Neutral-Neutral Reactions on Grains}}
    \label{tab:Grain}
    \begin{threeparttable}
    \centering
    \begin{tabularx}{\textwidth}{@{}
    l
    S[table-format=-1.2, table-alignment-mode=none]
    S[table-format=4, table-alignment-mode=none]
    l@{}}
    \toprule
    {Reactions{\tnote{1}}} & {$\alpha${\tnote{2}}} & {$E_a$} &  {Reference} \\
    \midrule
    {\ce{s-H + s-C3 -> s-c-C3H}} & 0.60 & 0 & \citet{loison_interstellar_2017} \\
    {\ce{s-H + s-C3 -> s-l-C3H}} & 0.40 & 0 & \citet{loison_interstellar_2017} \\
    {\ce{s-H + s-c-C3H -> s-c-C3H2}} & 0.80 & 0 & \citet{loison_interstellar_2017} \\
    {\ce{s-H + s-c-C3H -> s-l-C3H2}} & 0.20 & 0 & \citet{loison_interstellar_2017} \\
    {\ce{s-H + s-l-C3H -> s-c-C3H2}} & 0.80 & 0 & \citet{loison_interstellar_2017} \\
    {\ce{s-H + s-l-C3H -> s-l-C3H2}} & 0.20 & 0 & \citet{loison_interstellar_2017} \\
    {\ce{s-H + s-c-C3H2 -> s-CH2CCH}} & 1.00 & 0 & \citet{hickson_methylacetylene_2016} \\
    {\ce{s-H + s-l-C3H2 -> s-CH2CCH}} & 1.00 & 0 & \citet{loison_interstellar_2017} \\
    {\ce{s-H + s-CH2CCH -> s-CH3CCH}} & 0.50 & 0 & \citet{loison_interstellar_2017} \\
    {\ce{s-H + s-CH2CCH -> s-CH2CCH2}} & 0.50 & 0 & \citet{loison_interstellar_2017} \\
    {\ce{s-H2 + s-l-C3H -> s-CH2CCH}} & 1.00 & 3600 & \citet{hickson_methylacetylene_2016} \\
    \bottomrule
    \end{tabularx}

\begin{tablenotes}
    \item[1] {In addition to these reactions forming grain products, the model also accounts for chemical desorption of 1\% to form gas-phase products.}
    \item[2] The rate of reaction between two grain species $i$ and $j$ is calculated using the equation: $k=\alpha\kappa_{ij}\left(\dfrac{1}{t_{hop}(i)} + \dfrac{1}{t_{hop}(j)}\right)\dfrac{1}{N_{site}n_{dust}}$, where {$\alpha$ is the branching ratio and} $\kappa_{ij}$ is a parameter that depends on the activation barrier. For more information, refer to \citet{ruaud_gas_2016}
\end{tablenotes}
\end{threeparttable}
\end{table}

\clearpage

\section{Destruction rates of \ce{CH2CCH}, \ce{CH3CCH}, and \ce{CH2CCH2}}
Figure~\ref{fig:dest_rates} displays the rates of reaction for the major destruction mechanisms of \ce{CH2CCH}, \ce{CH3CCH}, and \ce{CH2CCH} as a function of time.

\begin{figure}[!hb]
    \centering
    \gridline{\fig{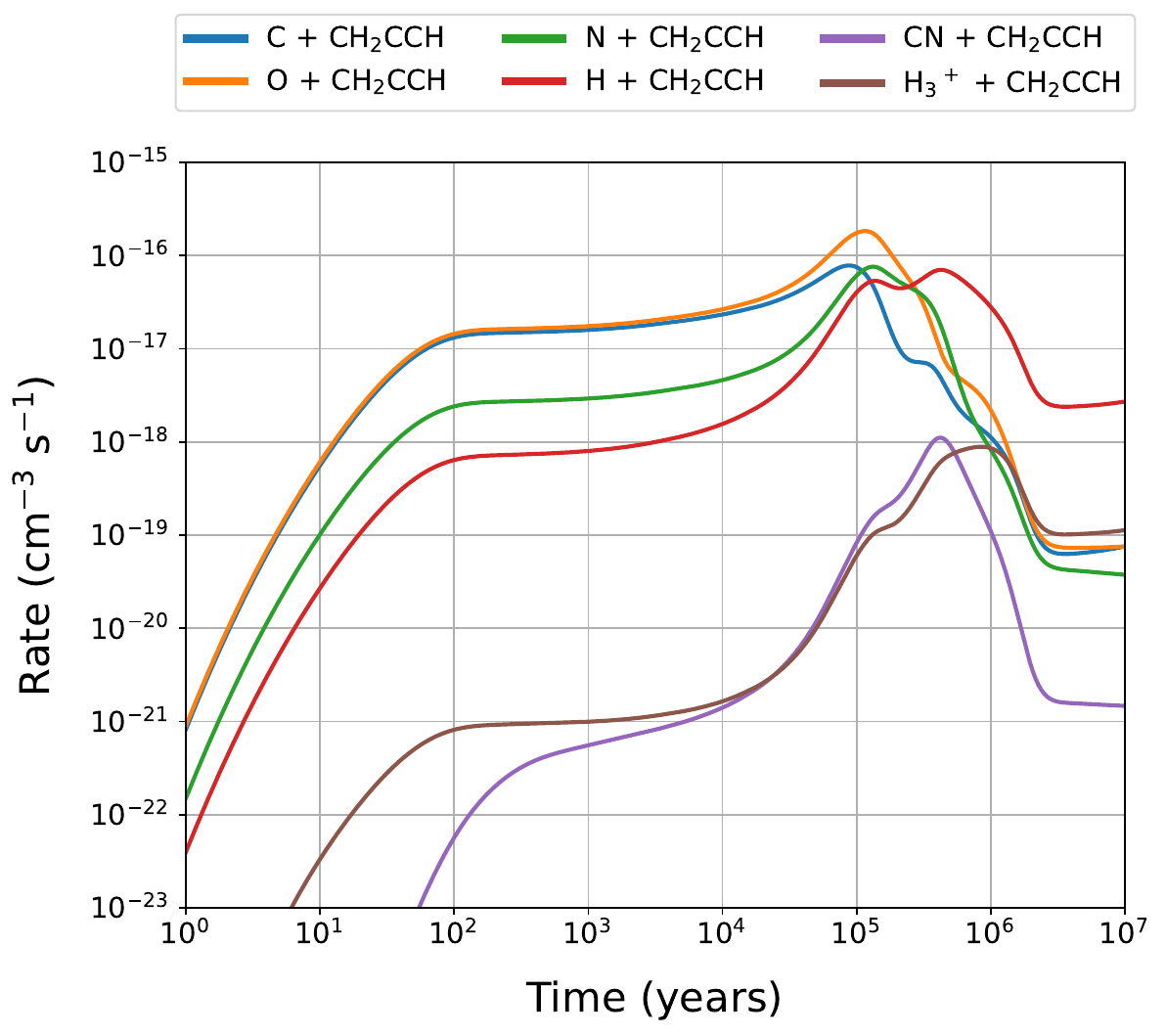}{0.5\textwidth}{(a)}
    \fig{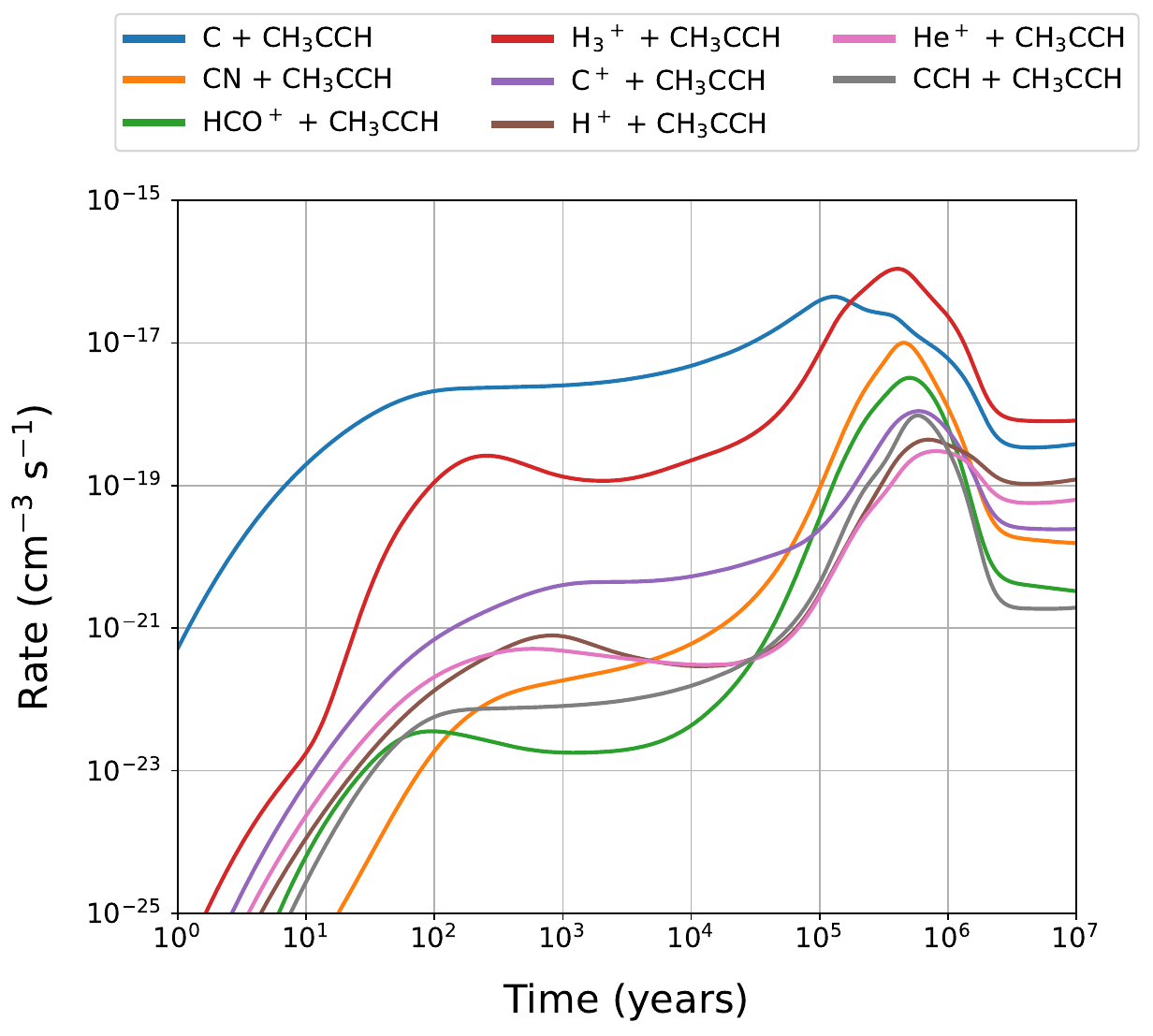}{0.5\textwidth}{(b)}}
    \gridline{\fig{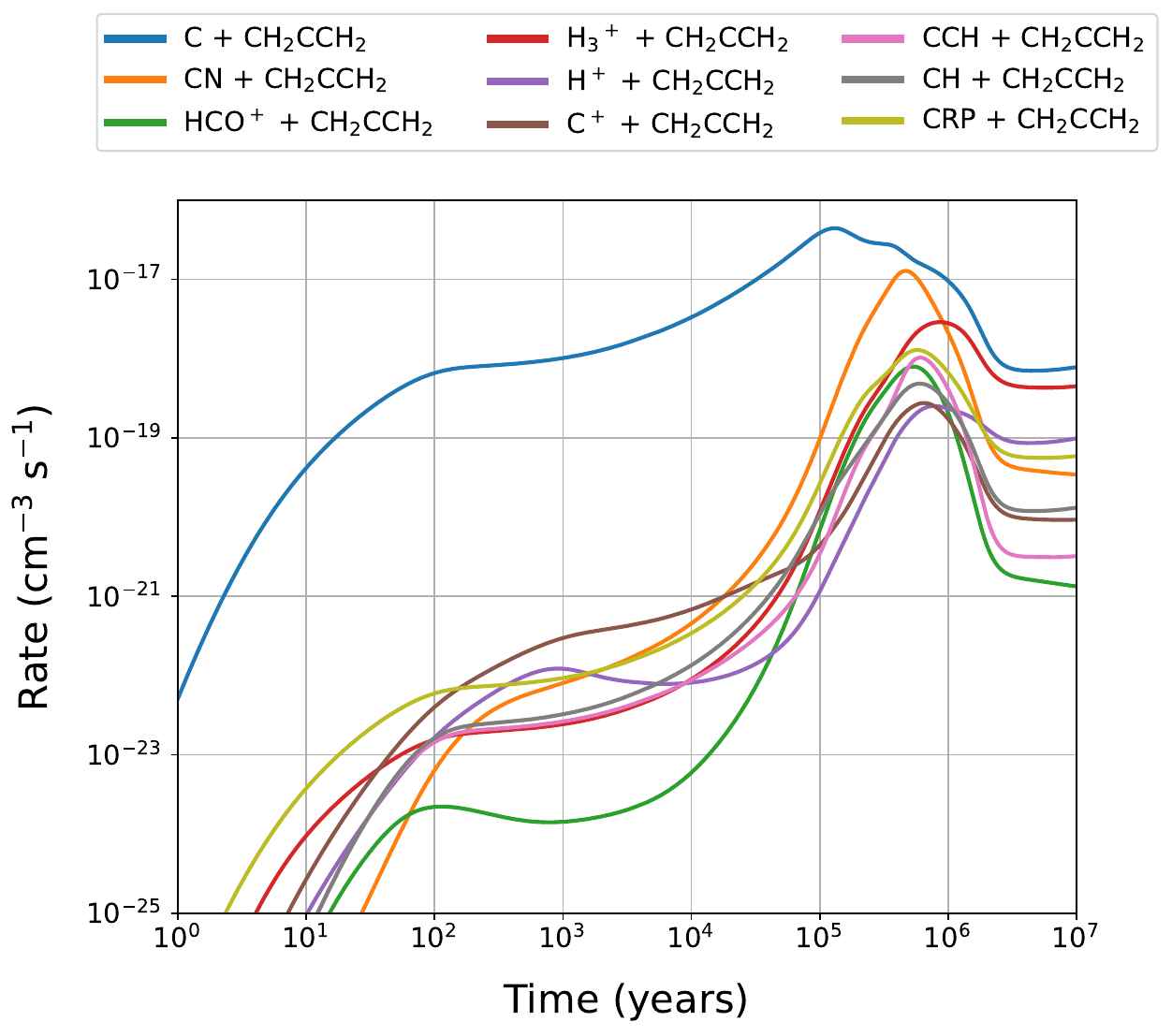}{0.5\textwidth}{(c)}}
    \caption{Reaction rates for key destruction pathways of (a) \ce{CH2CCH}, (b) \ce{CH3CCH}, and (c) \ce{CH2CCH2} as a function of time. For more information on these reactions, including product channels, refer to Tables \ref{tab:DR}-\ref{tab:Grain}.}
    \label{fig:dest_rates}
\end{figure}

\clearpage

\section{Sensitivity Analyses of Remaining Reactions}
Figures \ref{fig:SI_sens} and \ref{fig:prop_recomb_sens} display the sensitivity analysis results not shown in the manuscript. For each reaction, the rate constant was multiplied by a number of factors and a model performed for each factor.

\begin{figure}
    \centering
    \gridline{\fig{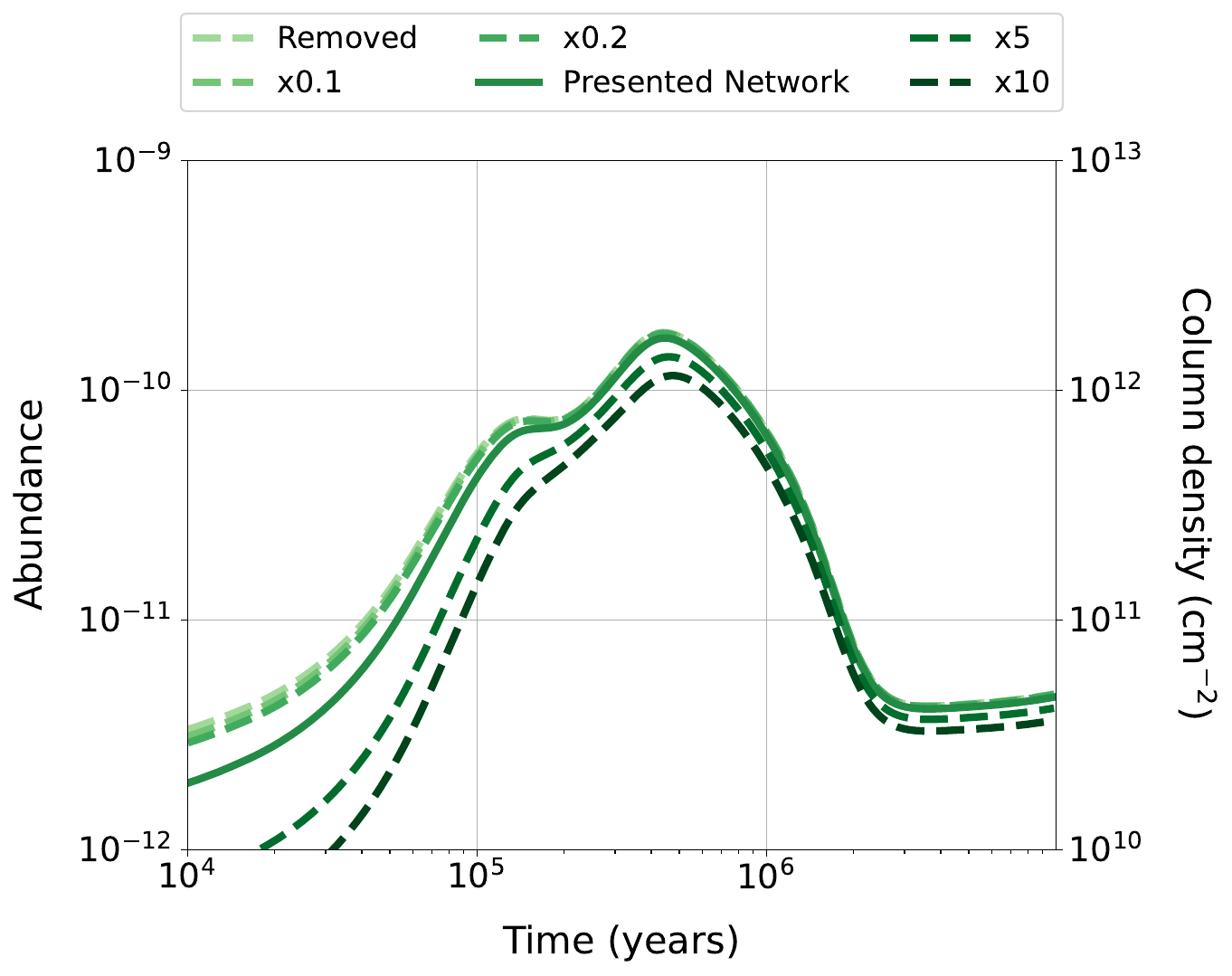}{0.4\textwidth}{(a)}
    \fig{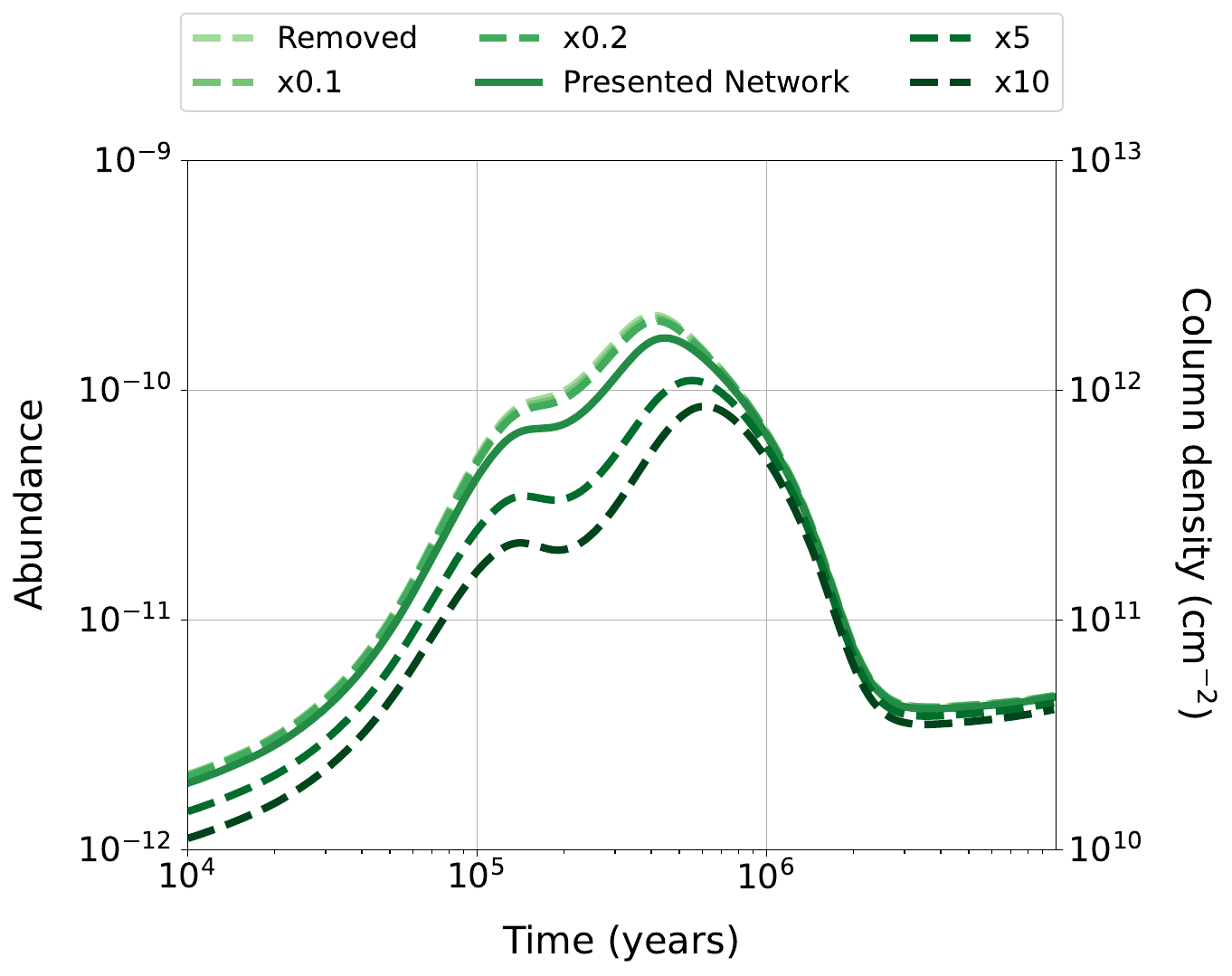}{0.4\textwidth}{(b)}}
    \gridline{\fig{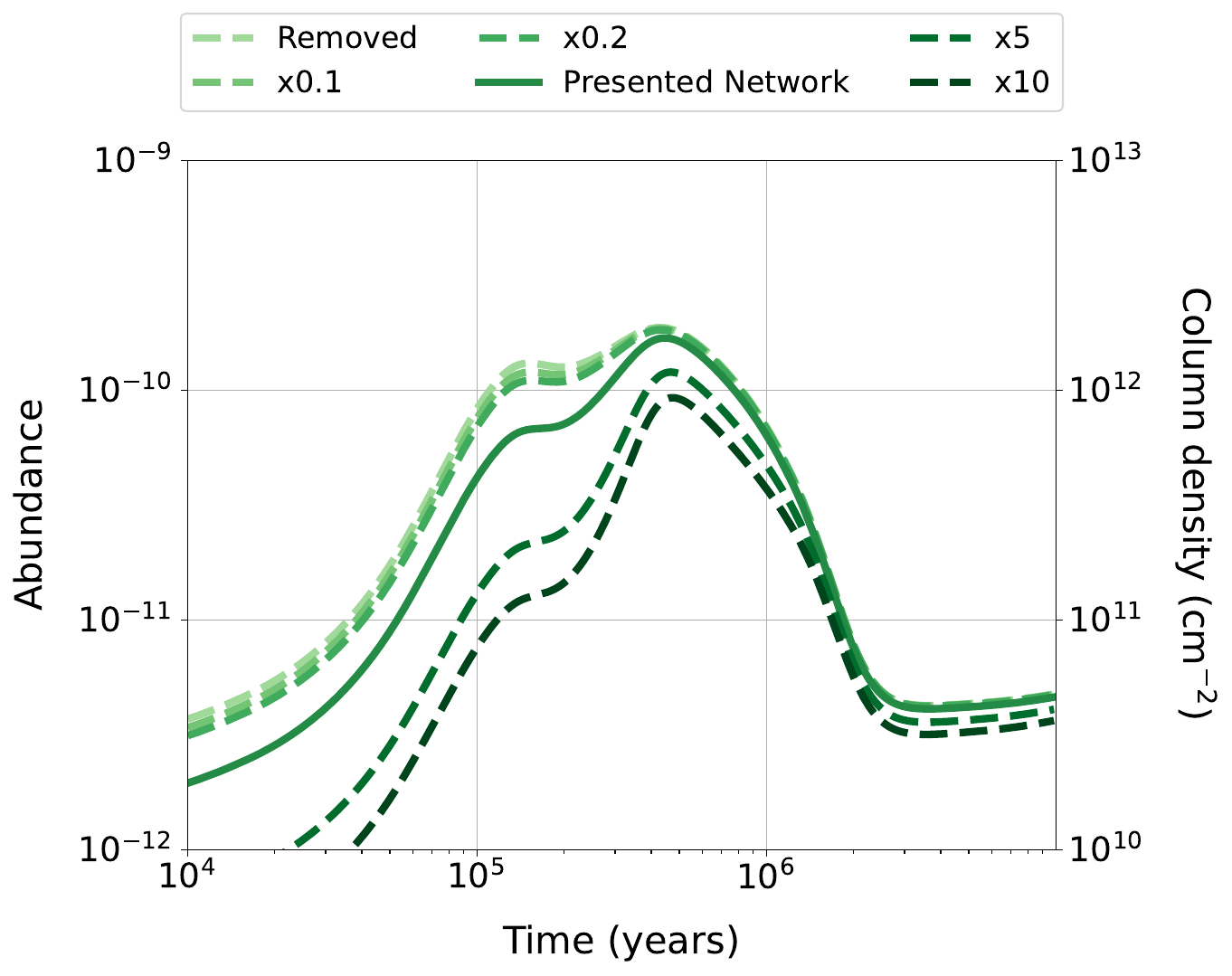}{0.4\textwidth}{(c)}
    \fig{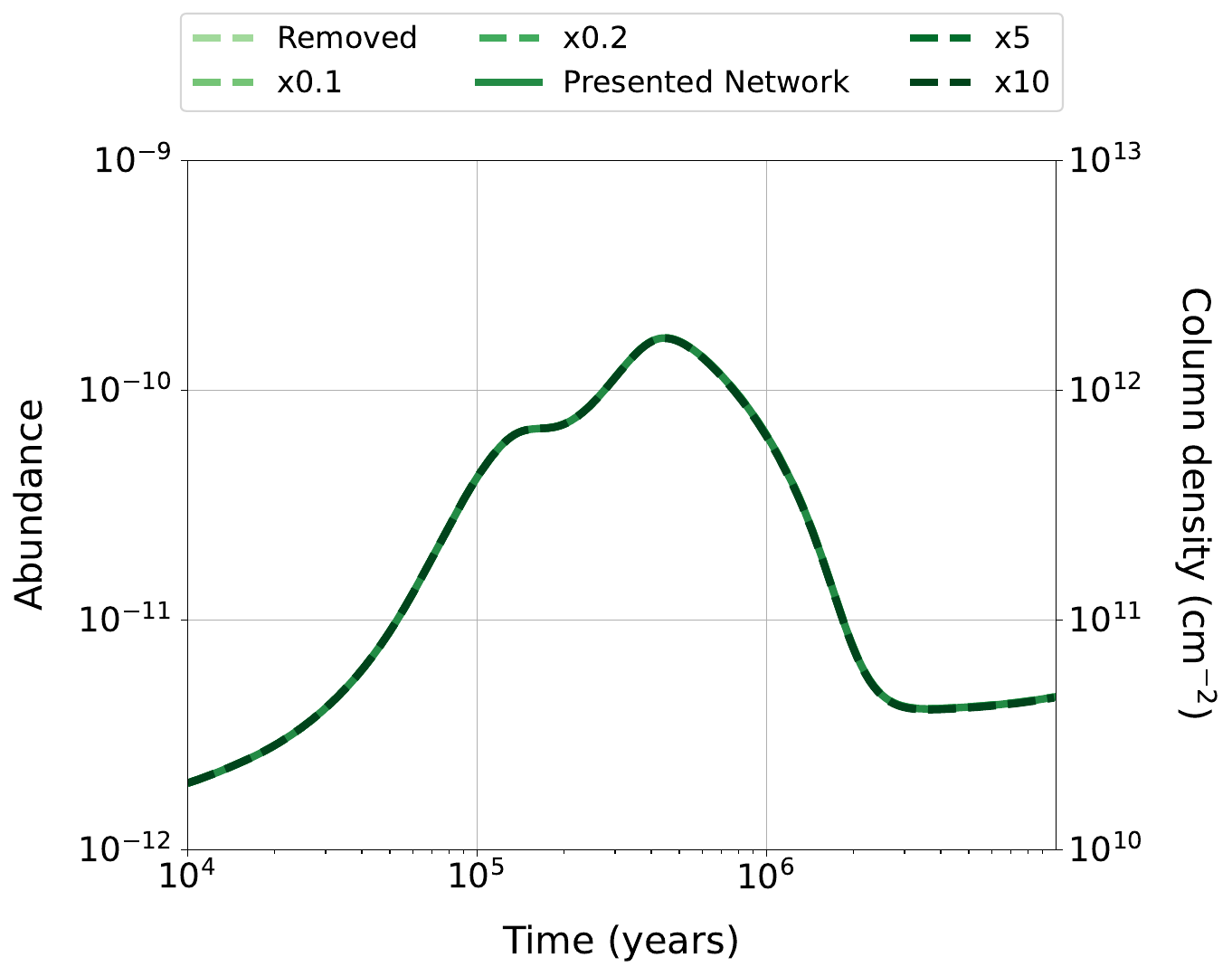}{0.4\textwidth}{(d)}}
    \gridline{\fig{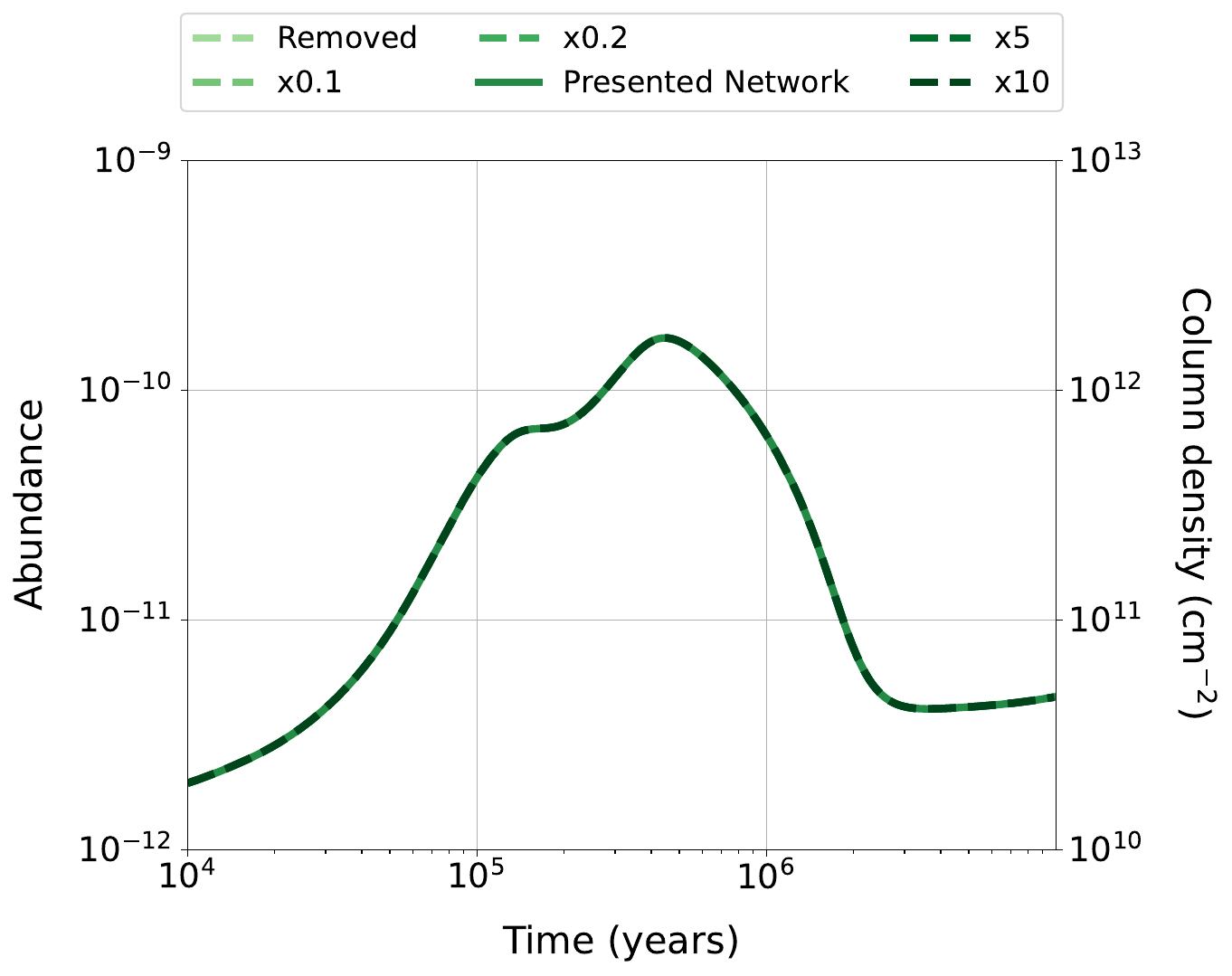}{0.4\textwidth}{(e)}
    \fig{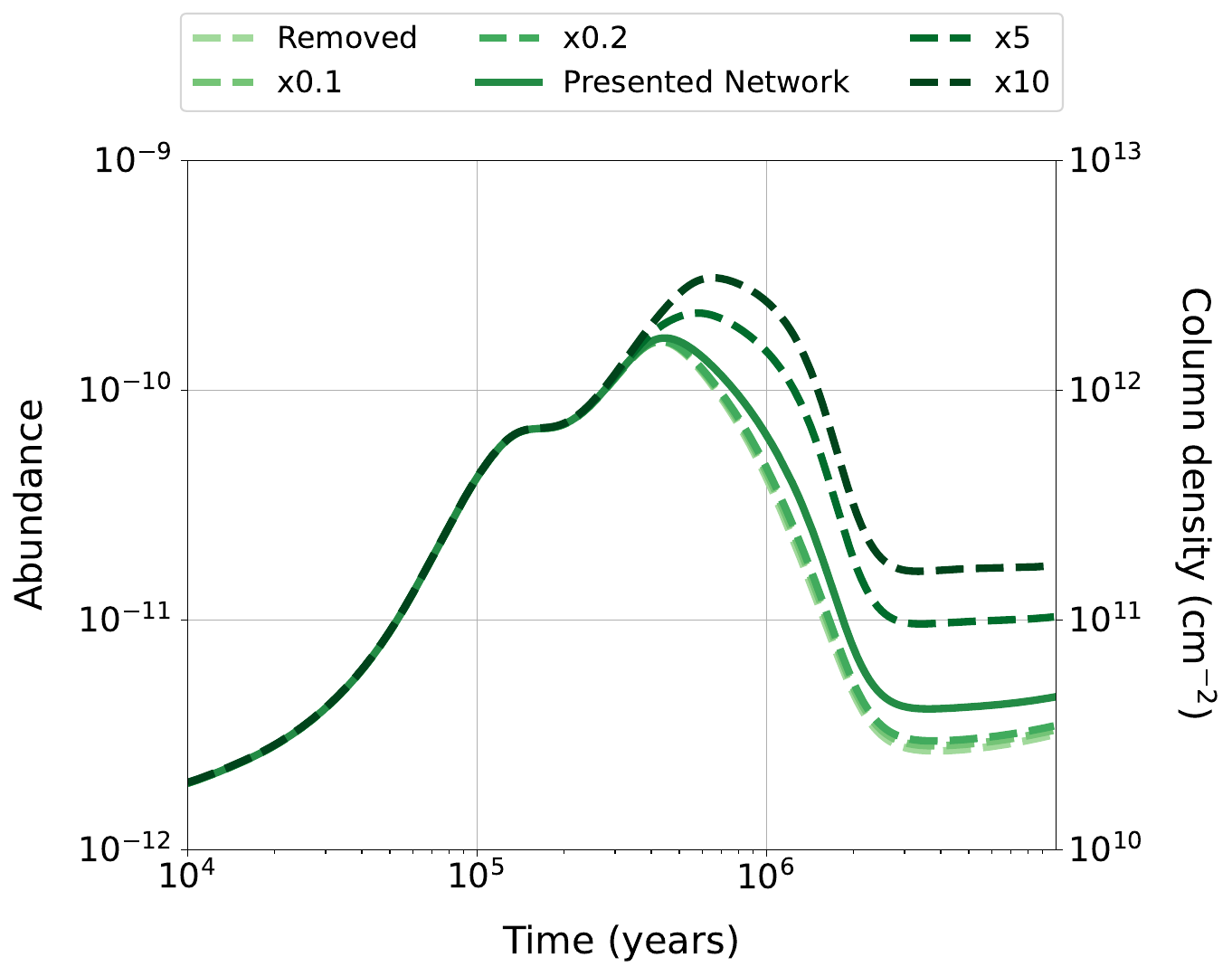}{0.4\textwidth}{(f)}}
    \caption{Modeled abundances and column densities of \ce{CH2CCH} for different values of the (a) \ce{CH2CCH + C}, (b) \ce{CH2CCH + N}, (c) \ce{CH2CCH + O}, (d) \ce{CH2CCH + OH}, (e) \ce{CH2CCH + CN}, and (f) \ce{CCH + CH3} rate constants. The dashed lines are networks where this rate constant has been modified by the factors in the legend. The base network (presented in this paper) is shown for comparison as a solid line, as well as a network where this reaction is removed entirely. The color gradient corresponds with the magnitude of the factor of change.}
    \label{fig:SI_sens}
\end{figure}

\begin{figure}
    \centering
    \gridline{\fig{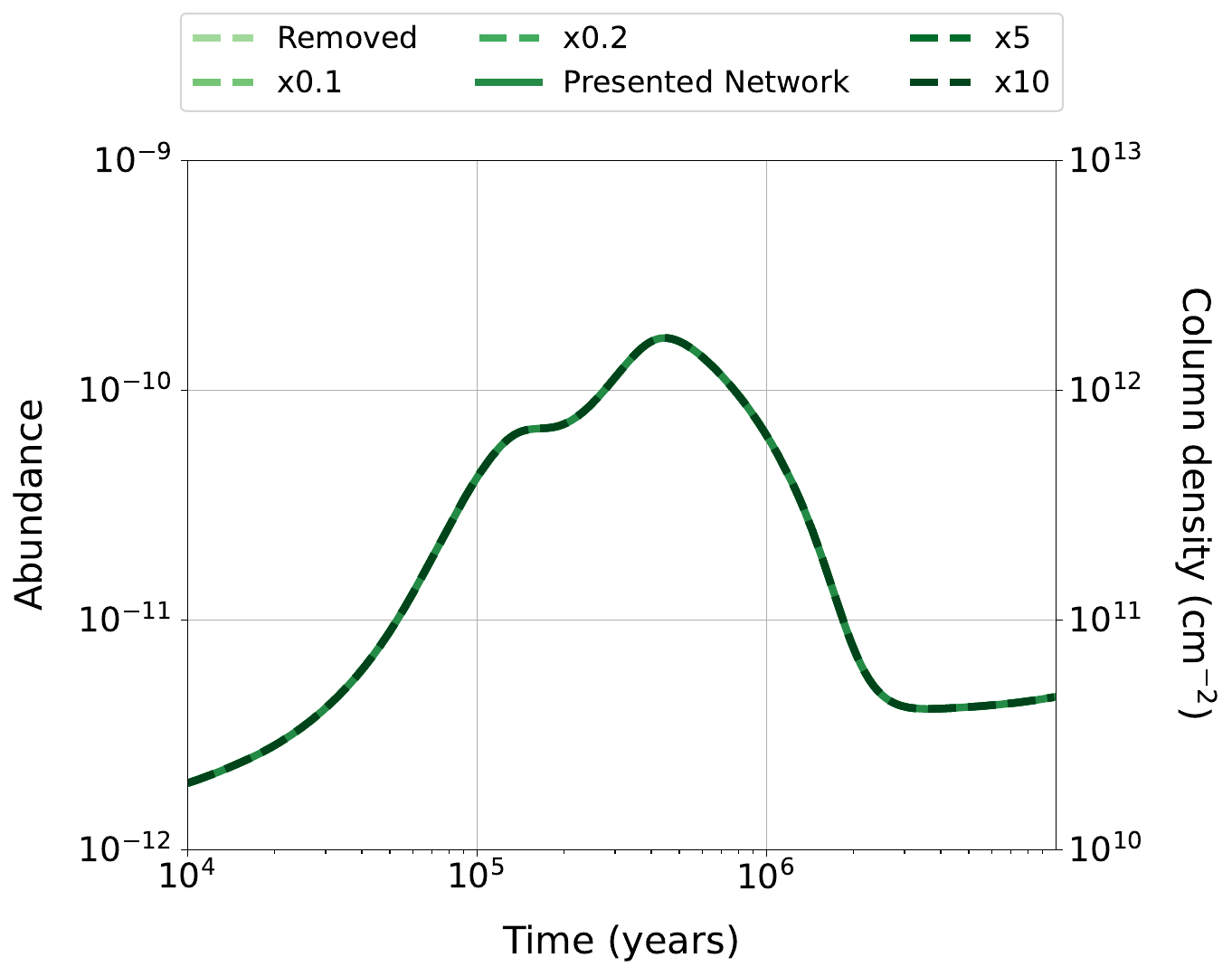}{0.5\textwidth}{(a)}
    \fig{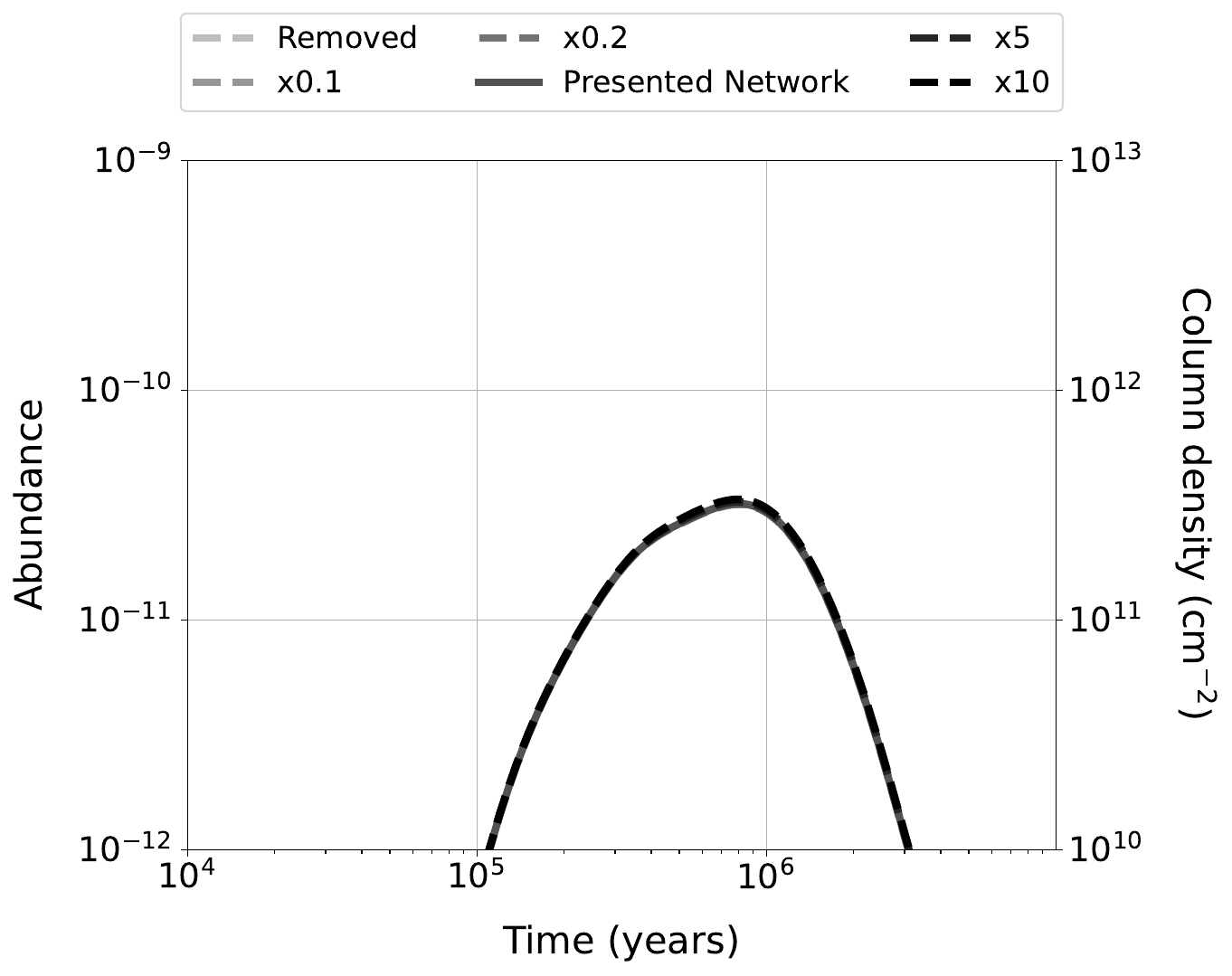}{0.5\textwidth}{(b)}}
    \caption{Modeled abundances and column densities of (a) \ce{CH2CCH} and (b) \ce{C6H5} for different values of the \ce{CH2CCH + CH2CCH} rate constant. {The dashed lines are networks where this rate constant has been modified by the factors in the legend. The base network (presented in this paper) is shown for comparison as a solid line, as well as a network where this reaction is removed entirely. The color gradient corresponds with the magnitude of the factor of change.}}
    \label{fig:prop_recomb_sens}
\end{figure}

\begin{figure}
    \centering
    \gridline{\fig{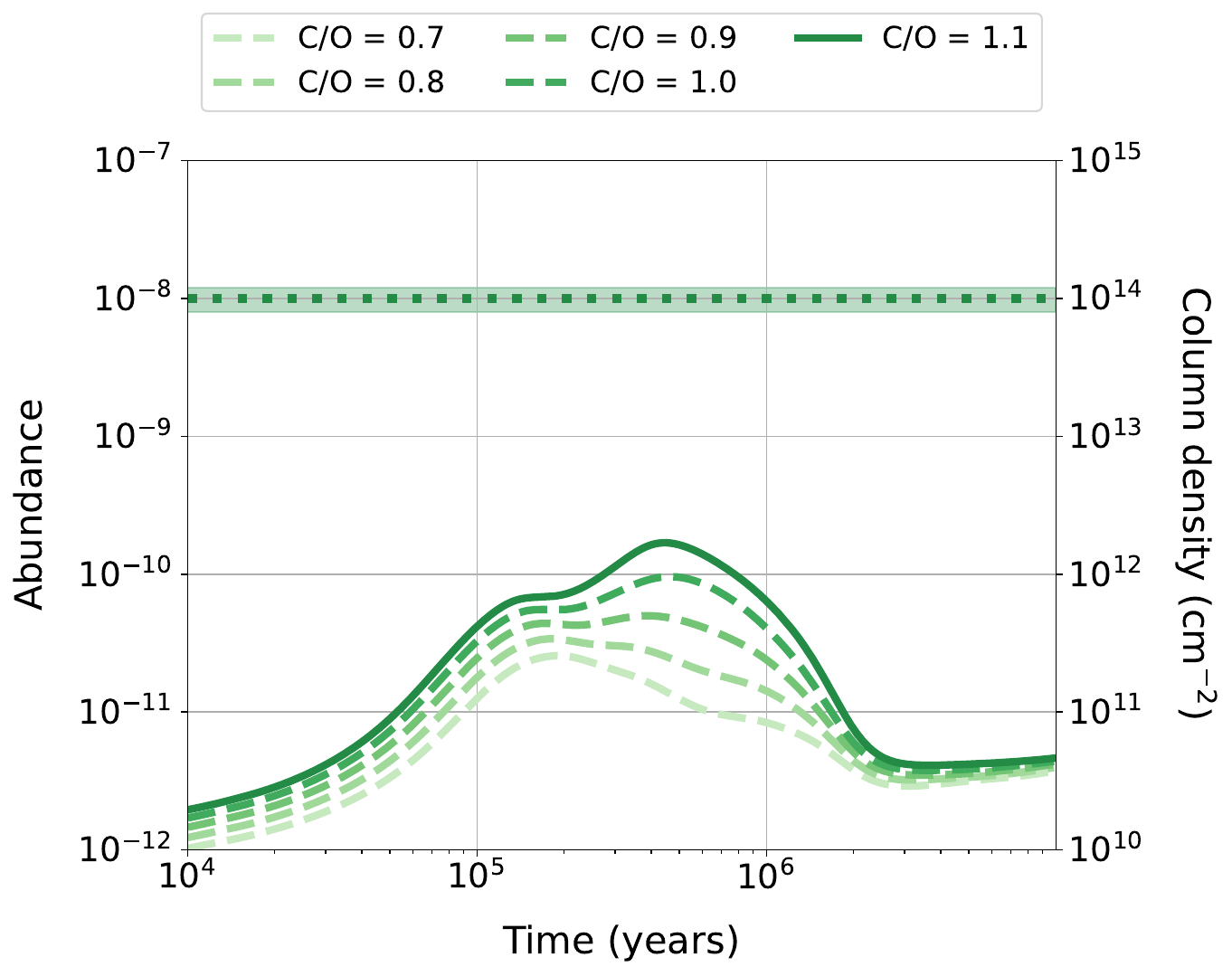}{0.5\textwidth}{(a)}
    \fig{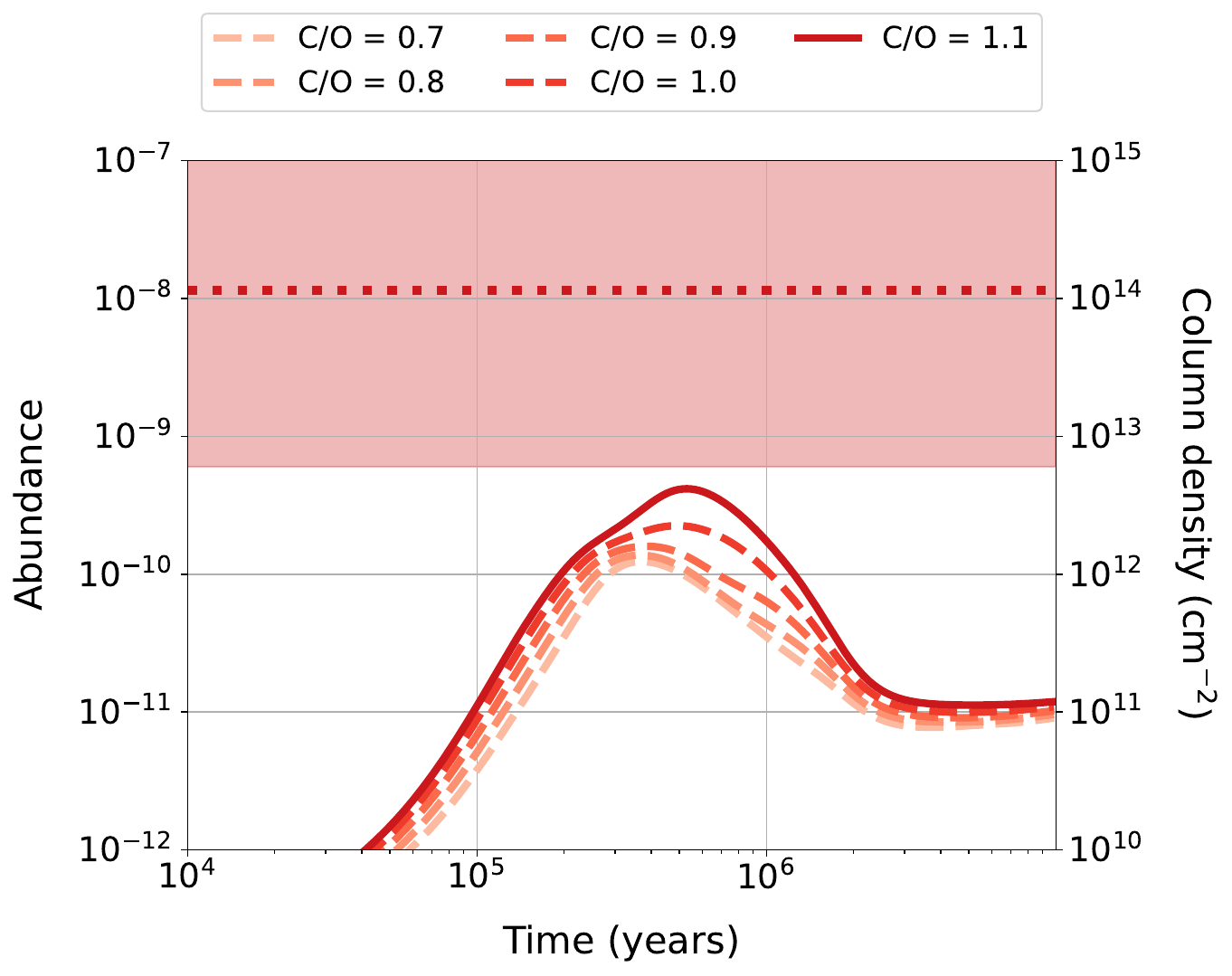}{0.5\textwidth}{(b)}}
    \gridline{\fig{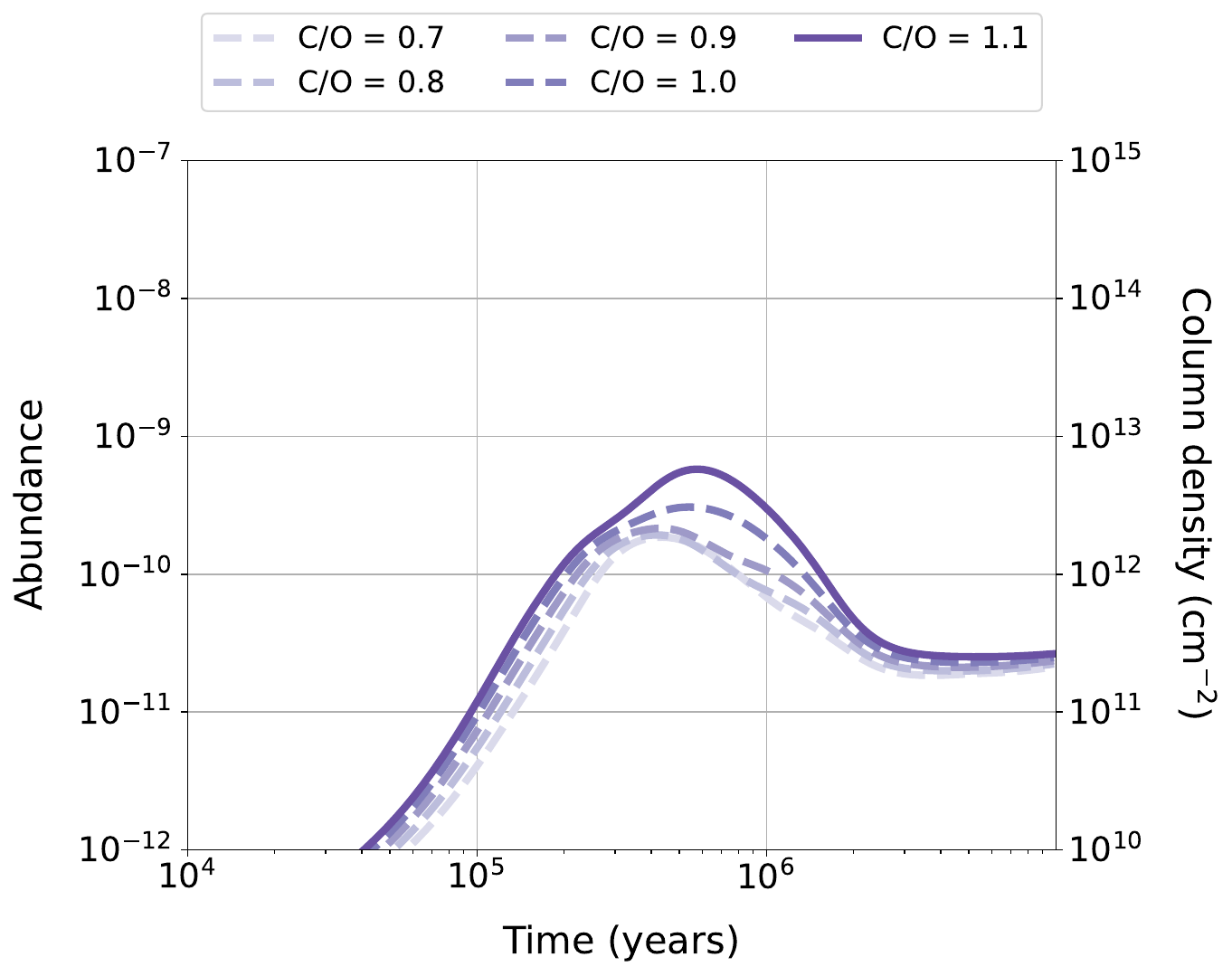}{0.5\textwidth}{(c)}}
    \caption{Modeled abundances and column densities of (a) \ce{CH2CCH}, (b) \ce{CH3CCH}, and (c) \ce{CH2CCH2} for different initial abundances of elemental oxygen. The solid line indicates the base model with a C/O ratio of 1.1, while the dashed lines indicate models with variations in C/O ratio. The color gradient corresponds with the magnitude of the C/O ratio. The dotted line represents the observed column density in TMC-1, with the shaded area signifying an error of 1 $\sigma$}
    \label{fig:COtest}
\end{figure}

\end{document}